\documentclass[iop,apjl,numberedappendix,appendixfloats,floatfix]{emulateapj}
\usepackage{lineno}
\usepackage{color}
\usepackage{amsmath}
\usepackage{url}
\usepackage{yfonts}
\usepackage[T1]{fontenc}

\newcommand{\solar}{$_{\odot}$}

\newcommand{\tco}{$^{12}$CO}
\newcommand{\ttco}{$^{13}$CO}
\newcommand{\ceto}{C$^{18}$O}
\newcommand{\hcop}{HCO$^+$}

\newcommand{\joz}{$J$=1$\rightarrow$0}

\newcommand{\kms}{\,km\,s$^{-1}$}
\newcommand{\degree}{$^{\circ}$}
\newcommand{\fdeg}{$^{\circ}$\hspace{-1mm}.}

\newcommand{\fsec}{$''$\hspace{-1mm}.}

\newcommand{\tex}{$T_{\rm ex}$}

\newcommand{\vlsr}{$V_{\rm LSR}$}
\newcommand{\htwo}{H$_2$}
\newcommand{\hii}{H\,{\sc ii}}

\newcommand{\ico}{$I_{\rm ^{12}CO}$}

\def\lapp{\ifmmode\stackrel{<}{_{\sim}}\else$\stackrel{<}{_{\sim}}$\fi}
\def\gapp{\ifmmode\stackrel{>}{_{\sim}}\else$\stackrel{>}{_{\sim}}$\fi}

\shorttitle{Magnetic Fields and Gas Structures in BYF\,73}
\shortauthors{Barnes et al.}

\begin{document}

\title{SOFIA and ALMA Investigate Magnetic Fields and Gas Structures in \\ Massive Star Formation:  The Case of the Masquerading Monster in BYF\,73}

\email{pbarnes@spacescience.org}

\author{Peter J. Barnes\altaffilmark{1,2}, Stuart D. Ryder\altaffilmark{3,4}, Giles Novak\altaffilmark{5,6}, Richard M. Crutcher\altaffilmark{7}, \\
Laura M. Fissel\altaffilmark{8}, Rebecca L. Pitts\altaffilmark{9}, and William J. Schap III\altaffilmark{10} \\
}

\altaffiltext{1}{Space Science Institute, 4765 Walnut St. Suite B, Boulder, CO 80301, USA}
\altaffiltext{2}{School of Science and Technology, University of New England, Armidale NSW 2351, Australia}
\altaffiltext{3}{School of Mathematical \& Physical Sciences, Macquarie University, NSW 2109, Australia}
\altaffiltext{4}{Astronomy, Astrophysics and Astrophotonics Research Centre, Macquarie University, NSW 2109, Australia}
\altaffiltext{5}{Center for Interdisciplinary Exploration \& Research in Astrophysics (CIERA), 1800 Sherman Ave., Evanston, IL 60201, USA} %
\altaffiltext{6}{Dept.\ of Physics and Astronomy, Northwestern University, 2145 Sheridan Rd., Evanston, IL 60208, USA}
\altaffiltext{7}{Dept.\ of Astronomy, University of Illinois, 1010 W. Green St., Urbana, IL 60801, USA}
\altaffiltext{8}{Dept.\ of Physics, Engineering Physics \& Astronomy, Queen's University, 64 Bader Lane, Kingston, ON, K7L 3N6, Canada}
\altaffiltext{9}{Niels Bohr Institute, Centre for Star \& Planet Formation, University of Copenhagen, \O ster Voldgade 5-7, 1350 Copenhagen K, Denmark}
\altaffiltext{10}{Astronomy Dept., University of Florida, P.O. Box 112055, Gainesville, FL 32611, USA}

\begin{abstract}
\hspace{5mm}We present SOFIA+ALMA continuum and spectral-line polarisation data on the massive molecular cloud BYF\,73, revealing important details about the magnetic field morphology, gas structures, and energetics in this unusual massive star formation laboratory.  The 154$\mu$m HAWC+ polarisation map finds a highly organised magnetic field in the densest, inner 0.55$\times$0.40\,pc portion of the cloud, compared to an unremarkable morphology in the cloud's outer layers.  
The 3mm continuum ALMA polarisation data reveal several more structures in the inner domain, including a pc-long, $\sim$500\,M\solar\ ``Streamer'' around the central massive protostellar object MIR\,2, with magnetic fields mostly parallel to the east-west Streamer 
but oriented north-south across MIR\,2.  
The magnetic field orientation changes from mostly parallel to the column density structures to mostly perpendicular, at thresholds $N_{\rm crit}$ = 6.6$\times$10$^{26}$\,m$^{-2}$, $n_{\rm crit}$ = 2.5$\times$10$^{11}$\,m$^{-3}$, and $B_{\rm crit}$ = 42$\pm$7\,nT. 
ALMA also mapped Goldreich-Kylafis polarisation in \tco\ across the cloud, which traces in both total intensity and polarised flux, a powerful bipolar outflow from MIR\,2 that interacts strongly with the Streamer.  
The magnetic field is also strongly aligned along the outflow direction;  
energetically, it 
may dominate the outflow near MIR\,2,  
comprising rare evidence for a magnetocentrifugal origin to such outflows.  
A portion of the Streamer may be 
in Keplerian rotation around MIR\,2, 
implying a gravitating mass 1350$\pm$50\,M\solar\ for the protostar+disk+envelope; alternatively, these kinematics can be explained by gas in free fall towards a 950$\pm$35\,M\solar\ object.  
The high accretion rate onto MIR\,2 apparently occurs through the Streamer/disk, and could account for $\sim$33\% of MIR\,2's total luminosity via gravitational energy release. 
\end{abstract}


\section{Introduction}

Magnetic fields (hereafter $B$ fields) in astrophysical settings are very widespread and may play an important role in the evolution of the interstellar medium (ISM), stars, galaxies, and the universe.  Yet, they are technically challenging to measure, limiting our ability to understand the full physics within these settings.  This is because $B$ field measurements depend on accurate values for the polarised contributions to emission or absorption (e.g., the Stokes parameters $Q$, $U$, $V$), which are usually much weaker than the total intensity $I$, and then interpreting the data in terms of particular physical polarisation mechanisms, e.g., as explained by \citet{cru12} or \citet{BL15}.

In star formation (SF), the role and importance of $B$ fields is a long-standing problem \citep{mo07,cru12}.  This is largely due to observational challenges of high-quality $B$ field measurements in large cloud samples at high spatial dynamic range (SDR), and relating these to the clouds' other physical conditions.  Prior work on the Zeeman effect shows that, below a threshold density $n_0$ $\approx$ 300\,cm$^{-3}$, $B$ fields can support gas against gravity and have fairly uniform strength.  Above this level, studies suggest the line-of-sight component increases with density, $B_{||} \propto n^{0.65}$, and the ratio of magnetic to gravitational forces is close to critical \citep{cru12}.

Confirming the higher-density behaviour is important to SF theory, since SF is not observed in low-density gas \citep{L15}.  Tracking local variations in the transition density $n_0$ is also significant, since this could change the SF efficiency and/or initial mass function.  Catching massive protostars, especially, in the act of formation is even more difficult compared to low-mass protostars, because of their greater distances, accelerated timescales, and rapid alteration of initial conditions.

The plane-of-sky component $B_{\perp}$ has recently begun to be mapped at high SDR via linear polarisation of mm--$\mu$m continuum emission or absorption \citep[e.g.,][]{pc16}.  This probably arises from non-spherical dust grains aligned by radiative torques to the $B$ field: while not all alignment mechanisms are magnetic, non-magnetic mechanisms are not thought to be dominant \citep{L07}.  If the alignment {\em is} magnetic, statistical methods can convert turbulent variations in field orientation $\theta_{B_{\perp}}$ to estimates of $|B_{\perp}|$ 
\citep[][hereafter DCF]{d51,cf53}.  Although approximate, DCF methods have been effectively used from cloud (10\,pc) to core (0.1\,pc) scales \citep{mg91,BL15} to meaningfully constrain the importance of $B$ fields in different situations.  

Large-scale maps of FIR/submm polarisation from {\em Planck} and other missions coupled with new analysis methods and high-quality molecular gas data \citep{faa16,saa17,LY18} permit new insights into the role of $B$ fields in SF.  In Vela C, for example, the alignment of $\theta_{B_{\perp}}$ with dense structures changes from parallel to perpendicular near the same threshold $n_0$ as in the Zeeman data \citep{faa19}.  However, data on massive cluster-scale clumps, where most massive protostars likely also form, are very sparse: we need to precisely measure both $|B|$ and $n$ in a wider variety of clouds and environments to test these results.

As part of a long-term project to systematically investigate the physics of $B$ fields and dense gas in a uniform sample of CN-bright, massive molecular clumps that are likely sites of high-mass star formation \citep{sh19}, we obtained observing time with both the Stratospheric Observatory For Infrared Astronomy (SOFIA) and Atacama Large Millimeter/submillimeter Array (ALMA) to map the first few targets in this sample.  We used the polarimetric far-infrared (FIR) High-resolution Airborne Wideband Camera-plus \citep[HAWC+;][]{hrd18} aboard SOFIA and ALMA's full-polarisation mode in both the 3\,mm continuum and spectral line observations.

We report here the first results for this project, an analysis of the $B$ field properties in the molecular cloud BYF\,73 with the most massive protostellar inflow rate known \citep{b10}, and following up recent multi-wavelength work on the same cloud \citep[][hereafter P18]{p18} from {\em Gemini} with T-ReCS, SOFIA with FIFI-LS, and ATCA.  P18 found that, of the 8 mid-IR point sources imaged with T-ReCS, MIR 2 seems to be the overwhelmingly dominant protostellar source in terms of mass (240\,M\solar) and luminosity (4700\,L\solar), yet comprises only $\sim$1\% of the cloud mass.  After ruling out gravitational energy release from the inflow and other forms of mechanical or thermal energy, it was not clear what MIR\,2's energy source is.  MIR 2 also seems remarkably young, perhaps only 7000 yr old at the very high mass accretion rate (0.034\,M\solar\ yr$^{-1}$) in the cloud \citep{b10}, making it potentially the most massive and youngest Class 0 protostar known.

Our intent was to map the global (5$'$) $B$ field structure and gas kinematics across this exceptional cloud exhibiting such large-scale mass motions, at a high enough resolution (13\farcs6 and 2\farcs5 for SOFIA and ALMA, resp.) to potentially constrain the role of the $B$ field, gas dynamics, and energy balance in this very unusual context.

This paper is structured as follows.  In \S\ref{observ} we describe the observational and data reduction approach, briefly overview the continuum data, and compare their calibration with prior studies.  In \S\ref{continuum} we explore features of the FIR and 3\,mm continuum emission globally and in detail, including the polarisation data and inferred $B$ field morphology.  In \S\ref{cubes} we present the velocity-resolved 3\,mm spectral-line mosaics and polarisation products, including key insights into the significance of the continuum features based on the lines' kinematic and dynamical information.  In \S\ref{analysis} we use two standard statistical methods, one in a new way for the spectroscopy, to analyse our polarisation data and obtain constraints on the role of $B$ fields in this cloud.  We discuss all these results in \S\ref{disc} in order to highlight new insights from the data as well as their limitations.  We present our conclusions in \S\ref{concl}.

\begin{figure*}[ht]
\vspace{-2mm}\centerline{
\hspace{-0.3mm}\includegraphics[angle=0,scale=0.14]{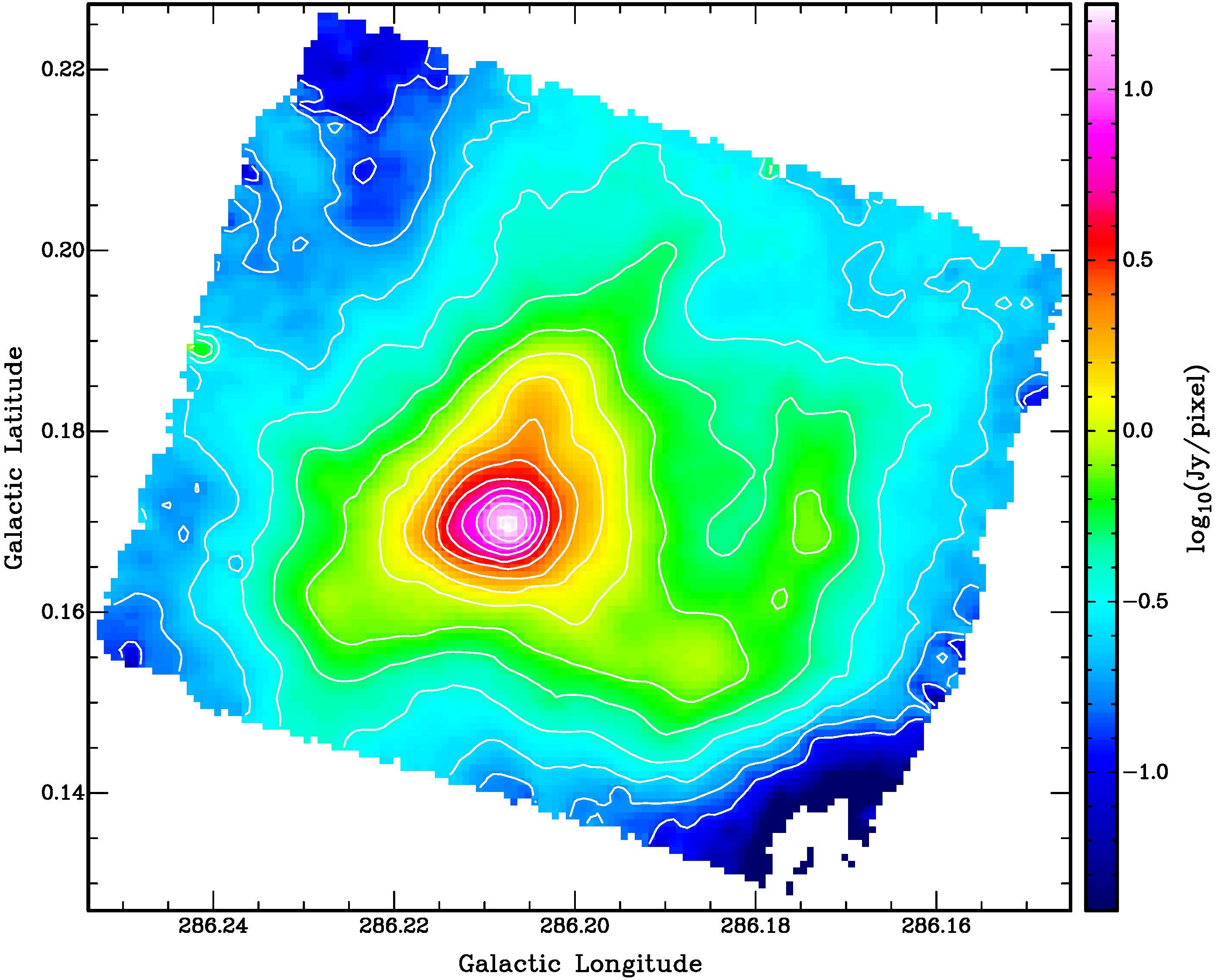}}\vspace{-112mm}
\hspace{33mm}\includegraphics[angle=-90,scale=0.655]{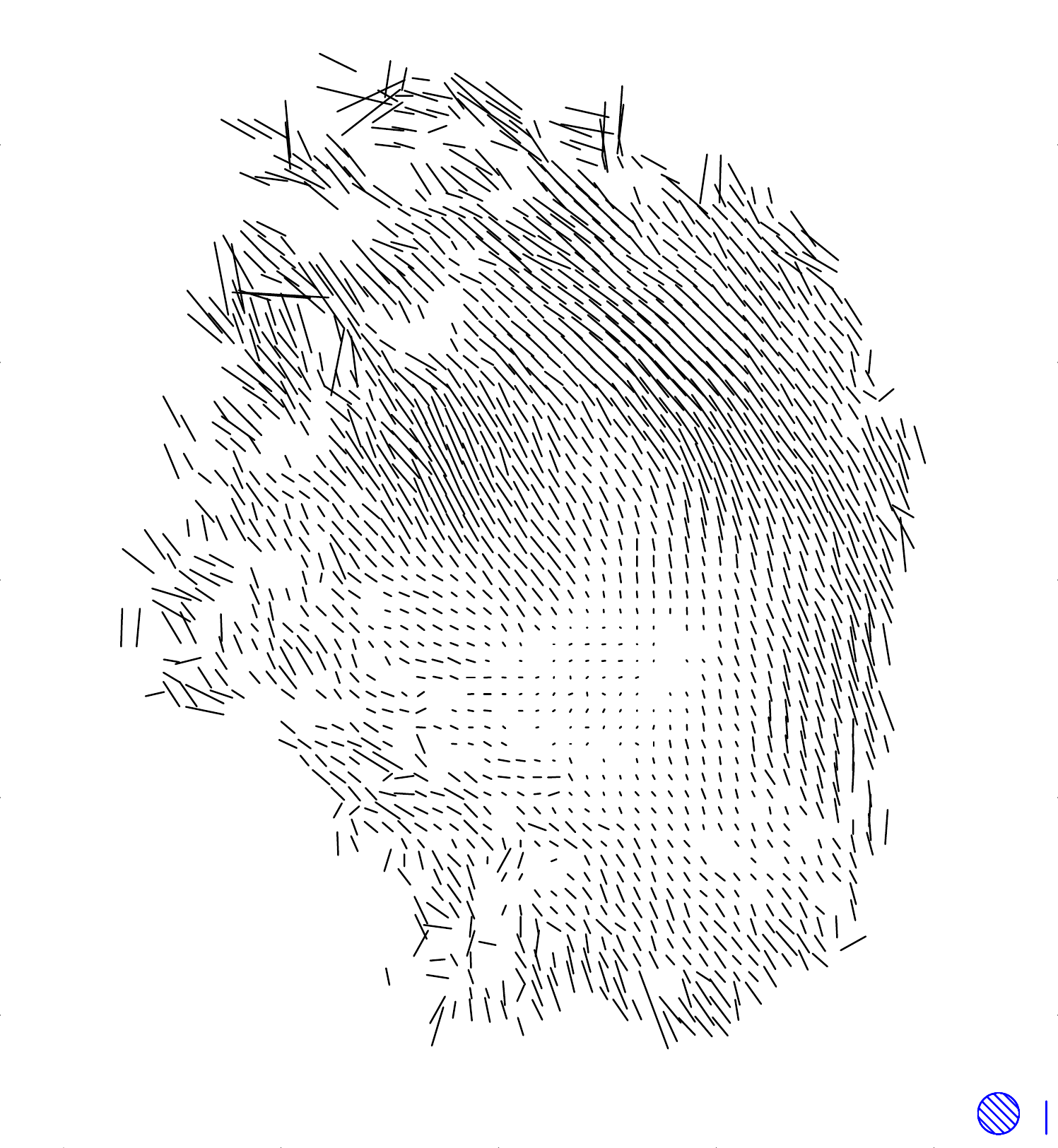}

\vspace{4mm}
\centerline{\includegraphics[angle=0,scale=0.14]{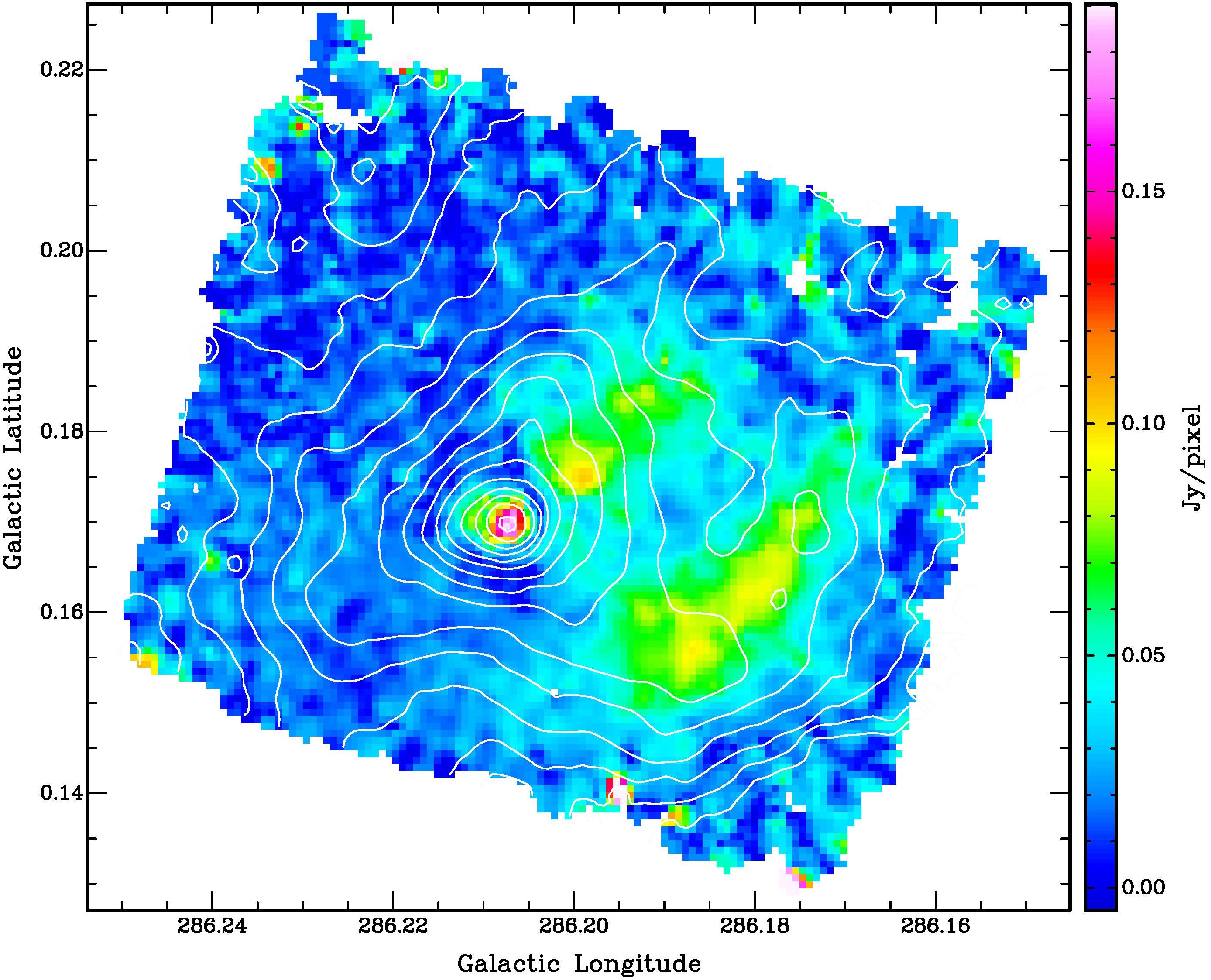}}\vspace{-112mm}
\hspace{33mm}\includegraphics[angle=-90,scale=0.655]{byf73-gbl-IPmask-eps-converted-to.pdf}

\vspace{-205mm}\hspace{97mm}{HAWC+ 154\,$\mu$m Stokes $I$ image}

\vspace{-0.5mm}\hspace{96mm}{$I$ contours: 0.125($\sqrt{2}$)16 Jy/pixel}

\vspace{0mm}\hspace{105mm}{Polo. data selection:}

\vspace{0mm}\hspace{98mm}{$I$ $>$ 0.25 Jy/pixel, $P'$/$\sigma_{P'}$ $>$ 2}

\vspace{80.2mm}\hspace{40mm}{beam FWHM}

\vspace{1mm}\hspace{42mm}{$p'$ = 10\%}

\vspace{7mm}\hspace{90.5mm}{HAWC+ 154\,$\mu$m Polarised Flux image}

\vspace{-0.5mm}\hspace{94mm}{$I$ contours: 0.125($\sqrt{2}$)16 Jy/pixel}

\vspace{0mm}\hspace{103mm}{Polo. data selection:}

\vspace{0mm}\hspace{96mm}{$I$ $>$ 0.25 Jy/pixel, $P'$/$\sigma_{P'}$ $>$ 2}

\vspace{80.2mm}\hspace{40mm}{beam FWHM}

\vspace{1mm}\hspace{42mm}{$p'$ = 10\%} 

\vspace{8mm}\caption{
{\bf ({\em Top})}  SOFIA HAWC+ band D (154$\mu$m) total intensity (Stokes $I$) image of BYF\,73 on a logarithmic scale, overlaid by white contours as labelled.  (In all figures, we use the notation x(y)z for contours running from level x in steps of y to level z.) 
All HAWC+ band D images have 2\fsec75 pixels, or 0.2$\times$ the 13\fsec6 beam. 
At every 2nd pixel (0.4 beam) satisfying the indicated selection criteria, we also display black ``vectors'' showing the measured polarisation percentage ($p'$) and position angle (with the usual $\pm\pi$ degeneracy) of the plane-of-sky $B$ field component (i.e., rotated 90\degree\ from the observed polarisation direction).  The peak $I$ intensity is 17.58 Jy/pixel with a typical rms error in the interior of the image 2--3 mJy/pixel, rising to 4--6 mJy/pixel around the image boundary due to the dither pattern of the observations; the peak S/N in the $I$ image is $>$5000.  Inside the $I$ = 0.5 Jy/pixel contour, nearly all $p'$ vectors have S/N ranging from $\sim$5 to $>$30; for 0.25$<$$I$$<$0.5 Jy/pixel, displayed vectors have S/N$\sim$2--6. 
\\ {\bf ({\em Bottom})}  Same as the top panel except with the HAWC+ band D debiased polarised flux $P'$ image on a linear scale; the peak is 189 mJy/pixel, with a typical error 4--5 mJy/pixel and S/N behaviour as for the $p'$ vectors.  
}\label{hawcmaps}\vspace{0mm}
\end{figure*}

\section{Observations and Data Reduction}\label{observ}

\subsection{SOFIA/HAWC+}

We mapped BYF\,73 on 2019 July 17 at 0832--0905 UT with HAWC+'s band D ($\lambda$154\,$\mu$m) filter.\footnote{See the HAWC+ description at https://www.sofia.usra.edu/ instruments/hawc, its Data Handbook at https://www.sofia.usra. edu/sites/default/files/Instruments/HAWC\_PLUS/Documents/ hawc\_data\_handbook.pdf, and the Cycle 7 Observer's Handbook at https://www.sofia.usra.edu/sites/default/files/Other/Documen ts/OH-Cycle7.pdf for details of the observing modes.} 
Chopping and nodding were done asymmetrically due to the nearby FIR emission to the Galactic west and south.  The total on-source integration time was 784.4\,s.  Pipeline processing with HAWC-DRP produced final Level 4 quality image products which were downloaded from the SOFIA archive.  This processing produces data that has all known instrumental and atmospheric effects removed, giving an absolute Stokes $I$ calibration uncertainty of 20\%, a relative polarisation uncertainty of 0.3\% in flux and 3\degree\ in angle, and astrometry which should be accurate to better than 3$''$ \citep{hrd18}.  However, we found the HAWC+ L4 astrometry was still consistently offset $\sim$2$''$ to the Galactic south compared to the {\em Gemini} 10\,$\mu$m, {\em Herschel} 70\,$\mu$m, and ALMA \& ATCA 3mm maps, all of which are strongly and consistently peaked on the massive protostellar core MIR 2 (allowing for MIR 1's proximity to MIR 2 in the {\em Gemini} data), so we inserted this correction by hand into the HAWC+ data files.

At a distance of 2.50$\pm$0.27\,kpc \citep[near NGC\,3324;][]{b10,s21}, the scale for BYF\,73 is 0\fdeg01 = 36$''$ = 0.44\,pc, or 0.1\,pc = 8\farcs25 = 0\fdeg0023.  Thus, HAWC+ band D gives us a useful spatial dynamic range from 0.16 to 3.6\,pc, a linear factor of 22 and almost 500 resolution elements in area. 
The resulting full-field images in both total intensity Stokes $I$ and the debiased polarised flux $P'$ = $\sqrt{Q^2+U^2-n_{\rm rms}^2}$, where $n_{\rm rms}$ is the combined instrumental and sky noise, are shown in {\color{red}Figure \ref{hawcmaps}}, overlaid also with the inferred $B$ field polarisation vectors.  

\begin{figure*}[ht]
\vspace{0mm}
\centerline{\includegraphics[angle=-90,scale=0.59]{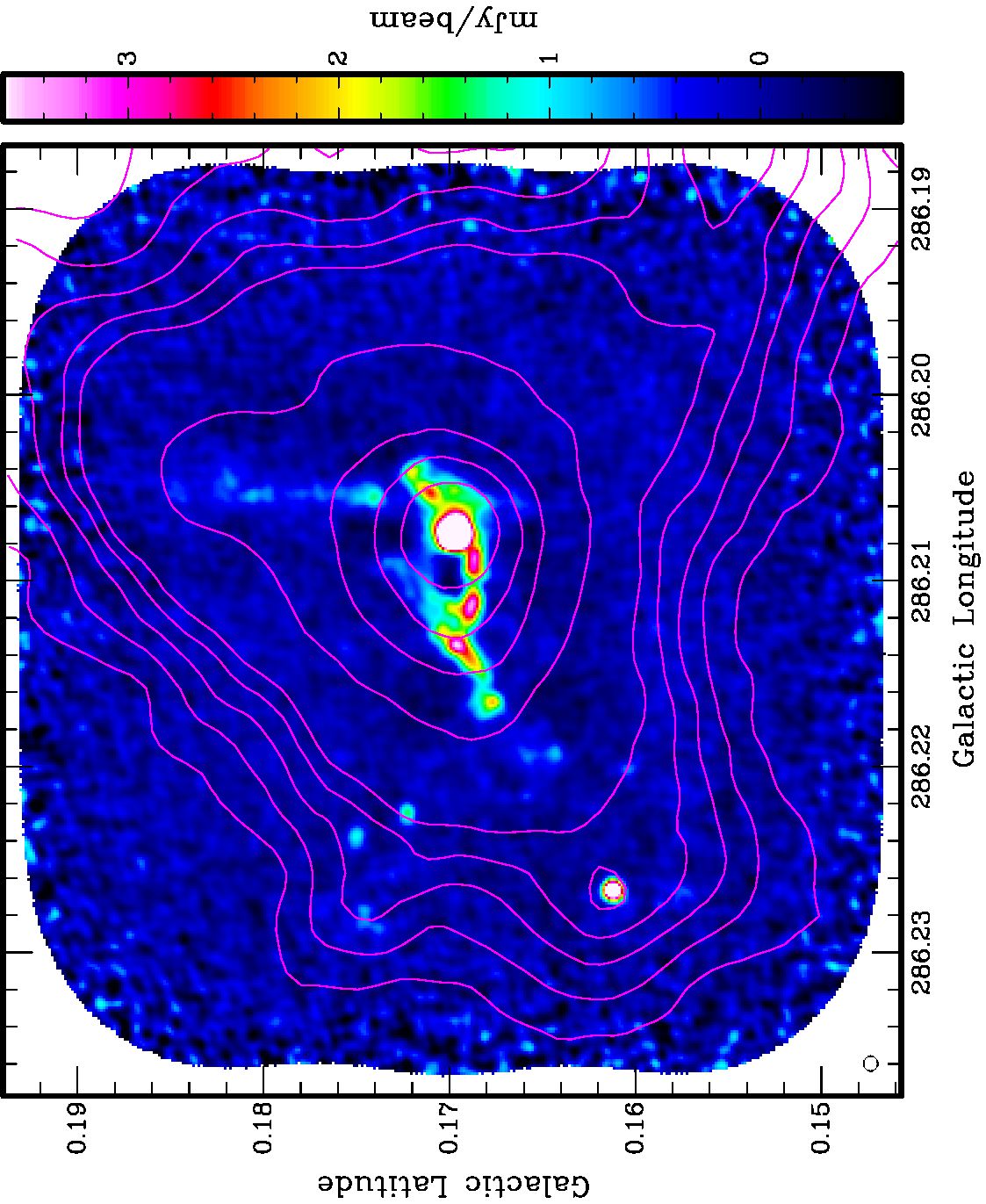}\hspace{8mm}}\vspace{-104mm}\hspace{48mm} {\color{white}{ALMA 3mm $I$ image}}

\hspace{49mm}{\color{magenta}{HAWC+ $I$ contours}}
\begin{picture}(1,1)
\thicklines
{\color{yellow}
	\put(-14,-131){\framebox(95,58)}
	\put(-15,-133){\line(-3,-4){99}}
	\put(82,-133){\line(3,-4){99}}
	}
\end{picture}

\vspace{90.5mm}
\centerline{\includegraphics[angle=-90,scale=.51]{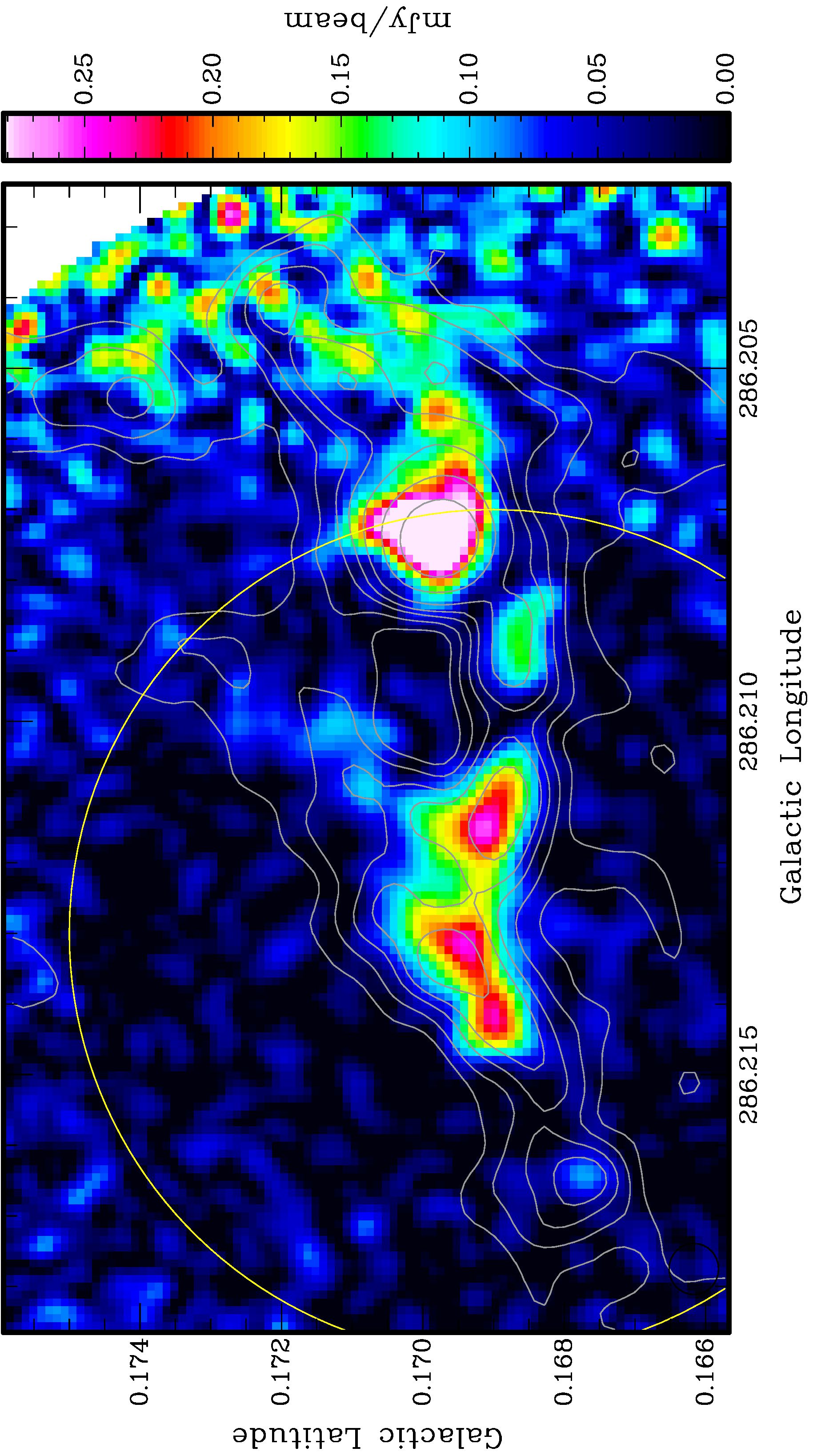}}\vspace{-79.2mm}
\hspace{30.6mm}\includegraphics[angle=-90,scale=.456]{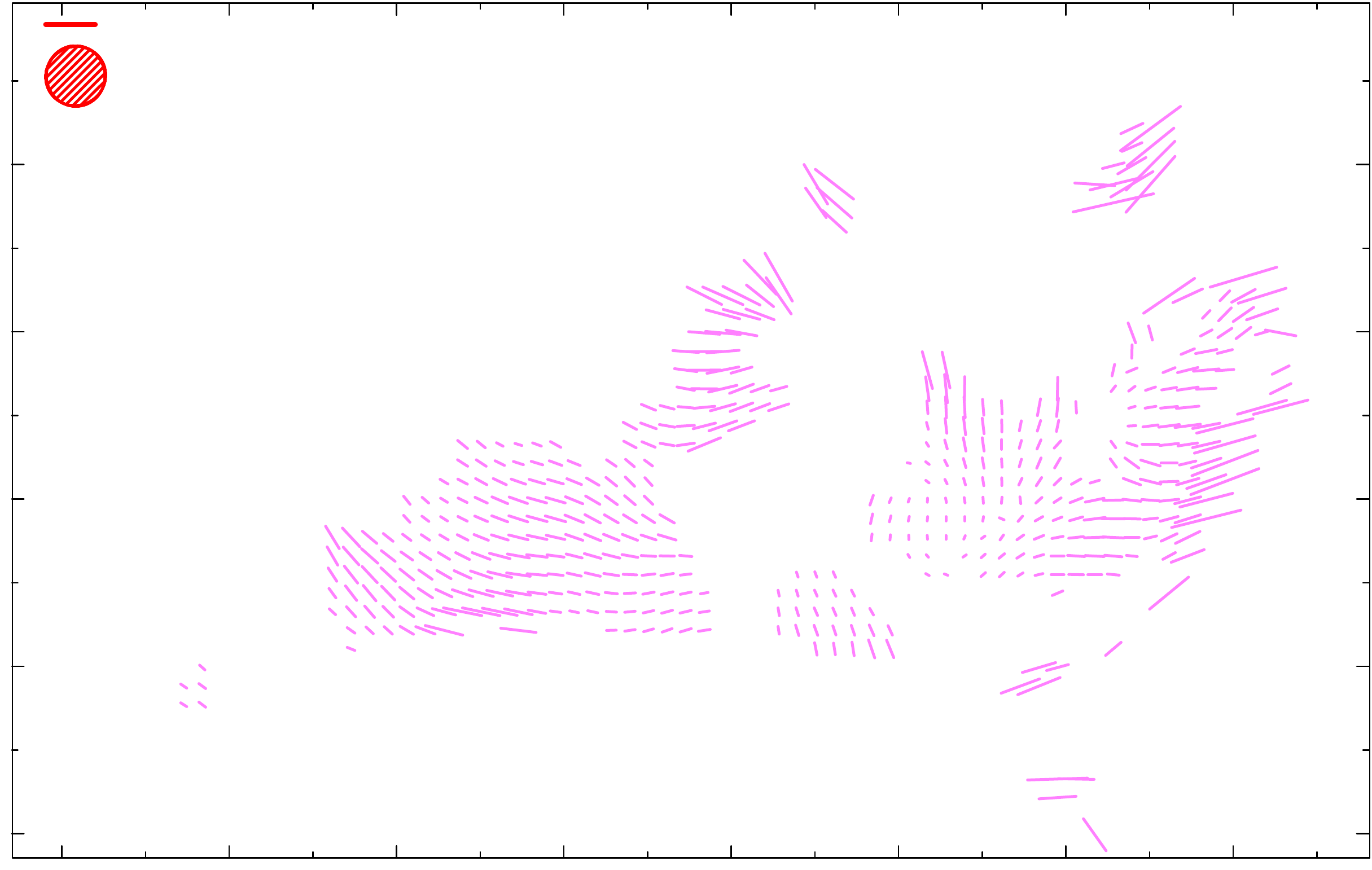}

\vspace{-71mm}\hspace{42mm}{\color{white}{$p'$ = 30\%}

\vspace{1mm}\hspace{43mm}{FWHM}}

\vspace{56mm}\hspace{36mm}{\color{white}{ALMA 3mm $P'$ image with $I$ contours}}

\hspace{38mm}{\color{magenta}{Polo. data selection: $P'$/$\sigma_{P'}$ $>$ 2.5}} 

\vspace{8mm}\caption{
{\bf ({\em Top})}  ALMA 13-pointing mosaic of 3\,mm continuum emission from BYF\,73, in a 3.5\,GHz-wide band centred at an effective frequency of 102.1346\,GHz.  
The contrast is enhanced to bring out the fainter structures, in particular the E-W Streamer, as indicated by the colour bar to the right.  The point sources MIR\,2 and 11 (see Fig.\,\ref{irac}) peak at 21 and 7\,mJy/beam, respectively.  The synthesised beam (2\fsec93$\times$2\fsec74 @ --38\fdeg0) is shown in the bottom-left corner, and the noise $\sigma_{\rm rms}$ = 0.13\,mJy/beam where the primary beam correction is small, away from the map edge, which is at a primary beam cutoff of 0.2.  This gives a peak S/N of 170.  For reference, magenta contours are overlaid from the HAWC+ 154$\mu$m Stokes $I$ data, at levels 0.44(0.10)0.84, 1.5, 3, 5, and 9 Jy/pixel (as in Fig.\,\ref{irac}). \\
{\bf ({\em Bottom})}  Zoom in to all detectable 3\,mm continuum polarised emission within a deeper, single ALMA pointing of BYF\,73's central structures, framed by the yellow box in the top panel.  The image is the debiased polarised flux on the colour scale to the right, peaking at 0.55\,mJy/beam for MIR\,2 (S/N = 24, $\sigma_{\rm rms}$ = 23\,$\mu$Jy/beam).  The debiased percent polarisation vectors are overlaid in magenta, rotated by 90\degr\ to show the $B$ field orientation at every second pixel in $l$ and $b$ (as in Fig.\,\ref{hawcmaps}).  Away from MIR\,2, most vectors shown have S/N $>$ 5 with typical noise $\sigma_{\rm rms}$ = 4\% in amplitude and 5\degr\ in angle.  The grey contours here (at 0.2, 0.6, 1.1, 1.6, 2.5, 5, 10\,mJy/beam) show the ALMA Stokes $I$ from the mosaic in the top panel. 
The single-field $I$ map has noise $\sigma_{\rm rms}$ = 85\,$\mu$Jy/beam for a peak S/N =240 at MIR\,2, slightly deeper than the mosaic.  The noisy polarisation features near the N-S ionisation front west of MIR\,2 are probably real, but are not accurately calibrated outside the roughly 1/3 FWHM primary beam limit 
(20$''$, large yellow circle) of ALMA's polarisation mode in Cycle 7. 
The synthesised beam (2\fsec61$\times$2\fsec52 @ 21\fdeg0) is shown in the top-left corner with a 30\% polarisation scale bar.
}\label{almaCont}
\end{figure*}

\subsection{ALMA}

BYF\,73 was observed with ALMA at 3\,mm wavelength 
on 2020 January 1 in the C-43 array (baselines 15--314\,m) and in two correlator setups and mapping modes; the total on-source integration time was $\sim$7500\,s.  The first mode mapped a standard, 13-pointing mosaic of size 2\farcm8 centered on the peak molecular line emission as measured in the Mopra maps \citep{b10}, similar in extent to the ATCA mosaic reported by P18, in both the 3\,mm cold dust continuum plus the $J$=1$\rightarrow$0 lines of \ttco\ \& \ceto\ and $N$=1$\rightarrow$0 line of CN.  The second mode was a single-pointing, full-polarisation, deeper integration at the peak emission position near MIR\,2, to map the $B$ field strength \& structure with (1) the cold dust continuum, and potentially with (2) the Goldreich-Kylafis effect in the line wings of \tco\ \citep{gk81}, and/or (3) the Zeeman effect in the hyperfine structure of CN \citep[e.g.,][]{hc11}.  Mosaicking with ALMA was not available for polarisation modes in Cycle 7.

Standard reduction pipelines were applied to the data, including bandpass, complex gain, flux, and polarisation calibration on nearby quasars; images were formed by a joint deconvolution for the mosaics with cleaning and restoration as implemented in the CASA task {\sc tclean}; and primary beam correction was made within a cutoff at 0.2.  The resulting science-ready FITS files were either downloaded from the ALMA Science Archive for the pipeline-reduced data, or the ALMA-North America servers for manually-reduced data not included in the automated pipeline processing.

The continuum mosaic is shown in {\color{red}Figure \ref{almaCont}} (top panel), with a maximum recoverable scale (MRS) $\sim$55$''$.  This is larger than the nominal single-field ALMA MRS due to the joint deconvolution for a mosaic, which recovers some of the larger spatial scales missed in a single pointing (see caption and \S\ref{comps}).  The mosaic also produced data cubes of the \ttco, \ceto, and CN emission at 0.16\,\kms\ velocity resolution, but with slightly smaller beams and MRS compared to the continuum, due to the higher frequencies; see \S\ref{cubes} for results and details.

In the top panel of Figure \ref{almaCont} it is clear that, except for the extensions off-frame to the NW and SW (i.e., into the \hii\ region), the HAWC+ and ALMA continuum $I$ maps seem to generally trace the same structures, including the weaker point sources to the E and along the ionisation front to the N, despite the 20- and 5-fold difference (resp.) in wavelength and resolution.

In the single polarisation field, the MRS = 28$''$ for all polarisation products, and the synthesised beams are also the same.  
An image of the debiased polarised continuum flux $P'$ is shown in Figure \ref{almaCont} (bottom panel), but vignetted to exclude spurious features due to missing short spacings along the N and S field boundaries.  The $P'$ image is also overlaid with the inferred $B$ field polarisation vectors.  

At 2.5\,kpc, the ALMA mosaics of BYF\,73 give a useful spatial dynamic range from 0.030 to 0.67\,pc (6,000--140,000 AU): coincidentally with the HAWC+ maps, this is a linear factor of 22 as well. 

\section{Features of the Continuum Emission}\label{continuum}

\subsection{Comparison with ATCA and Herschel}\label{comps}

It is instructive to compare the earlier 90\,GHz ATCA data (P18), which only detected MIR\,2 within the mapped mosaic of BYF\,73, with the 50$\times$ more sensitive ALMA 102\,GHz images.  The inferred flux density of MIR\,2 was 50\% higher (34\,mJy) in ATCA's $\sim$2$\times$ larger synthesised beam than in Figure \ref{almaCont}, but on convolving the ALMA data to the ATCA resolution, we recover the identical flux density for MIR\,2.  Further, P18 found that MIR\,2's 90\,GHz continuum and 89\,GHz \hcop\ line flux were $<$30\% of the values expected from the $\sim$arcminute-resolution SED fit to BYF\,73's {\em Herschel} data \citep[$\sim$120\,mJy;][]{p19} and the Mopra single-dish line flux \citep{b10} respectively, which P18 attributed to a smooth overall structure in BYF\,73 that ATCA apparently resolved out.  This turns out to be very close to half-true: we find the total flux density in our mosaic to be $\sim$70\% of the projected SED value at 102\,GHz, despite similar shortest baselines in both interferometers.  

These results are explained by the mJy-level structures around MIR\,2, which were too weak to be separately detected in the ATCA map (noise $\sigma_{\rm rms}$ = 7\,mJy/beam) but raised the measured flux density in the unresolved structure of MIR\,2; also, these structures together contribute $\sim$half the additional flux expected from the SED fit to the ALMA map.  With this insight, we see that the older ATCA and current ALMA data are completely consistent with each other, allowing for their respective sensitivities.

We also note that the deconvolved size of MIR\,2 in the ALMA data is measured to be 3\farcs2$\times$2\farcs8 in both the mosaic and deeper polarisation field, which is only slightly smaller than the 4\farcs2$\times$3\farcs0 derived from the ATCA map despite its $\sim$2$\times$ larger synthesised beam ($\sim$4$\times$ in area).  Therefore it seems MIR\,2's protostellar structure is close to being resolved at this scale, 3$''$ = 7500\,AU, and future sub-arcsecond imaging may reveal useful information about its mass distribution. 

The spatially-resolved SED fit to {\em Herschel} data of \cite{p19} not only provides the missing short-spacing flux information as above, but also allows calculation of a merged single-dish and ALMA 3\,mm continuum image.  The SED fits were used to project how BYF\,73 would look at 3\,mm with {\em Herschel}'s $\lambda$500\,$\mu$m resolution of 36$''$, assuming that at 3\,mm, MIR\,2 has a negligible contribution from free-free emission in an unresolved UC\hii\ region, since that would tend to push MIR\,2's flux density above the SED fit.  The derived image was combined with the ALMA map via the {\sc Miriad} task {\sc immerge} \citep{st95} to recover the missing flux density in Figure\,\ref{almaCont} that resides in larger angular scales.  The result changes the appearance of the image very little, nor the brightness of the individual small-scale structures, except to fill in the broad ($\sim$50$''$) but shallow ($\sim$0.3\,mJy/beam) negative bowl underlying MIR\,2 and its environs.  This shows that the missing 30\% of the flux density is distributed very smoothly across BYF\,73 after all.

\begin{figure*}[ht]
\vspace{-1mm}
\centerline{\includegraphics[angle=-90,scale=0.4]{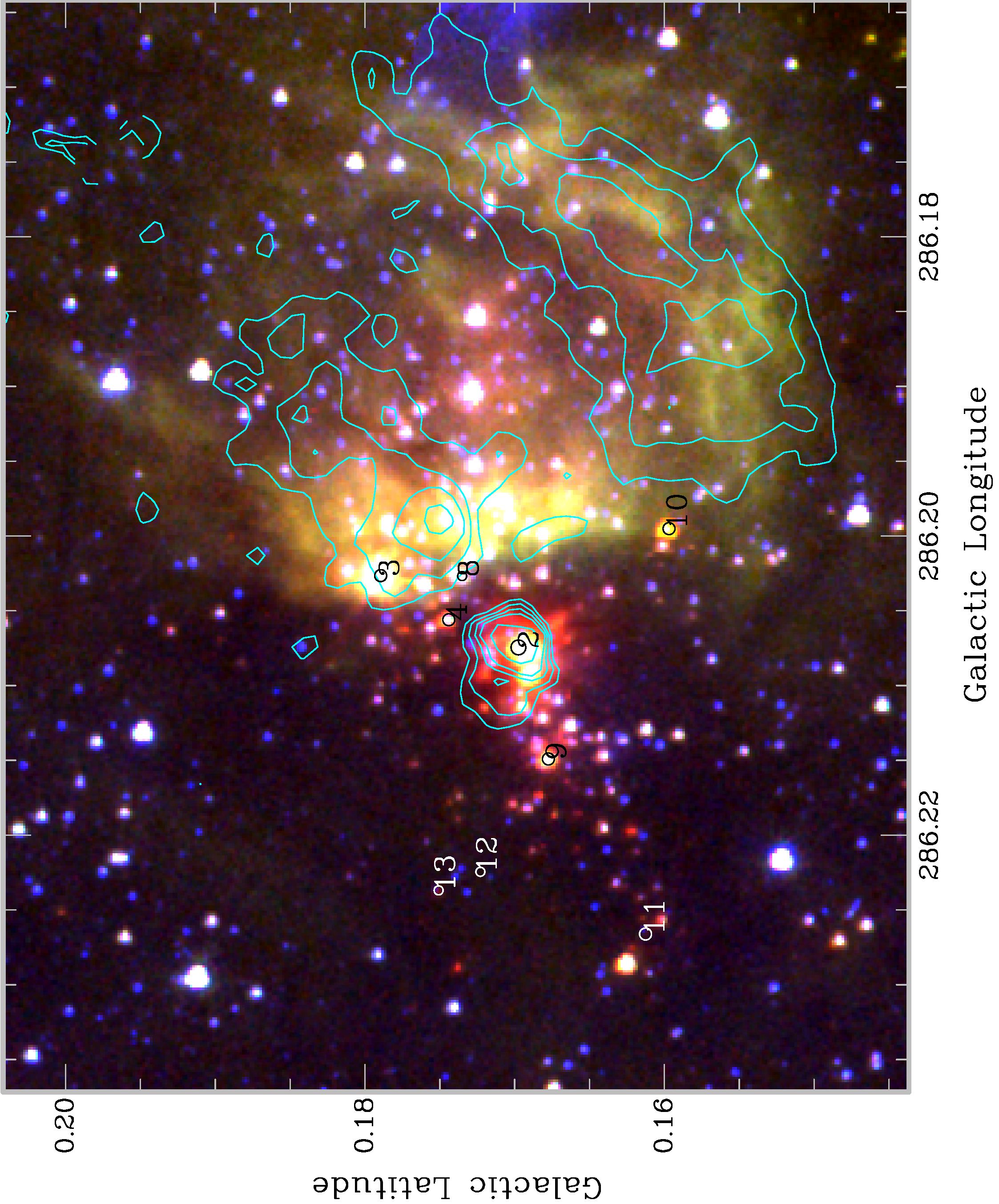}\hspace{-181mm}
		\includegraphics[angle=-90,scale=0.4]{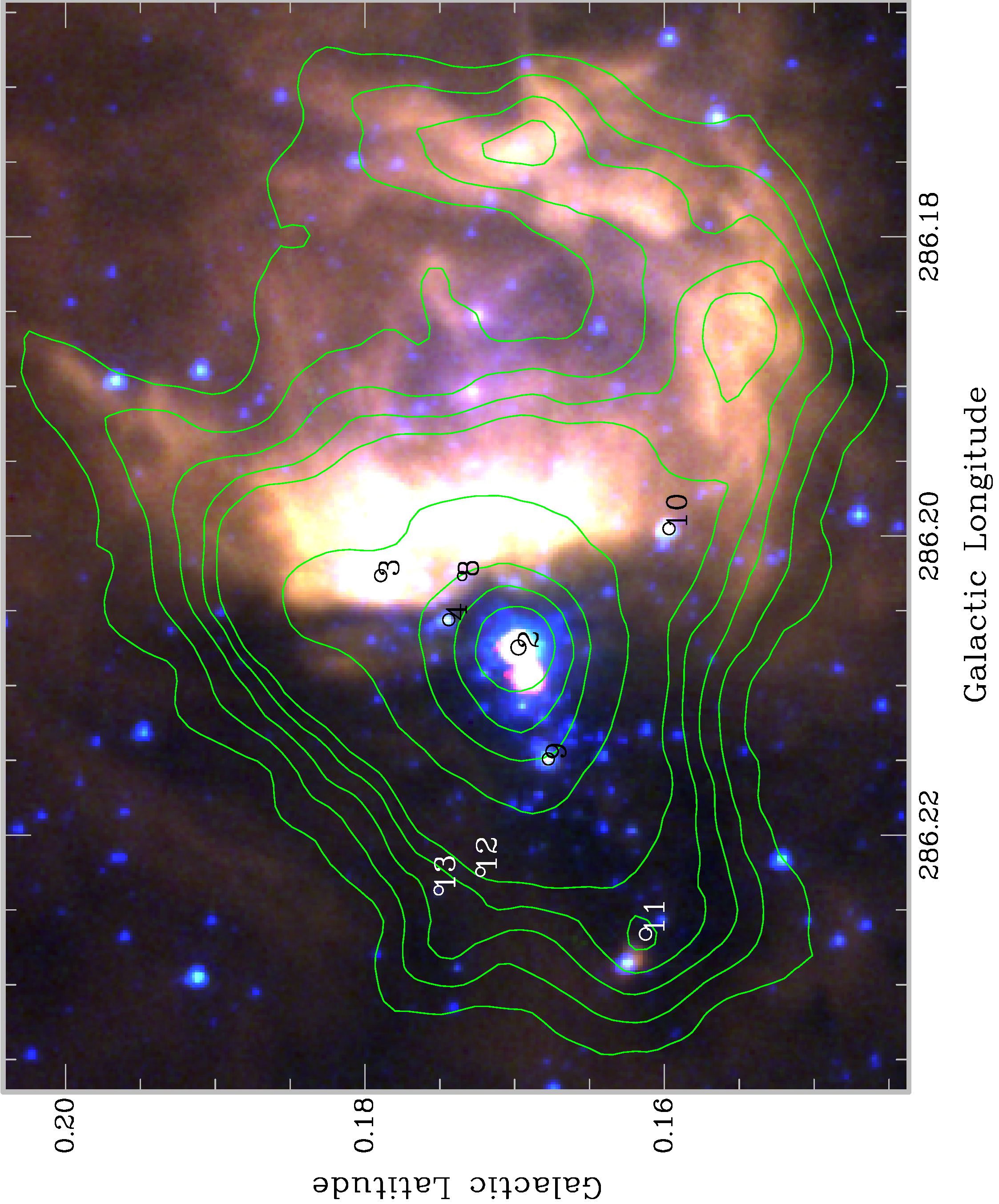}}

\vspace{-74mm} 
{\color{green}\hspace{12.5mm}{HAWC+ $I$ contours: 0.44(0.10)0.84, 1.5, 3, 5, 9 Jy/pixel}}
{\color{cyan}\hspace{20mm}{HAWC+ $P'$ contours: 50(16)98, 140 mJy/pixel}}

\vspace{4mm}
{\color{red}    \hspace{71mm}{IRAC band 4}	\hspace{11mm}{IRAC band 2}} \\
{\color{green}\hspace{71mm}{IRAC band 3}	\hspace{11mm}{IRAC band 1}} \\
{\color{cyan} \hspace{71mm}{IRAC band 2}	\hspace{14mm}{AAT K}}

\vspace{46mm}{\color{white}\hspace{12mm}{(a)}			\hspace{83mm}{(b)}}

\vspace{3mm}
\centerline{\includegraphics[angle=-90,scale=0.4]{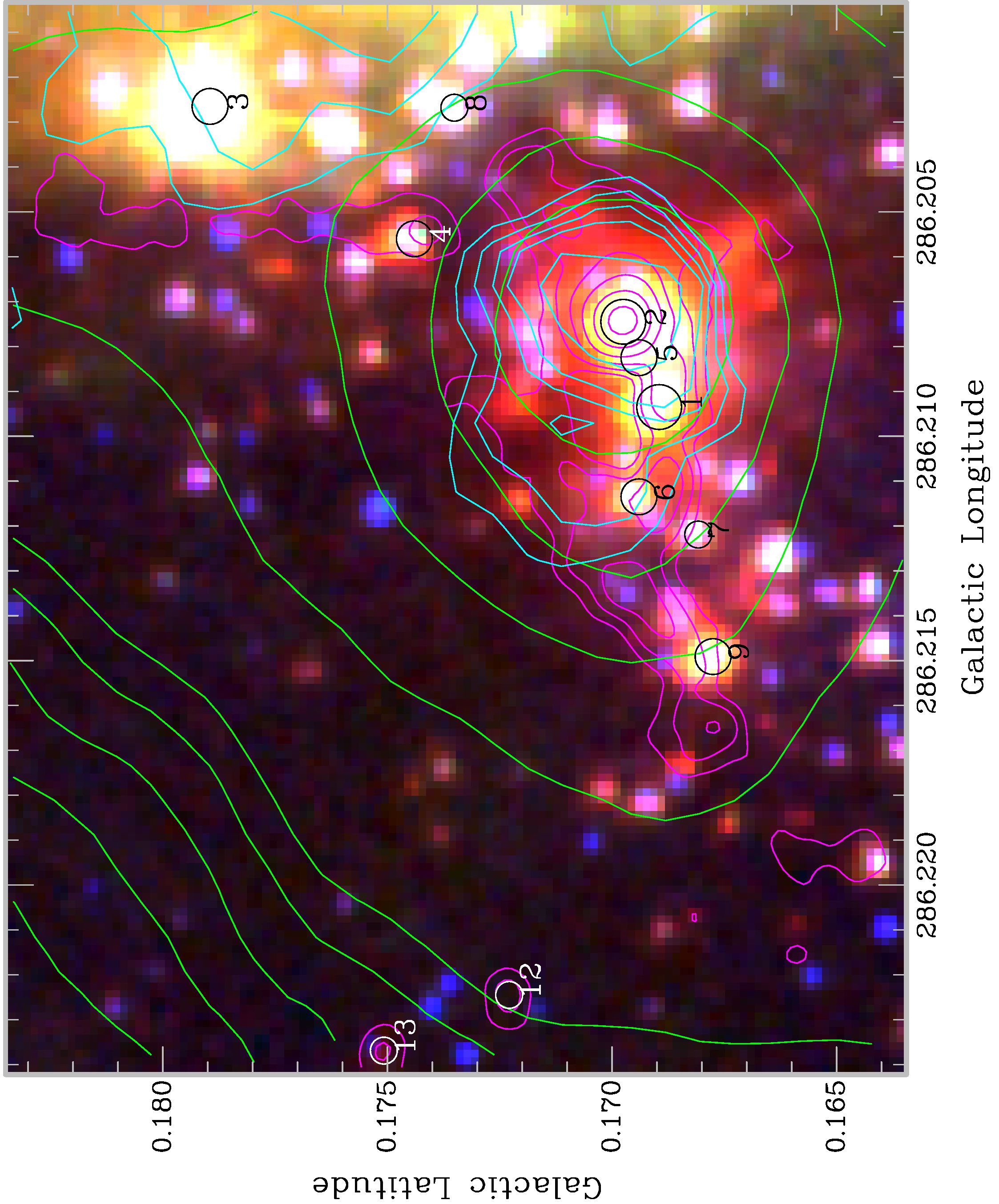}\hspace{-181mm}
		\includegraphics[angle=-90,scale=0.4]{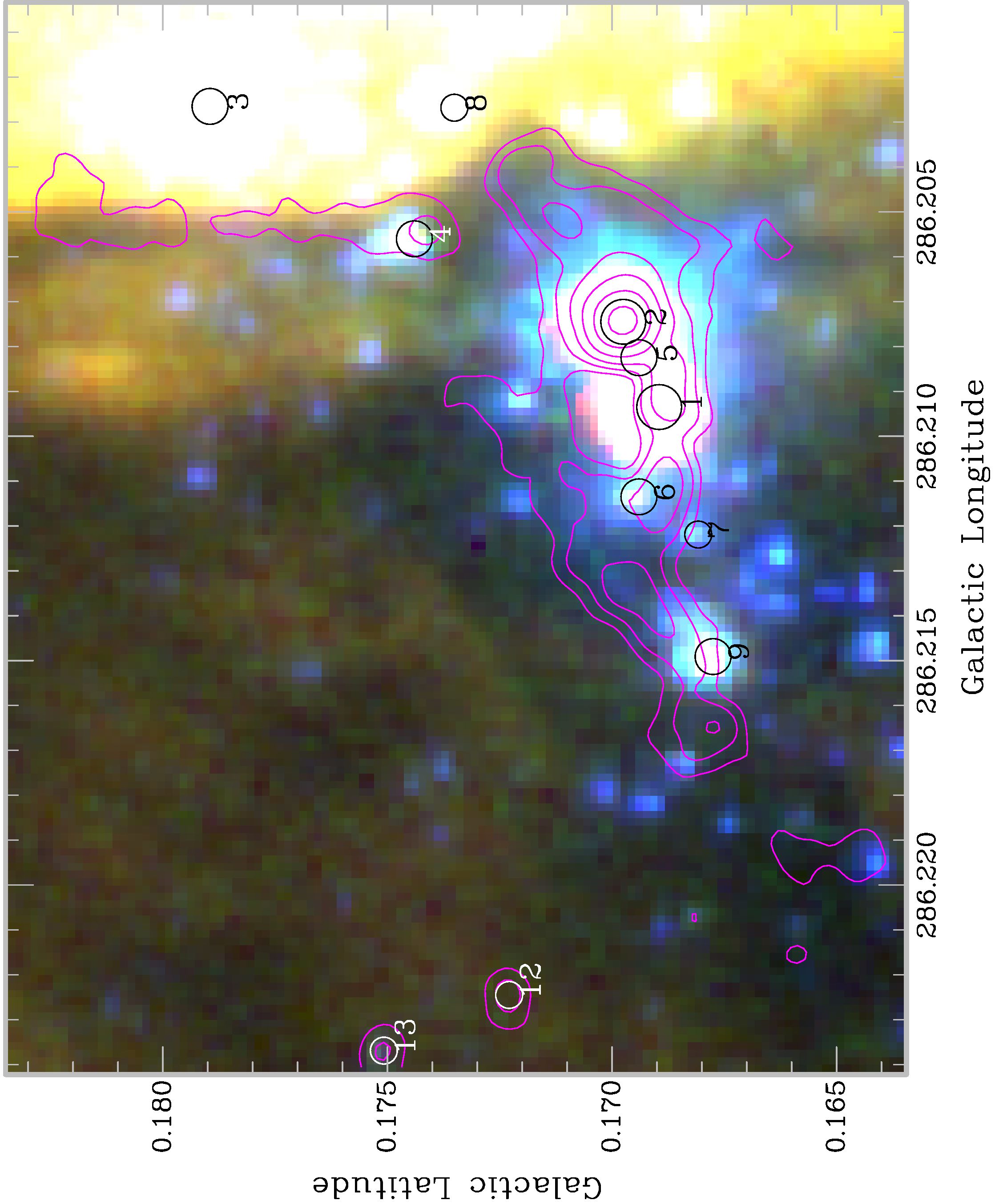}}

\vspace{-74mm}
{\color{magenta}\hspace{22mm}{ALMA $I$ contours: 0.4, 1, 2, 4, 8, 16\,mJy/beam}}
{\color{green}	\hspace{32mm}{HAWC+ $I$,}}		{\color{cyan}{HAWC+ $P'$,}}
{\color{magenta}			{ALMA $I$}}		{\color{white}{contours}}

\vspace{4mm}
{\color{red}    \hspace{14mm}{IRAC band 4}	\hspace{70mm}{IRAC band 2}} \\
{\color{green}\hspace{14mm}{IRAC band 3}	\hspace{70mm}{IRAC band 1}} \\
{\color{cyan} \hspace{14mm}{IRAC band 2}	\hspace{73mm}{AAT K}}

\vspace{46mm}{\color{white}\hspace{12mm}{(c)}	\hspace{84mm}{(d)}}

\vspace{3mm}
\centerline{\includegraphics[angle=-90,scale=0.34]{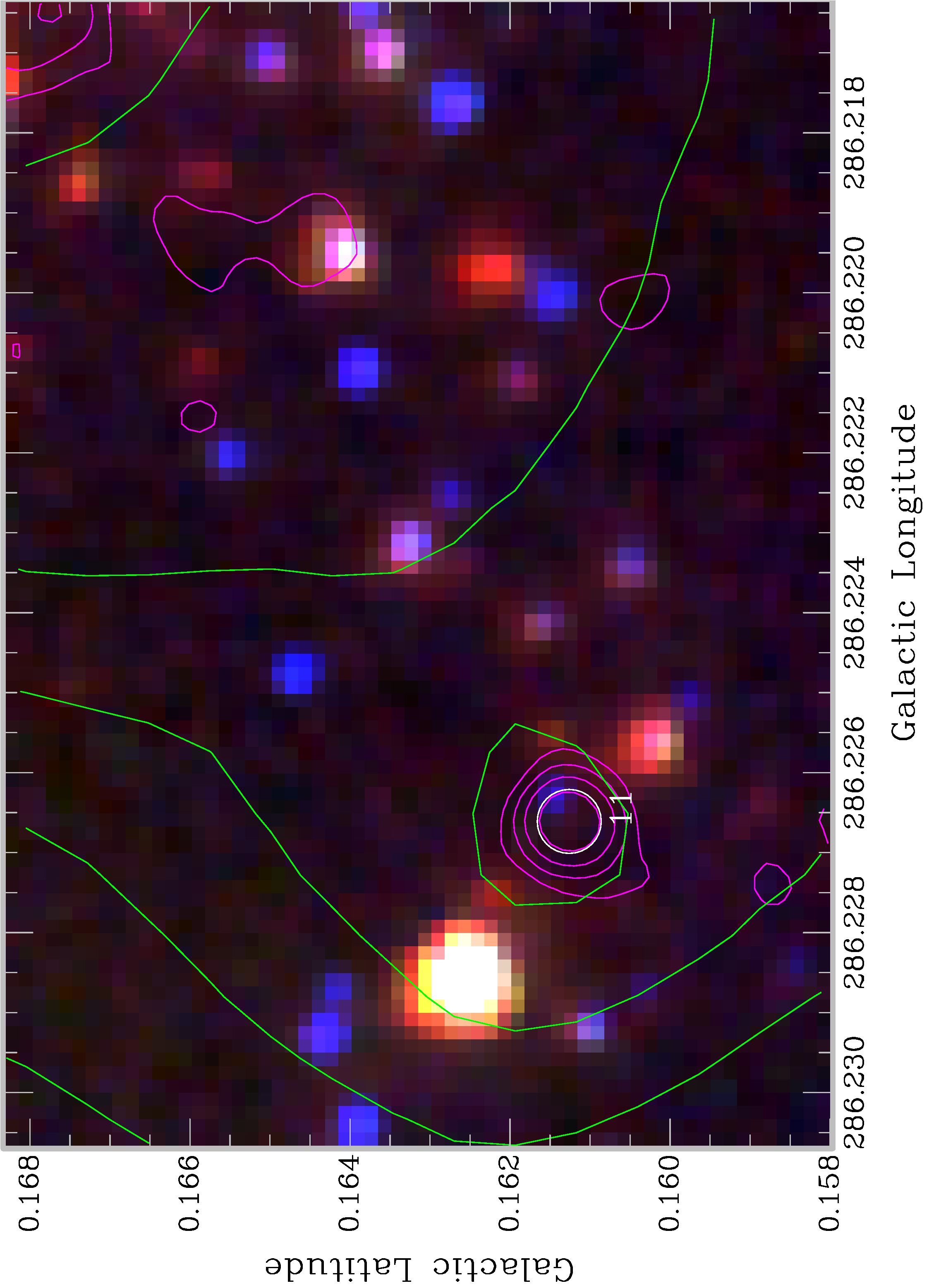}\hspace{-181mm}
		\includegraphics[angle=-90,scale=0.34]{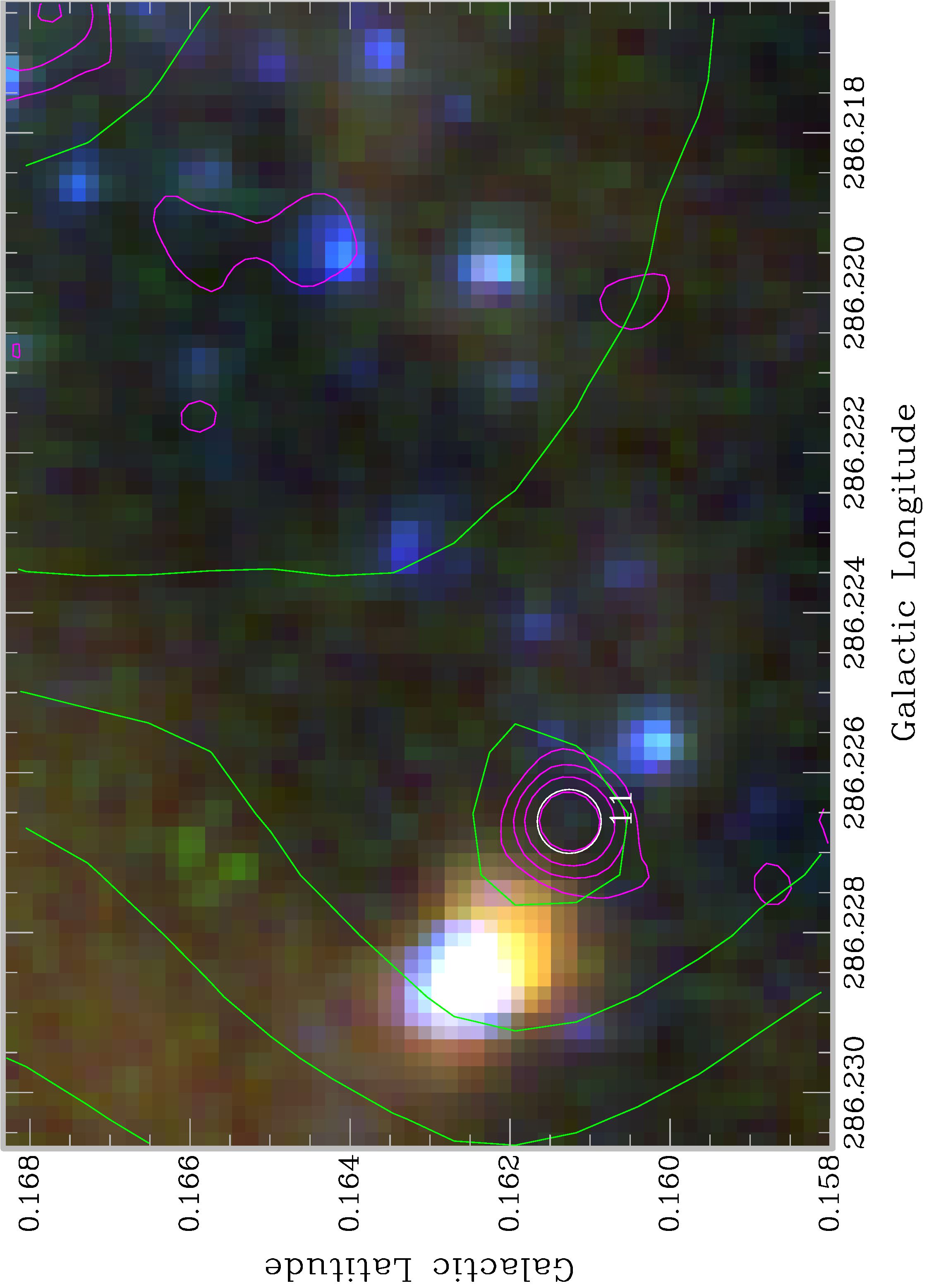}}

\vspace{-62mm}
{\color{magenta}\hspace{30mm}{ALMA $I$,}}  {\color{green}{HAWC+ $I$}}  {\color{white}{contours}}
{\color{magenta}\hspace{50mm}{ALMA $I$,}}  {\color{green}{HAWC+ $I$}}  {\color{white}{contours}}

\vspace{3mm}
{\color{red}    \hspace{14mm}{IRAC band 4}	\hspace{71mm}{IRAC band 2}} \\
{\color{green}\hspace{14mm}{IRAC band 3}	\hspace{71mm}{IRAC band 1}} \\
{\color{cyan} \hspace{14mm}{IRAC band 2}	\hspace{74mm}{AAT K}}

\vspace{35mm}{\color{white}\hspace{12mm}{(e)}	\hspace{86mm}{(f)}} 

\vspace{7.5mm}\caption{
({\em All panels}) Composite RGB images of {\em Spitzer} and Anglo-Australian Telescope data \citep[][P18]{b13} as labelled in each panel, plus locations of MIR sources 1--8 from P18 and new designations MIR\,9--13 in this work.  Contours are also colour-coded and labelled at the top of each panel.  Panels in the left column are IRAC 4-3-2 composites, with $c$,$e$ being more saturated than $a$ in order to bring out fainter features.  Panels in the right column are IRAC 2-1 + AAT K composites.  Panels on the same row are on the same scale to facilitate comparisons; the top row is a wider view, other rows are successive zooms. 
}\label{irac}\vspace{0mm}
\end{figure*}

\subsection{Overall Geometry}

To provide context, we show in {\color{red}Figure \ref{irac}} composite mid-IR images (similar to that in P18), overlaid with selected contours of the HAWC+ 
and ALMA data from Figures \ref{hawcmaps} and \ref{almaCont}.  We have added 5 more mid-IR point-source designations to the 8 identified by P18.  MIR\,9 and 10 are mid-IR bright in the IRAC images, especially band 2 (4.5\,$\mu$m), and would have been easily detected with T-ReCS (8--18\,$\mu$m) if the imaged area had been slightly larger.  MIR\,11--13 are really mm-continuum sources, but we use the mid-IR designations to avoid new nomenclature that might be confusing. 

At 7\,mJy, MIR\,11 is the second-brightest point source in the ALMA mosaic, and is also clearly detected by HAWC+ (154\,$\mu$m).  However, in IRAC bands 2--4 (4.5--8.0\,$\mu$m), it is detected not as a point source, but as a biconical nebula around the 3mm position (Figs.\,\ref{irac}$e$,$f$), with a bright NE lobe and fainter SW lobe; evidently the central object is deeply embedded and undetected shortward of 10$\mu$m.  The NE lobe also shows redder mid-IR colours at locations closer to MIR\,11 itself.  These are all classic hallmarks of another (massive) protostar.
MIR\,13 is weakly detected at $K$ and IRAC bands 2--4, while MIR\,12 is not detected shortward of 10 $\mu$m, like MIR\,11.  Together, however, MIR\,12+13 are marginally detected in the HAWC+ $I$ image as distinct extensions to the FIR emission; in combination with their 3mm continuum detections at S/N$\sim$10, we consider them to also be probable (lower-mass) protostars.  In the spectral-line data (\S\ref{cubes}), while MIR\,11--13 are all outside the single \tco\ field of view, the mosaics show interesting features near MIR\,11 consistent with its protostar status (\S\ref{mir11}).  In contrast, the mosaics near MIR\,12 \& 13 are completely unremarkable.  Based on the spectroscopy, none of MIR\,11--13 seem to have any impact on the wider cloud's evolution.

This extensive multi-wavelength data, showing a relative paucity of mm-wave point sources and almost-as-scarce mid-IR (i.e., 8--18\,$\mu$m) point sources, supports P18's inference that most of the plethora of near-IR (i.e., 1--5\,$\mu$m) stars are likely to be in the foreground of the BYF\,73 cloud.  That is, while scores of stars within the T-ReCS field show embedded near-IR colours \citep{a17}, most of these cannot be {\em deeply} embedded, since P18 only directly detected 8 of them with T-ReCS, suggesting a lack of embedding envelopes.  Based on comparisons with their near-IR visibility, T-ReCS would likely only have detected 2 more sources outside the observed mosaic, MIR\,9 and 10, at P18's sensitivity level.

Even among these 10 mid-IR bright sources, only MIR 2 is detectable at all in the 3\,mm continuum; specifically, even the very mid-IR-bright stars MIR\,1, 3, 9, and 10 are not detected with ALMA.  By comparison with MIR\,2, this suggests that these other four mid-IR bright stars have very minimal (if any) protostellar dust envelopes, of mass $<$3--4\,M\solar\ (ALMA's 3$\sigma$ detection limits in the two observing modes).   Therefore, it is reasonable to conclude that, among all these point sources, only MIR\,11--13 have similarly ``cold'' spectral energy distribution to MIR\,2, and are in a similarly early stage of protostellar evolution.  Scaling MIR\,11's 3mm flux density to MIR\,2, which is 3$\times$ as bright, suggests that its mass may very approximately be 80\,M\solar, still large by protostellar standards.  Similarly scaling MIR\,12+13's 3mm flux densities yields dust masses $\sim$7\,M\solar\ and 10\,M\solar, respectively.

For the extended emission, both the polarised and unpolarised 154\,$\mu$m structures simultaneously trace two very different dust populations, each in their own way: (1) the warm dust permeating and surrounding the \hii\ region, arcing out to the west and northwest from the molecular clump, and seen well in {\em Herschel} and {\em Spitzer} images at 70\,$\mu$m and shorter wavelengths, and (2) the cold dust in the massive (2$\times$10$^4$\,M\solar) molecular clump to the east, traced well by the usual mm-wave molecular lines and longer wavelength ($\ge$250\,$\mu$m) continuum.

Similarly, the 3\,mm emission mostly traces the cold molecular structures, but apparently also some high-density warm dust associated with the N-S ionisation front (IF) between the molecular cloud and \hii\ region.  Circumstantially, MIR\,3 appears to be the main driver of the IF in Fig.\,\ref{irac}$a$--$d$; MIR\,4 also lies close to the southern end of the IF, but seems not to have as much impact on its surroundings.  The main extended features in the 3\,mm continuum are the rather striking arcs of emission running mainly east and west of MIR\,2, which for lack of a better term, we call the ``Streamer''.\footnote{We resist calling it a filament since that term has a specific meaning in SF studies, which we do not wish to pre-judge.  For example, the IF also looks filamentary, but as is common with such interfaces, it is likely only prominent in Fig.\,\ref{almaCont} due to its flat geometry being viewed edge-on.}  There is also a notable $\sim$5$''\times$10$''$ gap (the ``Hole'') in the 3\,mm emission within the Streamer, immediately adjacent to MIR\,2 on its eastern side.  It is unclear from its continuum properties whether this is a true lack of emission due to an absence of material in the Streamer, or whether it is the shadow of an extremely cold, high-optical-depth component in the foreground of the Streamer, completely absorbing the 3\,mm emission beyond it.  The spectral-line maps, however, resolve this question; see \S\S\ref{outflow},\ref{EPLcont}.


\subsection{Magnetic Field Structures in the Molecular Core: HAWC+}\label{hawcIBL}

\begin{figure}[t]
\centerline{
\hspace{0mm}\includegraphics[angle=-90,scale=0.36]{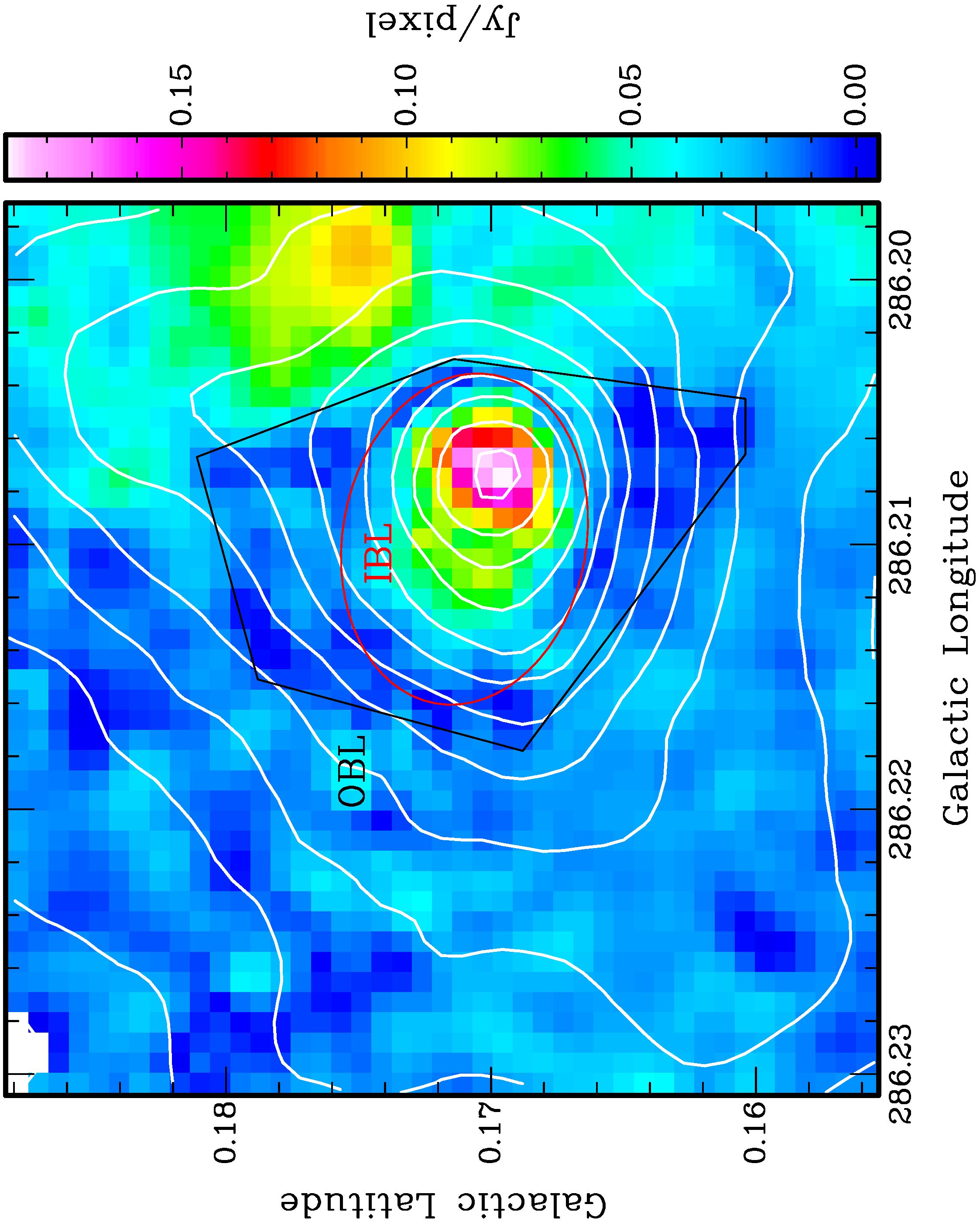}}\vspace{-69mm}
\hspace{8.8mm}\includegraphics[angle=-90,scale=0.39]{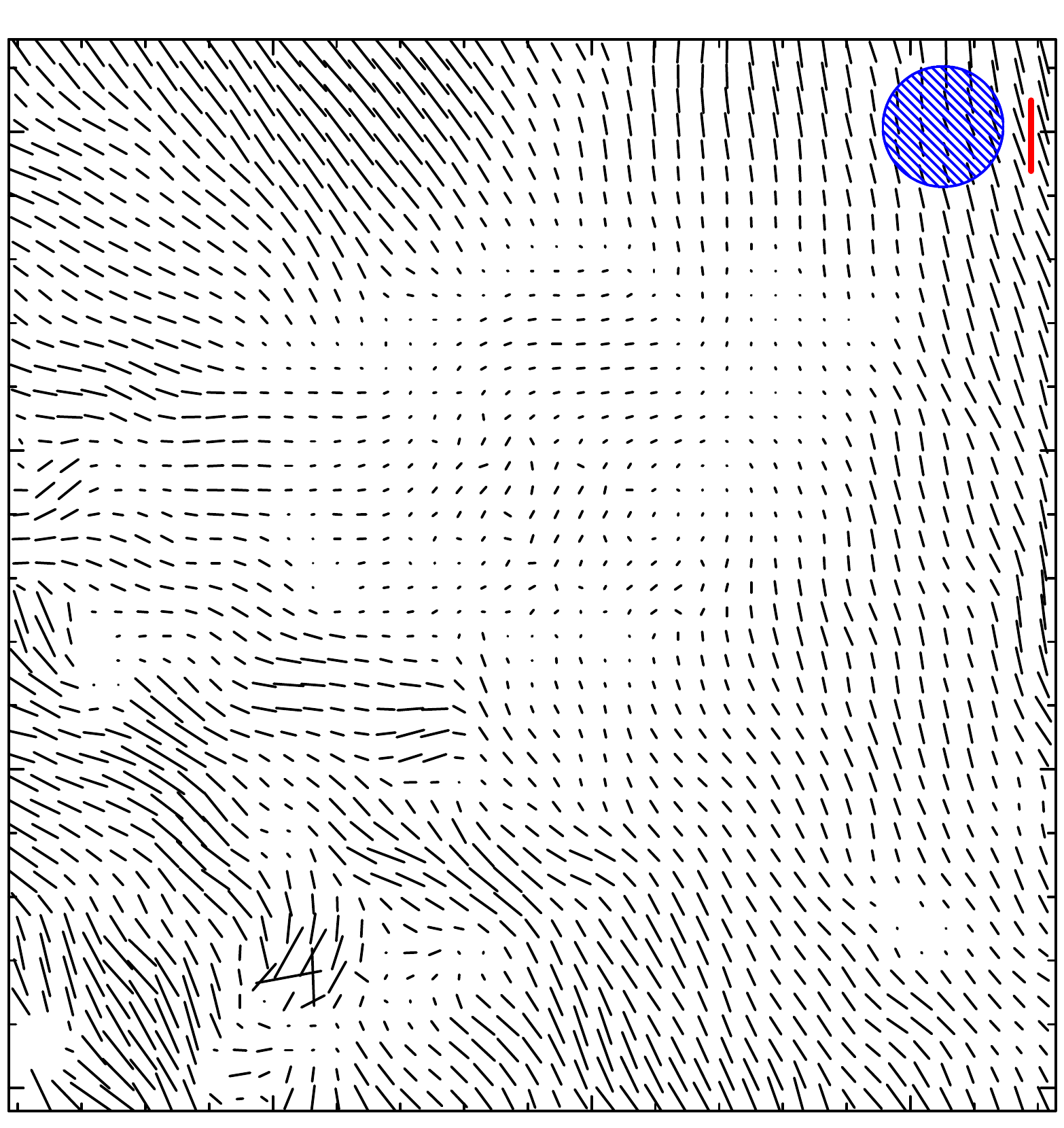}





\vspace{5mm}\caption{ 
Zoom in to $P'$ image from Fig.\,\ref{hawcmaps}'s lower panel on the indicated colour scale, for all pixels with $I$ $>$ 0.25 Jy/pixel, plus white $I$ contours at 0.25($\sqrt{2}$)16 Jy/pixel, blue HAWC+ beam, and red 10\% $p'$ vector scale.  We also show the $p'$ vectors at every pixel (0.2 beam) unmasked by S/N, since here low-S/N $p'$ vectors are very small anyway.  Also shown are positions of the outer \& inner boundary layers (OBL, black; IBL, red), described in the text.
}\label{nullflux}\vspace{0mm}
\end{figure}

We begin our exploration of the molecular core's $B$ field as revealed by HAWC+.  Zooming in to the inner portion of the molecular cloud near the massive protostar MIR\,2 (P18), a very striking feature of the polarised emission stands out immediately -- see {\color{red}Figure \ref{nullflux}}.  There is a strong, narrow null in $P'$ curving around the western side of the molecular peak, between it and the \hii\ region, in particular the darker blue colours indicating very low $P'$.  The EW width of this null is quite small, apparently only 1 pixel or $\lesssim$7000 AU (i.e., much less than the angular resolution of HAWC+, but we show this is not unphysical below).  Further, this null can be traced most of the way around the molecular peak, although it broadens out somewhat to the N, S, and E, to an approximate width of 3--6 pixels, or 0.1--0.2\,pc.

The area enclosed by this boundary layer, signified by where $P'$ rises above 0 on the inside, is an approximately elliptical zone of size 0.55$\times$0.40\,pc at PA $\approx$ 80\degree\ (labelled IBL for {\em inner boundary layer} in Fig.\,\ref{nullflux}).  The {\em outer boundary layer} of the null is a vaguely elliptical polygon of approximate height 0.91\,pc and width 0.60\,pc, at PA $\sim$ 20\degree\ (labelled OBL in Fig.\,\ref{nullflux}).  The space between these boundary layers has essentially zero (i.e., S/N$<$3) FIR polarisation and $B_{\perp}$. 

\begin{figure}[h]
\vspace{0mm}
\centerline{\includegraphics[angle=-90,scale=0.36]{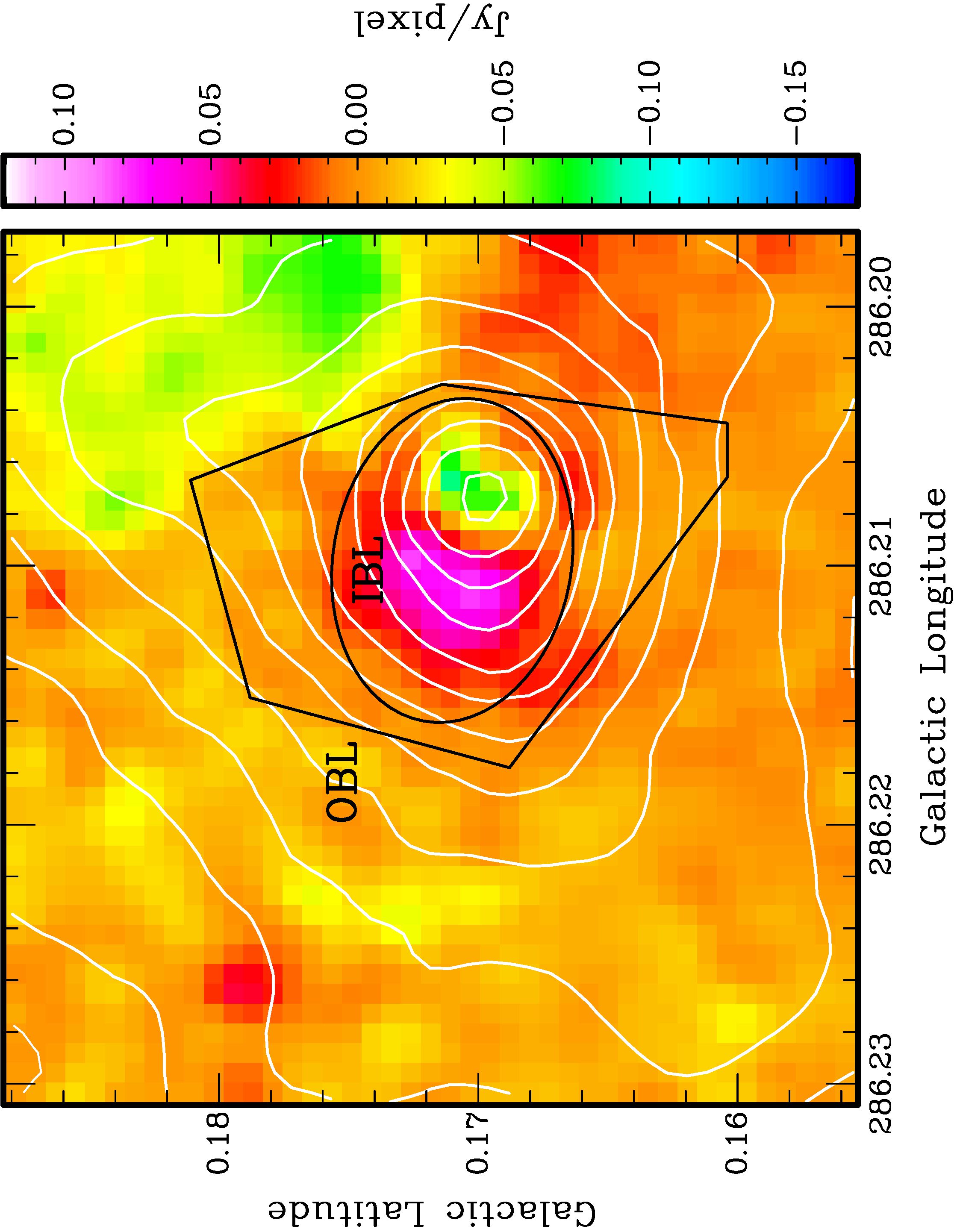}}\vspace{-5mm}
\centerline{\includegraphics[angle=-90,scale=0.36]{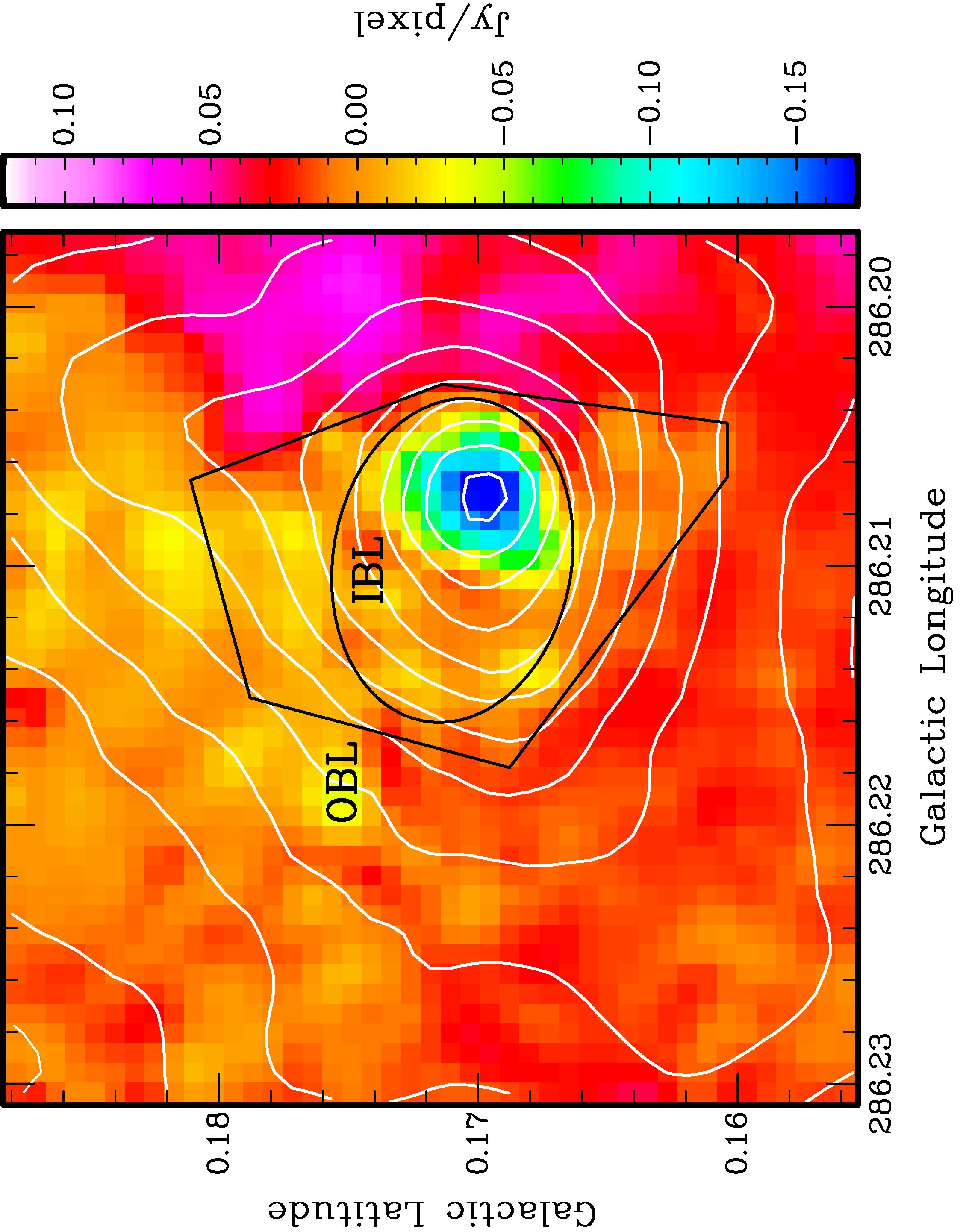}}

\vspace{-130mm}\hspace{15mm}{HAWC+ $Q$ image}

			\hspace{10mm}{$I$ cntrs: 0.25($\sqrt{2}$)16 Jy/pix}

\vspace{58mm}\hspace{15mm}{\color{white}{HAWC+ $U$ image}

			\hspace{10mm}{$I$ cntrs: 0.25($\sqrt{2}$)16 Jy/pix}} 

\vspace{57.5mm}\caption{ 
({\em Top}) Same area of the inner molecular clump as Fig.\,\ref{nullflux}, except showing only Stokes $Q$ data with $I$ contours.  ({\em Bottom}) Same as top panel but for Stokes $U$.  Both $Q$ and $U$ images are displayed on the same scale, with zero as orange.  
}\label{qunulls}\vspace{0mm}
\end{figure}

To see why the narrow western null is real, we show in {\color{red}Figure \ref{qunulls}} the Stokes $Q$ and $U$ maps in the same zoomed area as Figure \ref{nullflux}.  Carefully comparing the three images in Figures \ref{nullflux}-\ref{qunulls} on the western side of the MIR\,2 core, one sees that where $P'\approx0$ (e.g., close to the $I$ = 5.6 Jy/pixel contour), $Q$ drops from positive to negative values going into the centre, and $U$ behaves similarly ($Q$ and $U$ are both 0 where the images are orange).  Therefore, the $P'$ null is very narrow here because both $Q$ and $U$ are changing rapidly through 0 at the same locations (but in a manner consistent with HAWC+'s angular resolution), from larger positive to larger negative values as one traverses into the centre of the core.

In other words, each of these {\em components} of $P'$ is strongly reversing sign on this boundary, corresponding to a null in $P'$ and 90\degr\ change in direction (due to the definition of the Stokes parameters) for both the observed polarisation angle and inferred $B$ field direction $\theta_{B_{\perp}}$ across this boundary.  West of MIR\,2, this change is very sharp, much less than a beamwidth.  Indeed, this change of direction is visible in the $p'$ vectors themselves, as shown in Figure \ref{nullflux}.  The vectors just outside the null, i.e., in the \hii\ region and also east of the MIR\,2 core, are all oriented roughly east-west, while the vectors inside the null, especially close to the MIR\,2 core, are mostly oriented roughly north-south.  The change in PA is sharpest where $Q$+$U$ are reversing most sharply, just west of MIR\,2.  For the small-$P'$ locations between the two boundary layers to the N, E, and S of the MIR\,2 core, the changes in sign for $Q$ and $U$ are more gradual.  But they do change sign in each case, when looking from the outer areas of the cloud to the centre.

In the more gradually changing boundary N, E, \& S of the core, the $P'$ nulls seem to indicate an area where the 2D plane-of-sky $B$ field component ($B_{\perp}$) is dropping to zero, since the $P'$-null boundary's width is roughly beam-sized and there are few pixels with $P'$$>$3$\sigma$.  In contrast, the sharp $P'$-null boundary west of MIR\,2 is much thinner.  Rather than merely dropping to zero, this seems to be due to a 90\degr\ change in direction alone, 
i.e., that {\em $B_{\perp}$ is sharply changing in direction at the boundary layer around MIR\,2, producing a null in $P'$ without a corresponding null in $B_{\perp}$}.

An alternative situation at the IBL is that the 3D ${\bf B}$ vector is aligned very close to our line of sight there, i.e., that ${\bf B}$ consists only of $B_{||}$.  In other words, looking progressively from the \hii\ region (west) through the IBL and towards MIR\,2 and its immediate east, the 3D ${\bf B}$ field orientation points one way ($\sim$EW) in the \hii\ region, twists through 90\degr\ at the null so {\bf B} points directly towards or away from us at the IBL, and then twists another 90\degr\ to point in another 3D direction ($\sim$NS) nearer to MIR\,2 within the IBL, with $Q$ \& $U$ changing sign in the process.  We consider this a less likely proposition, however, since the $P'$ is so narrow, any $B_{||}$ would put additional unresolved structure into the IBL, and it would require a rather large flux annulus to be pointed right at us, a fairly significant ``finger of God'' effect in our view.  On the other hand, a pure 90\degr\ change to $B_{\perp}$ alone might require a discontinuity in $B_{\perp}$.  To resolve this, higher-resolution polarisation data could reveal more details to this narrow change in $B_{\perp}$ (see \S\ref{almaIBL}), while Zeeman measurements 
could measure $B_{||}$ directly, potentially distinguishing between a single 90\degr\ $B_{\perp}$ twist or additional $B_{||}$ in the IBL. 

A third possibility for the IBL's apparent null is that the polarisation signal comes from dust emission on the low-opacity (west, \hii\ region) side, but dust absorption on the high-opacity (east, molecular core) side, making the null more of a polarisation radiative transfer effect, without necessarily implying anything significant for the cloud's inherent $B$ field.  This would be similar to the FIR polarisation pattern in Sgr B2, although observed at a much coarser physical resolution of 1.5\,pc (35$''$ beam) with the KAO \citep{ndd97}.  In order to be relevant, the dust opacity in the core would need to be \gapp1; however, based on P18's SED fitting to {\em Herschel} and other data, we compute an effective peak $\tau$ $\approx$ 0.001 for BYF\,73 at 154\,$\mu$m and 37$''$ resolution.  Allowing for HAWC+'s  7.4$\times$ smaller beam area, it is unlikely that the peak $\tau$ within the IBL is more than 7$\times$ higher than this.  Even in ALMA's beam, $\sim$30$\times$ smaller in area than HAWC+'s, $\tau$ could rise to $\sim$0.2 at 154\,$\mu$m if all the flux were coming from MIR\,2, but this is still less than 1 and we know MIR\,2 contains less than half the flux there, \S\ref{comps}.  Therefore, we can discount this possibility. 
For now, we prefer the pure $B_{\perp}$-twist interpretation since it is the simplest.  

\begin{figure}[t]
\centerline{\includegraphics[angle=-90,scale=0.38]{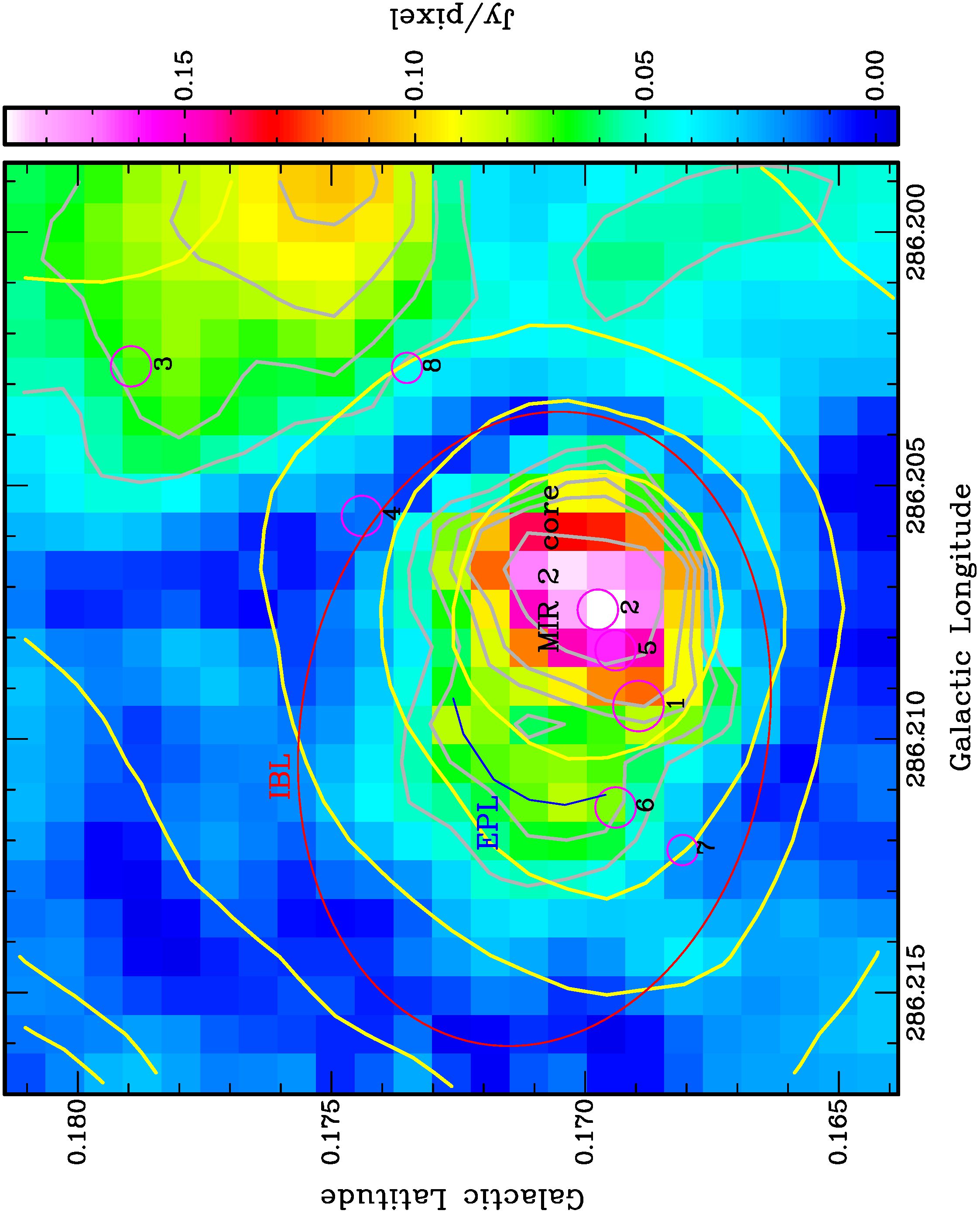}}\vspace{-3.7mm}
\centerline{\includegraphics[angle=-90,scale=0.38]{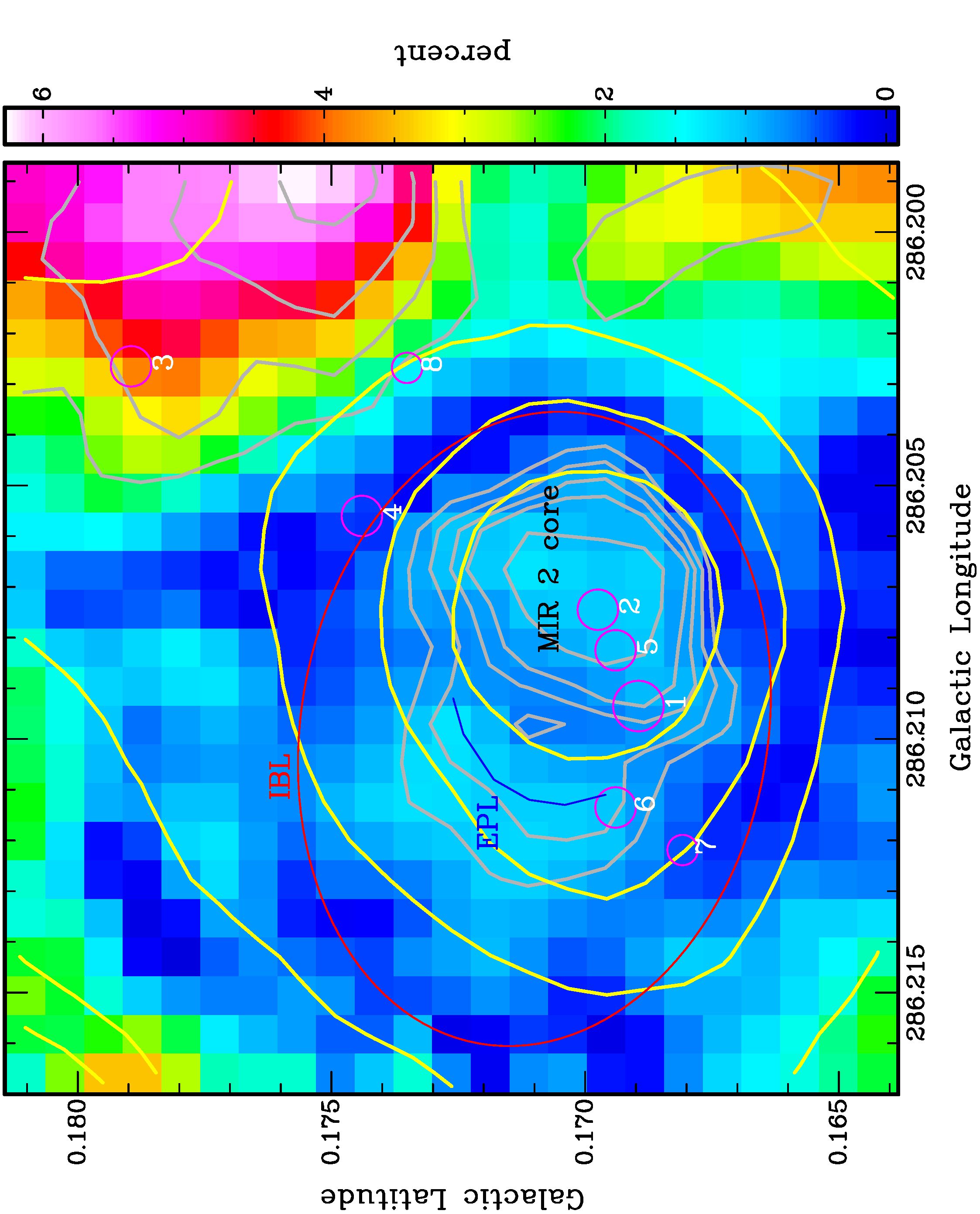}}

\vspace{-140mm}\hspace{12mm}{HAWC+ $P'$ image} \\
			{\color{yellow}\hspace{17mm}$I$ contours} \\
			{\color{white}\hspace{16mm}$P'$ contours}

\vspace{62mm}\hspace{12mm}{HAWC+ $p'$ image}

			{\color{yellow}\hspace{17mm}$I$ contours}

			{\color{white}\hspace{16mm}$P'$ contours} 

\vspace{60mm}\caption{ 
Further zoom-in (compared with Figs.\,\ref{nullflux}--\ref{qunulls}) to MIR\,2 area in $P'$ ({\em top}) and $p'$ ({\em bottom}), with the same contours as in Fig.\,\ref{irac} (i.e., $I$ = yellow at 0.44(0.10)0.84, 1.5, 3, 5, 9 Jy/pixel, $P'$ = grey at 50(16)98, 140 mJy/pixel).  Here we also label the locations of MIR\,1--8 from P18 (numbers with magenta circles), the MIR\,2 core (black), the eastern polarisation lobe (EPL, blue), and the inner $P'$ = 0 boundary layer (IBL, red).
}\label{pplobes}\vspace{0mm}
\end{figure}

Upon further inspection of the inner 0.5\,pc $\sim$ 0\fdeg01, one can see another significant feature around MIR\,2 in the polarisation maps, even where the central $I$ structure is very smooth.  While the $I$ and $P'$ emission peak exactly on MIR\,2's position, the $I$ morphology is slightly more extended to the east, compared to the sharper decline towards the west/\hii\ region.  This morphology is mimicked in the inner $P'$ distribution, i.e., where $P'$ $>$ 0 inside the IBL, except that in $P'$ the point-source nature of MIR\,2 is much more distinct, while the extension to the east is revealed as a semi-circular ring structure adjacent to MIR\,2.  We dub this the ``eastern polarisation lobe'' (hereafter EPL).  Morphologically, it seems unlikely that this lobe is associated with any of MIR\,1--8, as can be seen in {\color{red}Figure \ref{pplobes}}, which overlays their positions on both the $P'$ and $p'$ images.  This is underscored by indications from P18 that MIR\,1 and 3--8 are possibly on the near side of the BYF\,73 cloud, and not as deeply associated with the MIR\,2 core.

Intriguingly, the EPL shows a similar polarisation signature to the MIR\,2 core proper (see top panel of Fig.\,\ref{qunulls}) but inverted in $Q$, going from negative values outside the EPL to positive values across it.  This is equivalent to a sharp rotation of the inferred $B$ field between each structure as we will see in the next section, further suggesting that the EPL is distinct from the MIR\,2 core/protostar, each having its own physics, despite the much more amorphous appearance around MIR\,2 in Stokes $I$ (Fig.\,\ref{hawcmaps}).  
Identification of these features was based solely on the HAWC+ data, and before the ALMA maps (next) were in hand.

\begin{figure*}[t]
\vspace{0mm}
\centerline{
\hspace{-5mm}\includegraphics[angle=-90,scale=0.55]{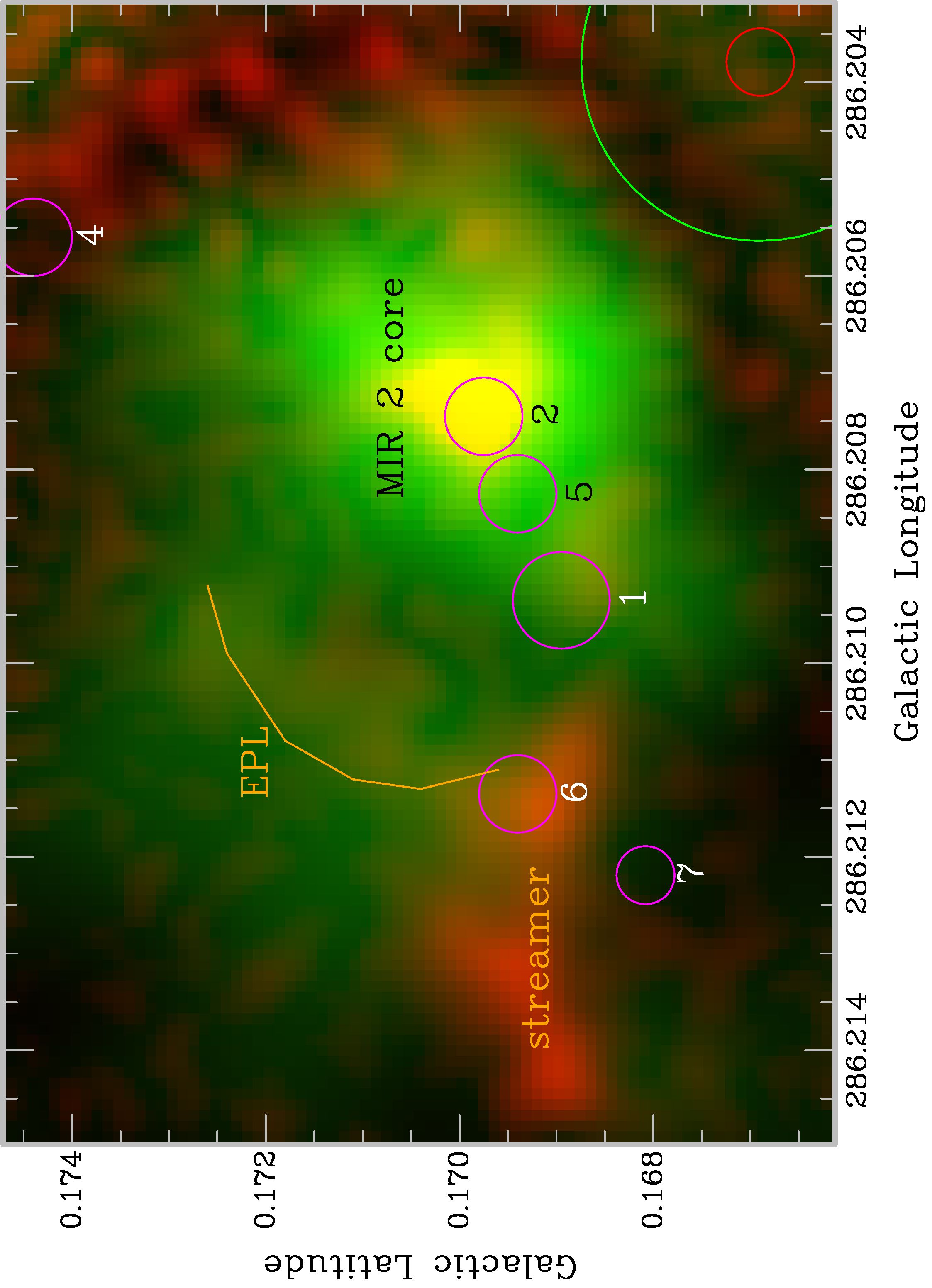}}\vspace{-102.8mm}
\hspace{36mm}\includegraphics[angle=-90,scale=0.556]{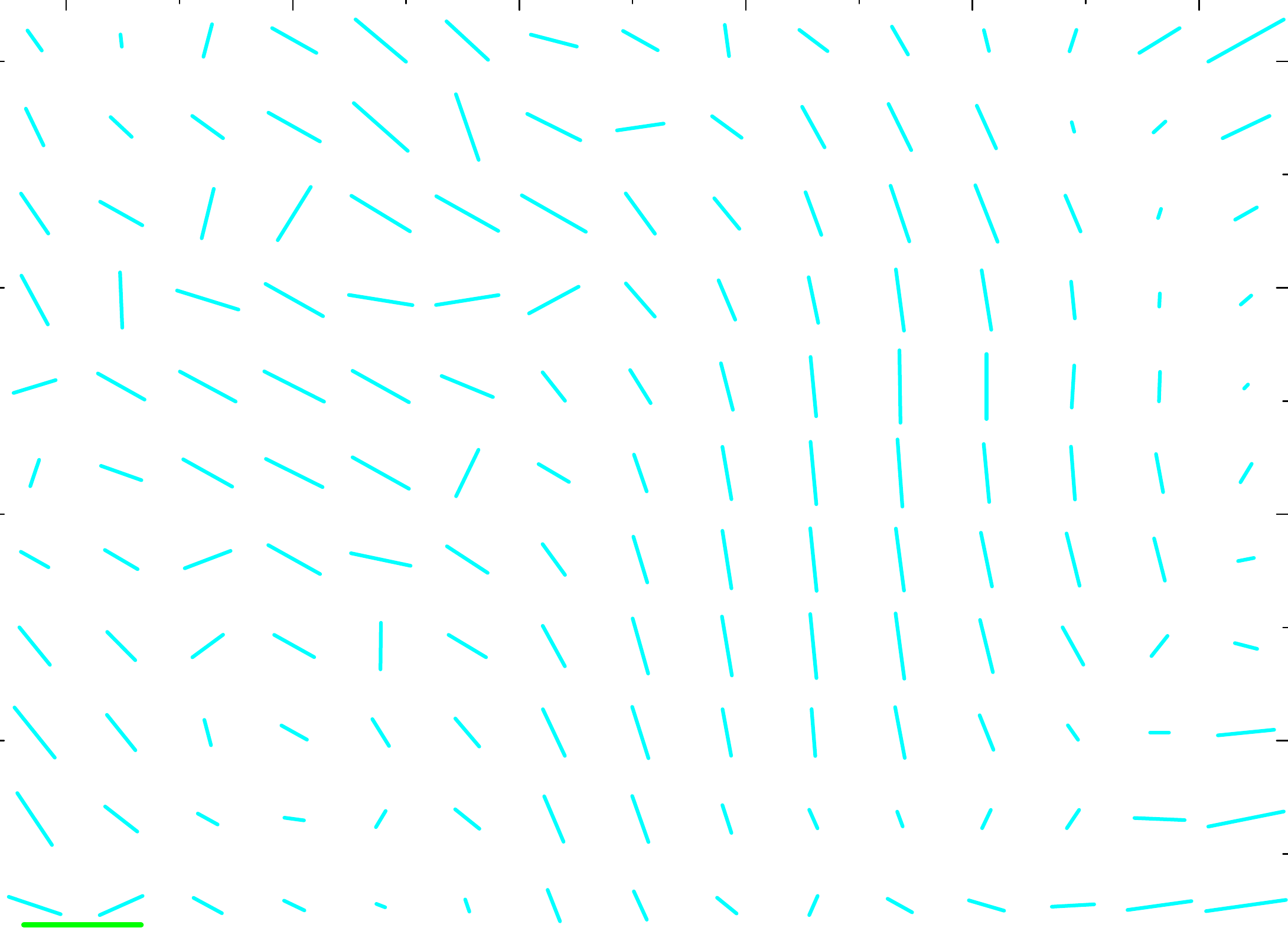}

\vspace{-89.6mm}\hspace{33mm}\includegraphics[angle=-90,scale=0.567]{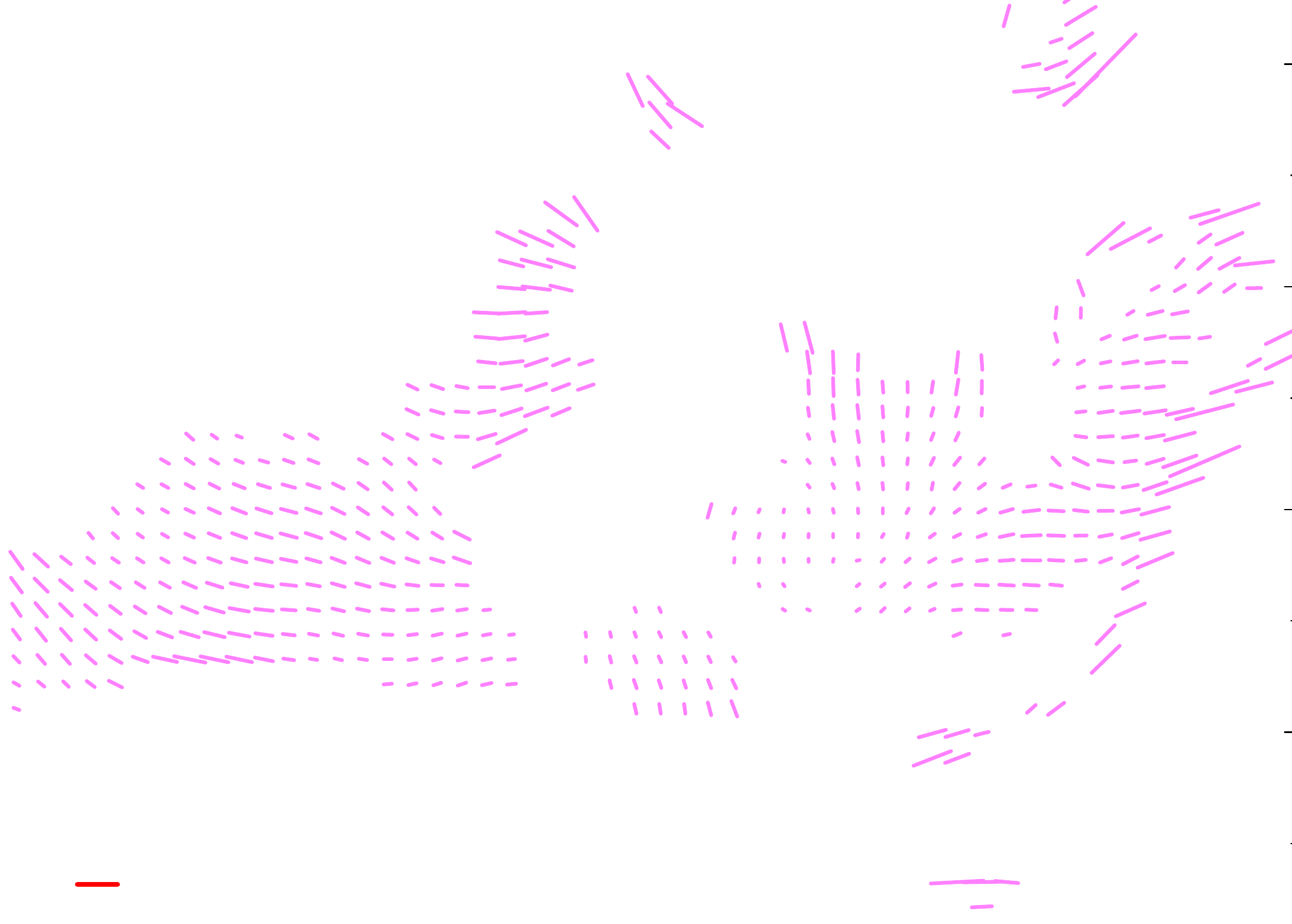}

\vspace{-84mm}\hspace{38mm}{\large {\color{green}HAWC+} {\color{red}ALMA} {\color{white}$P'$ images}}

\vspace{72mm}\hspace{39mm}{\color{green}2\%} {\color{red}30\%} 

\vspace{17mm}\caption{ 
Zoom in to all high-S/N polarisation data among the central structures of the BYF\,73 molecular core, from HAWC+ (green $P'$ image with displayed range 0--190\,mJy/pixel, and cyan rotated $p'$ vectors with maximum 1.47\%) and ALMA (red $P'$ image with displayed range 0--32\,mJy/beam, and pink rotated $p'$ vectors with maximum 65\%).  These are like Fig.\,\ref{nullflux} for HAWC+ and the bottom panel of Fig.\,\ref{almaCont} for ALMA, with coloured $p'$ vector scale bars in the bottom-left corner and coloured beam sizes in the bottom-right.  For HAWC+, the typical uncertainty above 50\,mJy/pixel is $\Delta P'$ = 5--6\,mJy/pixel and $\Delta\theta$ = 2\degree, for a S/N = 10--30.  For ALMA, the uncertainty is $\Delta P'$ = 23\,$\mu$Jy/beam, giving a typical $\Delta\theta$ = 5\degree and S/N = 5--10.  Six of the 8 MIR point sources from P18 are also shown for reference, as are labels for the MIR\,2 core, and in orange, the ALMA Streamer and the arc of the EPL.  
}\label{IBLpolo}\vspace{0mm}
\end{figure*}

\subsection{Magnetic Field Structures in the Molecular Core: ALMA}\label{almaIBL}

Turning now to the ALMA data, the only high-column-density dust structures seen in the 3\,mm continuum (Fig.\,\ref{almaCont}) are (i) MIR\,2; (ii) the 1--3\,mJy/beam structures east and west of it which we call the ``Streamer'' and ``Streamer-west,'' respectively; (iii) a $\sim$0.5\,mJy/beam linear feature aligned almost exactly N-S with the \hii\ region's ionisation front (IF); (iv) another $\sim$0.5\,mJy/beam patch NE of MIR\,2 aligned with the EPL; (v) some even weaker diffuse features $\sim$1$'$ to the east of MIR\,2; plus (vi) three other eastern point sources which we have designated MIR\,11-13 in Figure \ref{irac}.  The larger-scale features in Stokes $I$ at the two different wavelengths have a very nice overall correspondence, despite the different resolutions: the EW extension of the brightest core emission, the weaker emission extending north along the IF, and the new point sources MIR\,11--13 and diffuse emission extending to the east all look mutually consistent.  Even the EPL's structure, inferred from the HAWC+ data alone, is easily and gratifyingly verified in the ALMA $I$ images.  The detectable ALMA polarisation, however, is limited to a subset of these features, namely MIR\,2, both sides of the Streamer, the EPL, and possibly the southernmost parts of the IF (near MIR\,4).

The ALMA $P'$ map is therefore much more spatially compact than the HAWC+ $P'$ map.  However, the ALMA $p'$ values in the molecular core are quite large, typically 5--20\% or more in high S/N areas, as opposed to the more typical HAWC+ $p'$ values around 1\% within the IBL/molecular core (HAWC+ $p'$ is 10\% or more only in the \hii\ region, but does rise to $\sim$5\% in the diffuse, eastern extremes of the molecular cloud).  This lower percentage polarisation at shorter wavelengths could be due to two effects:

(A) The polarisation signal is being diluted in the larger HAWC+ beam due to its origin in small structures, such as those found in the ALMA maps, but which to some extent cancel each other out in the HAWC+ beam.  For example in the MIR\,2 core, the correspondence between HAWC+ and ALMA vectors is modest, and the ALMA vector PAs vary more strongly than the HAWC+ PAs.  However, due to the correspondence between both HAWC+ and ALMA inferred $B$ field morphologies described below, we discount this effect.

(B) More probably, in the denser parts of the cloud, the 3\,mm emission is more efficiently polarised by the cold dust than the 154\,$\mu$m emission: the ``polarisation spectral index'' (PSI) is $>$1.  This would run counter to the situation in the $\rho$ Oph cloud, where radiative torques from external illumination are thought to more efficiently align grains in the {\em less dense} parts of that cloud, giving a PSI $<$ 1 \citep{scd19}.  Here, we argue that MIR\,2's radiation could be aligning grains more efficiently in the cloud core, if radiative torques from {\em internal} illumination are the cause \citep{L07}.

We overlay both instruments' polarisation maps of the IBL, the peak column density area in all maps, in {\color{red}Figure \ref{IBLpolo}}.  Within the cold, high column density dust preferentially traced by the 3\,mm maps, we note two distinct magnetic domains comprised of five sub-structures, 
each with its own orientation and character:  

(1a) Close in to MIR\,2, the field is oriented mostly N-S, which is very similar to that inferred from the HAWC+ data, but with amplitudes $p'$$\sim$1\% for HAWC+ and $p'$\gapp3\% for ALMA.  We call this the ``MIR\,2 core.''

(1b) Just to the SE of MIR\,2, at the western end of the main Streamer, there is a small patch of polarisation with a similar N-S orientation, which we call the ``MIR\,2 extension.''

These two structures comprise the predominantly N-S magnetic domain inside the IBL.  The following three structures comprise a different magnetic domain, oriented mostly E-W or somewhat NE-SW.

(2a) Across the EPL, the uniformity is almost as good as in the MIR\,2 core, with most HAWC+ vectors $p'$$\sim$1\% @ N60{\degree}E, while the ALMA vectors range over $p'$=10--20\% and run mostly E-W, although some vectors turn towards $\sim$N60{\degree}E at the more distant fringe from MIR\,2.  

(2b) Along the main Streamer east of MIR\,2, ALMA vectors are $p'$$\sim$5--15\% while running mostly E-W nearer to MIR\,2, but again turning more towards $\sim$N60{\degree}E as they move away from MIR\,2.  HAWC+ does not detect high S/N polarised emission from the Streamer; thus, its vectors are somewhat jumbled in orientation there, but their alignment with the ALMA vectors is still reasonably good.

\begin{figure*}[ht]
\centerline{\includegraphics[angle=-90,scale=0.6]{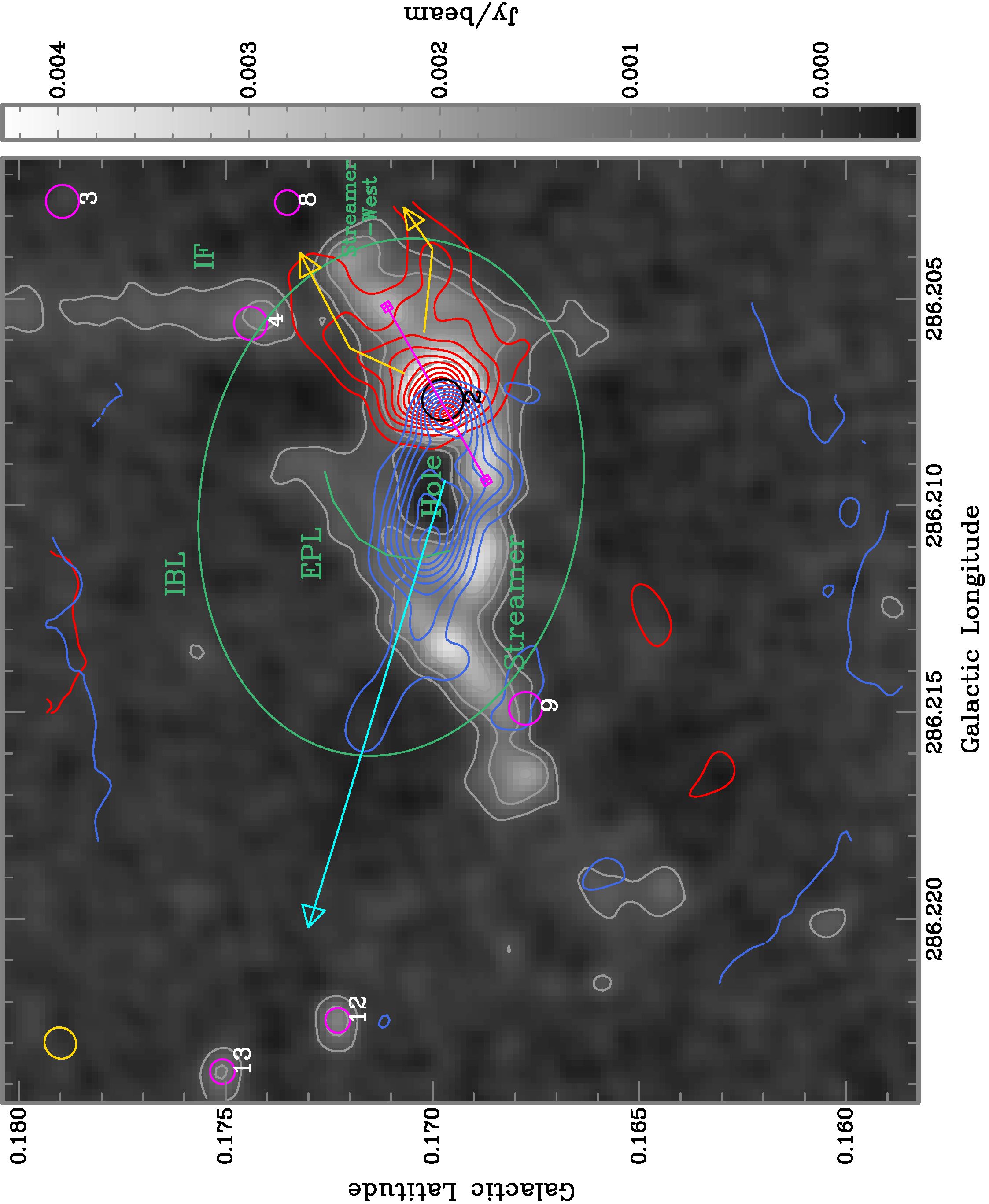}\hspace{0mm}}
\vspace{-1mm}\caption{ 
\tco\ outflow features overlaid on the 3\,mm Stokes $I$ continuum mosaic from Fig.\,\ref{almaCont}; the latter is displayed as a greyscale + grey contours at 0.4, 1, 2, \& 5\,mJy/beam.  The blue and red wings of the \tco\ Stokes $I$ emission are shown with blue and red contours at levels of 30(40)450\,K\kms\ in each case, integrated from --60 to --22\,\kms\ and --17.5 to +18\,\kms\ respectively, for all voxels above 5$\sigma$ using the smooth-and-mask (SAM) algorithm \citep{rbv90}.  The lowest red and blue contours also approximately indicate the circular boundary of the single ALMA polarisation field at the 20\% primary beam cutoff.  The average rms noise in the \tco\ line wing maps is 0.14 (blue) and 0.25 (red) K\kms, giving S/N peaks $\sim$2400 and 1850, respectively.  Green labels show various continuum features as discussed in the text, along with selected MIR point sources (Fig.\,\ref{irac}) in magenta with white labels, except for MIR\,2 which is shown in black.  The synthesised beam is 2\farcs83$\times$2\farcs66 @--35.4\degr, shown in the top-left corner as a gold ellipse.
}\label{almaCOjet} \vspace{0mm}
\end{figure*}

(3) In the Streamer-west, while the HAWC+ vectors continue to align N-S, the ALMA vectors turn E-W, but this is beyond the reliably-calibrated polarisation radius in the ALMA field, so the divergence may not be significant.  It is also possible that, because of the larger beam, the HAWC+ data are dominated by the bright emission from MIR\,2 (polarised N-S) further into the Streamer-W, IF, and \hii\ region than are the ALMA data, before HAWC+ finally picks up the E-W $B$ field orientation in the \hii\ region itself.  If true, this would make the polarisation signal from both instruments more consistent with each other here too, as per effect B above.  

In terms of the $P'$ null and sharp 90\degr\ $B_{\perp}$ twist seen in the HAWC+ data, Figure \ref{IBLpolo} seems to suggest that it is mostly an artifact of resolution and sensitivity, as per the pure $B_{\perp}$-twist explanation (\S\ref{hawcIBL}).  In other words, we can see the inferred $B_{\perp}$ direction change quickly between the MIR\,2 core and the Streamer-west, right under the edge of the N-S $B_{\perp}$ vectors in Figure \ref{IBLpolo}, even if the ALMA Streamer-west vectors are less reliable there.

In summary, the significant $B$ field structures in the molecular core of BYF\,73 are MIR\,2, the EPL, and the Streamer, where both the HAWC+ and ALMA inferred $B$ fields are broadly consistent.  The $B$ field structures seen by both facilities in the \hii\ region may also be consistent with each other.
We reserve discussion of the $B$ field structures in the \hii\ region for \S\ref{DCFstuff}.



\section{Features of the Spectral Line Emission}\label{cubes}

\subsection{A Strong Bipolar Outflow in \tco}\label{outflow}
The continuum structures seen in the molecular core at both the SOFIA/HAWC+ and ALMA wavelengths are intriguing, both in total intensity and polarised emission.  However, while apparently related to the dominance of MIR\,2, from their structure alone their physical significance is not entirely obvious, nor is how they are connected to BYF\,73's star-forming activity.

Not surprisingly, the ALMA spectroscopy provides important insights; perhaps more surprising is that it is the \tco\ data that provide the key.  Although the \tco\ spectral polarisation cubes only cover a single ALMA field as in the bottom panel of Figure \ref{almaCont}, the information they reveal about the nature of the continuum emission in BYF\,73 is pivotal.  First, the brightest \tco\ emission by far lies in the highly Doppler-shifted line wings, extending up to $\pm$35--40\,\kms\ from the cloud's systemic \vlsr, as illustrated in {\color{red}Figure \ref{almaCOjet}}.  Spatially, these line wings delineate a massive bipolar outflow clearly emanating from MIR\,2.  The opening angle appears small near MIR\,2, $\theta_{\rm open}$\lapp10\degr, and the outflow appears to impact the Streamer: the flow directions are apparently strongly affected by the large inertia of ambient cloud material in the Streamer, and deviate from their initial vectors.  Based on the small $\theta_{\rm open}$ and lack of overlap between the red \& blue wings, we estimate the outflow's inclination to the line of sight lies in 40\degr \lapp $\theta_{\rm incl}$ \lapp 80\degr.

From detailed inspection of the \tco\ Stokes $I$ cube, the intrinsic outflow direction from MIR\,2 is along the magenta line in Figure \ref{almaCOjet} at a Galactic PA = 120\degr, but this terminates at the magenta boxes at each end of that line.  The red-shifted outflow then deviates around both sides of the Streamer-West along the paths indicated by the gold arrows in Figure \ref{almaCOjet}.  The blue-shifted outflow is apparently deflected by the highest-density portion of the Streamer into the direction shown by the cyan arrow of Figure \ref{almaCOjet}.  Spectrally, the highest relative speeds appear to lie near MIR\,2 in the red wing, but are displaced from MIR\,2 by $\sim$14$''$ = 0.17\,pc downstream in the blue wing.  Apart from this offset, the outflow speeds are generally more modest as one looks further downstream.

In the case of the blue wing, the outflow direction along Galactic PA = 120\degr\ close to MIR\,2 is seemingly deflected by a clear 47\degr\ ``bounce'' into a single new direction along PA = 73\degr. The flow then continues to at least the eastern edge of the single polarisation field, 0.6\,pc away from MIR\,2.  This deflection is clearly seen in the individual \tco\ Stokes $I$ cube's channel maps, and is not an artifact, for example, of opacity-masking of a more southerly portion of the blue wing by the Streamer, hiding a continued blue outflow along PA = 120\degr.  This is because (1) the Streamer and outflow are well separated in \vlsr, so there is no opportunity for the Streamer to mask some parts of the blue wing (see also below); and (2) the \ttco\ line, which is much lower opacity than \tco, shows the same structural features coincident with the deflection 9$''$ east of MIR\,2.

In the case of the red wing, the initial outflow from MIR\,2 along PA = --60\degr\ appears to be somewhat blocked by the Streamer-west, such that the flow deviates to either side of this obstruction, before continuing to flow at the same PA to the western edge of the field, 0.3\,pc away from MIR\,2.  These deviations are similarly easy to see in the channel maps.

Note that the outflow widths, at $\sim$10$''$--15$''$, are well under the ALMA MRS in the single \tco\ field, so we believe we recover essentially all the outflow structure in the line wings.  It is difficult to say, however, if the outflow continues beyond these boundaries (e.g., into the \hii\ region), since the \tco\ data are limited by the field of view.  But it is fairly obvious that the outflow and Streamer interact strongly, one sculpting the other, including the appearance of the EPL and Hole.

The second noticeable feature of the \tco\ data, apart from where the outflow can be specifically traced in to MIR\,2, is that the rest of the cloud is {\em much} fainter in the line core between roughly --22 and --17\,\kms, with any non-outflow features being $<$10\% as bright.  This confirms that the cloud is extremely optically thick everywhere, and that except for the large outflow-driven Doppler shifts, virtually no \tco\ emission can escape from the cloud's interior.

Third, and even more interestingly, we detect the linearly-polarised Goldreich-Kylafis effect {\em almost everywhere in the outflow}, and at high S/N in $P_d$ in almost all channels which trace the outflow in $I$.  This is 
presented and analysed in \S\ref{GKstuff}.

\begin{figure*}[ht]
\centerline{\includegraphics[angle=-90,scale=0.55]{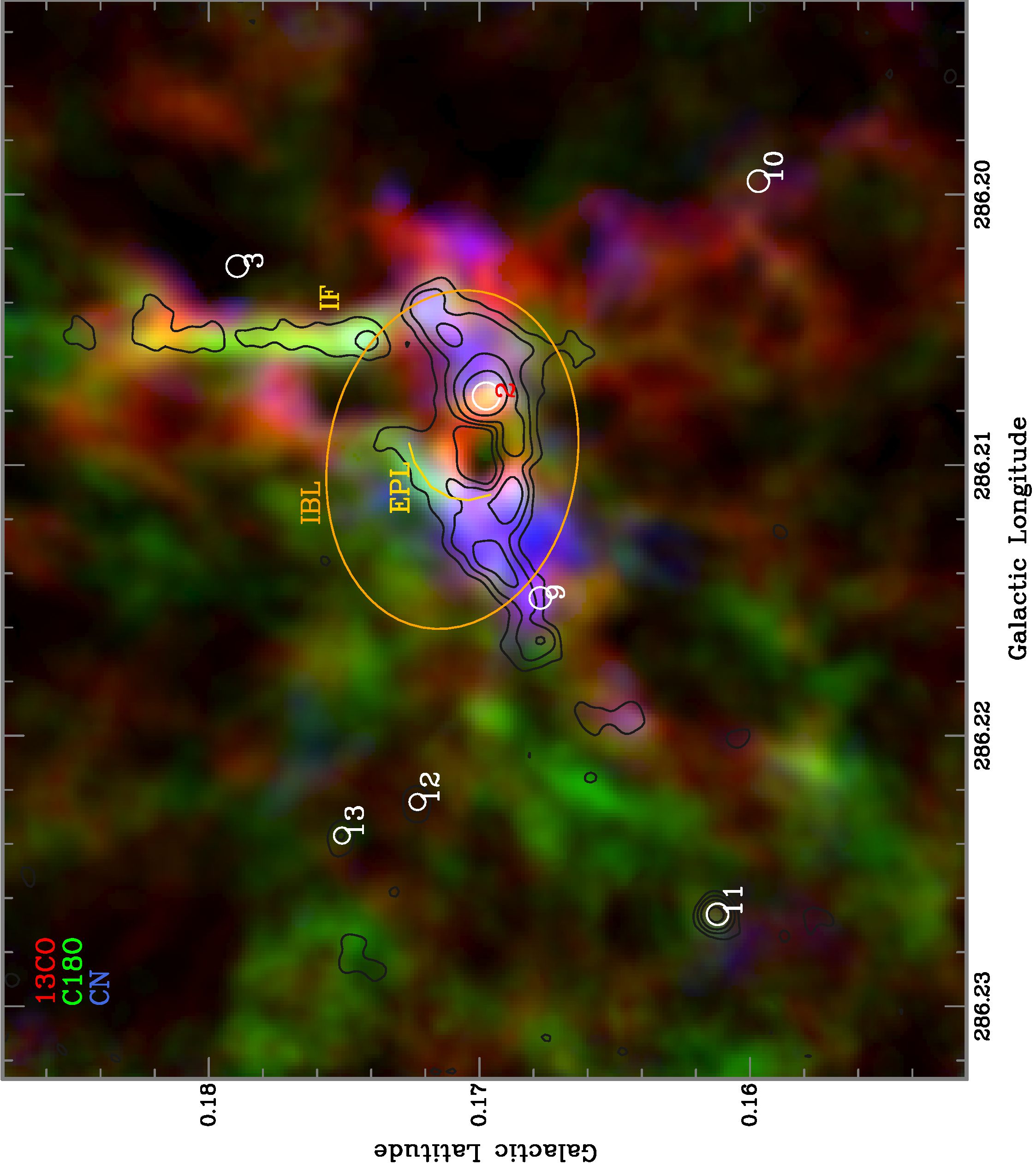}\hspace{1mm}}
\vspace{-1.5mm}\caption{ 
Slightly cropped composite RGB image of ALMA spectral line mosaics' total intensities \citep[species as labelled, integrated from --24.5 to --15.5\,\kms\ for all voxels above 5$\sigma$ using SAM;][]{rbv90} and overlaid by contours (at 0.4, 1, 2, and 5 mJy/beam) of the 3\,mm continuum (Fig.\,\ref{almaCont}) plus selected MIR point sources (Fig.\,\ref{irac}).  The IBL, EPL, and IF (Figs.\,\ref{pplobes}--\ref{almaCOjet}) are also labelled in yellow.  The brightness scales in each colour channel run from dark at 0 to saturation at 38.15, 11.71, \& 11.65\,K\kms, respectively; the respective overall peak levels, and error levels in areas away from the map edges, are 53.28$\pm$0.11, 13.42$\pm$0.08, and 16.66$\pm$0.21\,K\kms.  Thus, the \ceto\ and CN moment maps have typical S/N $\approx$ 20--60, while that of \ttco\ is 100--300 across much of the mosaic.  
}\label{almaCOrgb} \vspace{-1mm}
\end{figure*}

\begin{figure*}[ht]
\centerline{\includegraphics[angle=-90,scale=0.43]{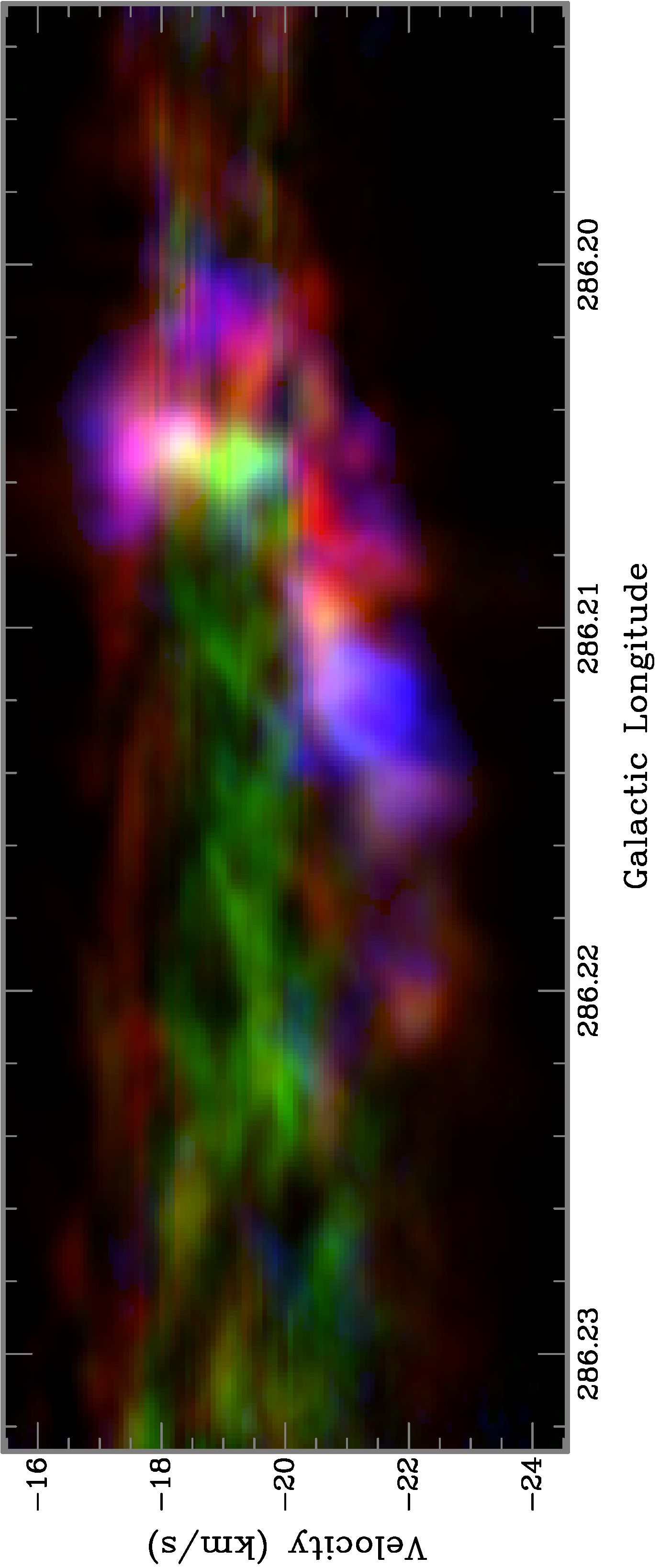}\hspace{1mm}}
\vspace{-1.5mm}\caption{ 
Longitude-velocity diagram of the same data presented in Fig.\,\ref{almaCOrgb}, i.e., integrated over the same latitude range as displayed therein.  The brightness scales are dark for 0\,K\,arcmin in each colour channel, and the saturation/peak $\pm$ uncertainty levels (in areas away from the map edges) are \ttco\ 10.42/12.77$\pm$0.04 (red), \ceto\ 4.14/5.08$\pm$0.04 (green), and CN 1.73/2.69$\pm$0.10\,K\,arcmin (blue).  
}\label{almaLVrgb} \vspace{0mm}
\end{figure*}

\subsection{Cloud Architecture from Spectral Line Mosaics}\label{moments}

The ALMA \ttco, \ceto, \& CN mosaics provide further details for analysis of the BYF\,73 cloud emission, but we focus here on kinematic features associated with the Streamer and MIR\,2, in order to shed further light on the $B$ field structures described above, and their dynamics.

In contrast to the very compact 3\,mm continuum emission (compared to the mosaic size, Fig.\,\ref{almaCont}), the \ttco, \ceto, \& CN mosaics illustrated in {\color{red}Figure \ref{almaCOrgb}} all show much more extended structure, although this emission is brightest near the continuum features, and for \ttco, across much of the IF visible in the {\em Spitzer} images as well (Fig.\,\ref{irac}).  The \ttco\ and \ceto\ emission fills most of the mosaic, seemingly even extending beyond it, in all directions for \ttco, and to the north and east for \ceto.  Even the CN is somewhat extended, although less so than the iso-CO lines.  The \ttco+\ceto\ extents include parts of the \hii\ region (the area west of the IF), presumably due to residual molecular gas on its near and far sides that has not yet been ionised by the UV field or swept up in the general \hii\ region expansion.  Some of this effect is more easily visible in the LV diagram of {\color{red}Figure \ref{almaLVrgb}}, where the \ttco\ is brightest at velocities slightly redward and blueward of the \ceto\ across the cold cloud.

In fact, in the data cubes this is very widespread: the effect can be seen at positions and velocities of nearly all structures, even deep within the molecular cloud.  There are many extended, often filamentary features with shallow velocity gradients and a distinct \ttco\ layer lying just westward of, and slightly red- or blue-shifted from, each bright \ceto\ structure.  Evidently, the cloud is actually somewhat porous to the UV field emanating from the \hii\ region, despite the cloud's more opaque appearance in the near-IR.  The \ceto\ structures then seem to delineate the colder, more shielded parts of the cloud's interior, while the cocooning \ttco\ around each feature may define its more excited side, facing the \hii\ region.  Curiously, the brightest CN emission seems to track better with bright \ttco\ in position and velocity rather than with \ceto, although widespread fainter CN does lie across the mosaic and various \ceto\ features.  The variation of line ratios with position and velocity is difficult to portray here, but Figures \ref{almaCOrgb} and \ref{almaLVrgb} give some idea of the complexity.

\begin{figure*}[t]
\centerline{	\includegraphics[angle=-90,scale=.095]{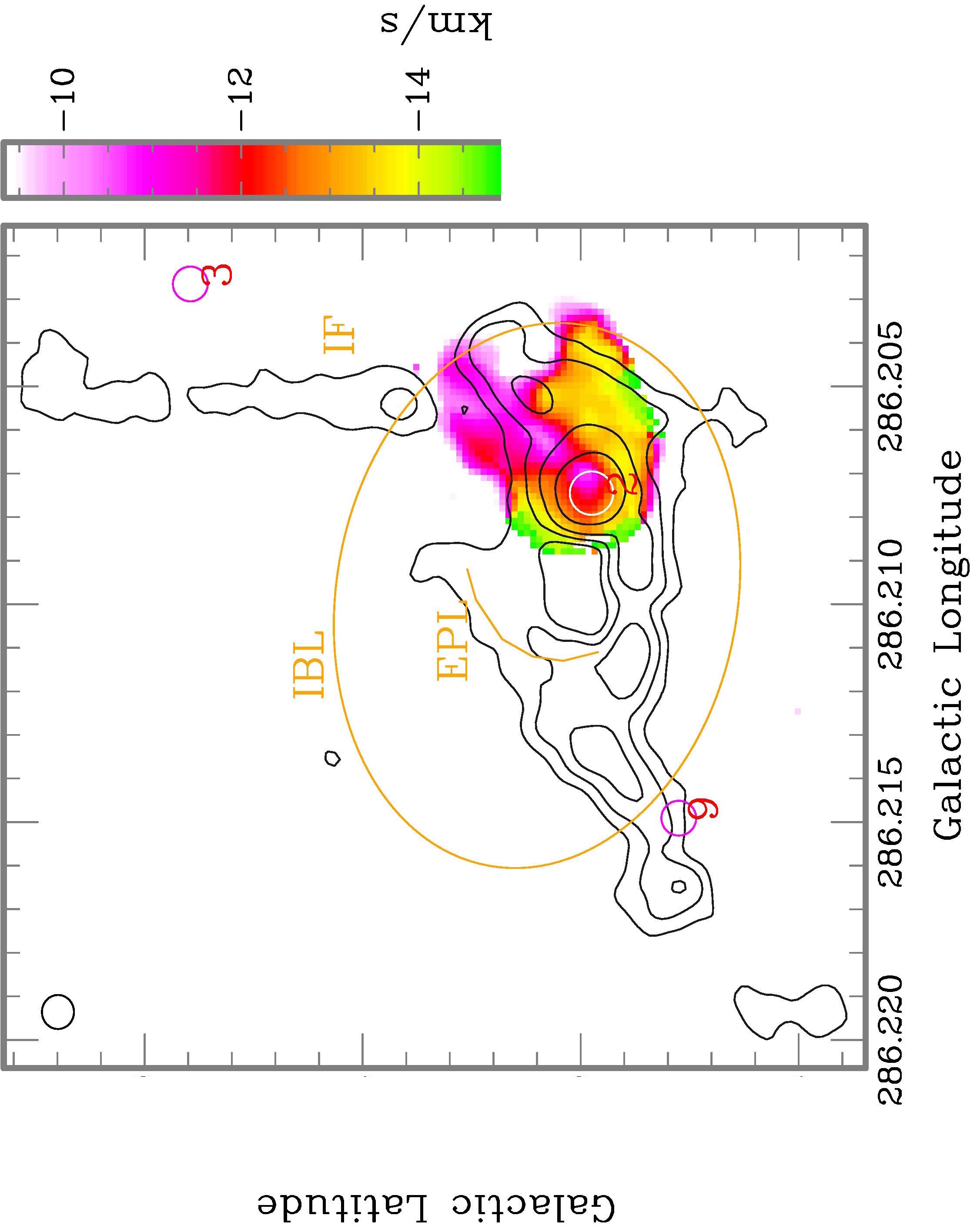}
\hspace{-177mm}\includegraphics[angle=-90,scale=.095]{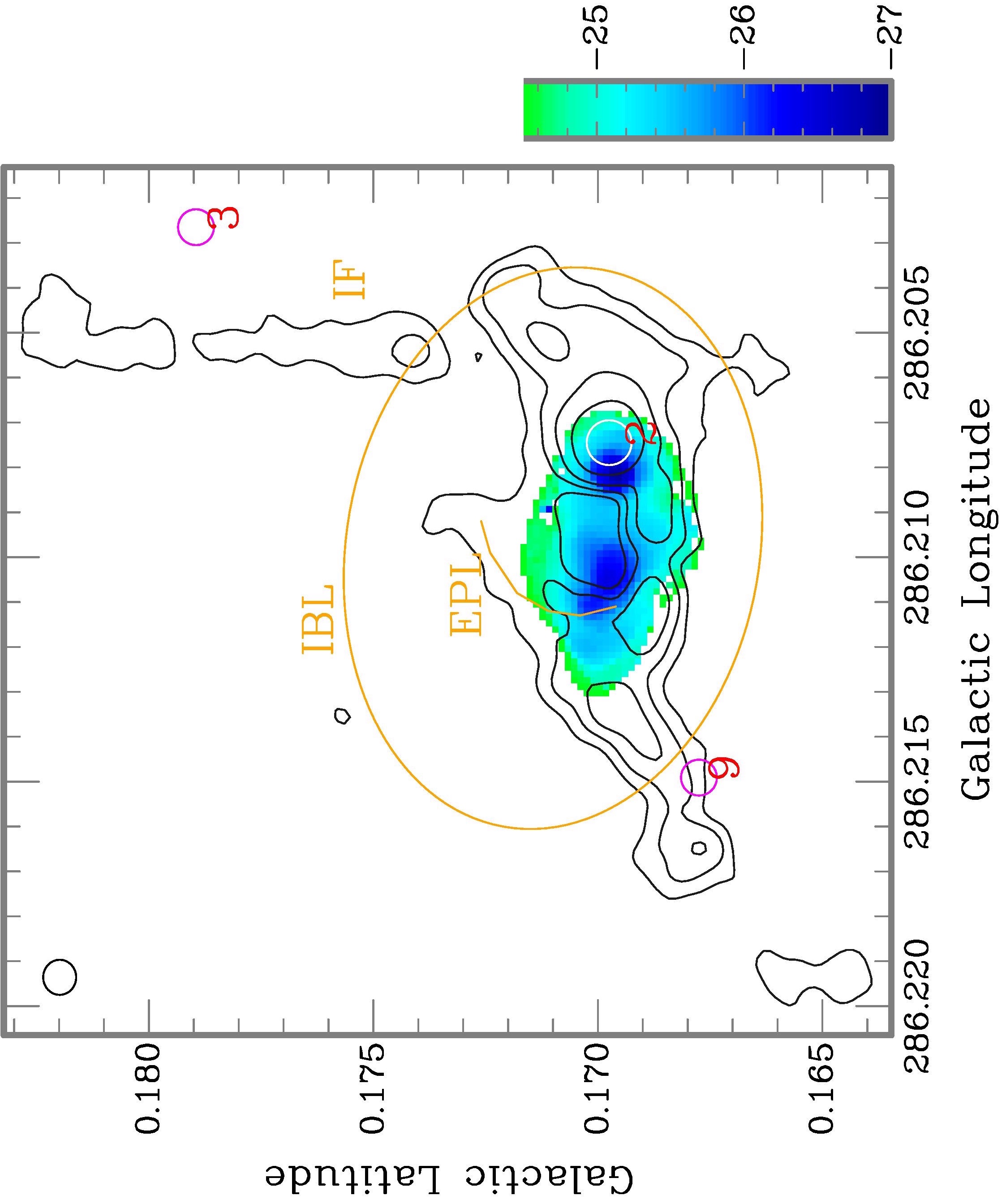}}

\vspace{-1mm}\caption{ 
Velocity fields (first moments 
with SAM) of \ttco\ line wings integrated from --31 to --24.5\kms\ 
(left, blue wing) and from --15 to --3.5\kms\ 
(right, red wing).  The colour scales match the corresponding truncated velocity wedges.  Overlaid are the same 3mm continuum contours as in Fig.\,\ref{almaCOrgb}, plus selected labels.  
}\label{COkep}\vspace{0mm}
\end{figure*}

The most noticeable kinematic feature in the spectral line cubes is an EW velocity gradient across the Streamer, one remarkable in several respects.  Near the bright continuum emission (and thus the brightest parts of each emission line), the gradient is consistent across all three species, reaching a maximum blueshift of --22.0 $\pm$0.1\,km s$^{-1}$ about 20$''$ = 0.25\,pc east of MIR\,2, and a maximum redshift of --17.0$\pm$0.2\,\kms\ around 11$''\pm$2$''$ = 0.13$\pm$0.02\,pc northwest of MIR\,2: see {\color{red}Figure \ref{COkep}}.  
The gradient is also at its sharpest exactly across the middle of MIR\,2 itself, $\sim$2--3\,\kms\ across only 1 ALMA beam ($\sim$6000\,AU = 0.03\,pc), or $\sim$75\,\kms\,pc$^{-1}$.  The steepest part of the gradient, defining a symmetry axis, lies on a nearly N-S curve, just like the $B$ field orientation in Figure \ref{IBLpolo}.  Moreover, this symmetry axis across the middle of MIR\,2 (straddling the width of the Streamer) looks identical in all 3 lines, underscoring its dynamical importance and strongly suggesting a flat NS feature within MIR\,2 as the origin for the outflow (more on this next).  Meanwhile, the full extent of the EW blue-to-red velocity gradient lies along a curved line across $l$$\sim$286\fdeg23--286\fdeg20, roughly 1.3\,pc.

These details suggest the possibility that the main and western extensions of the Streamer form part of a rather large structure (perhaps including a disk), in which inward flow to MIR\,2 must occur and there generate the outflow.  While the case is somewhat circumstantial so far, the evidence becomes much stronger upon closer inspection.

\subsection{High Velocity \ttco\ Emission: \\ A Massive Keplerian Disk or Freefalling Accretion?}\label{kepler}

For BYF\,73 as a whole, we estimate that its systemic velocity is $V_{\rm sys}$ = --19.6$\pm$0.2\,\kms\ on the LSR scale, based on inspection of the \ceto\ data cube.  As seen in Figures \ref{almaCOrgb}+\ref{almaLVrgb}, nearly all of the mosaics' line emission lies between --24 and --16\,\kms.  However, there is clear evidence in the \ttco\ cube of high-velocity line wings close to the position of MIR\,2: up to $\Delta$$V_{\rm blue}$ = --11.5\,\kms\ and $\Delta$$V_{\rm red}$ = +16\,\kms, for a total \vlsr\ range of 27.5\,\kms\ (i.e., from --31 to --3.5\,\kms).  This emission is quite small in spatial extent, with length $\sim$ 20$''$ = 0.25\,pc and width 10$''$--15$''$ = 0.12--0.18\,pc for each lobe: see Figure \ref{COkep}.

\begin{figure*}[t]
\centerline{\includegraphics[angle=-90,scale=0.42]{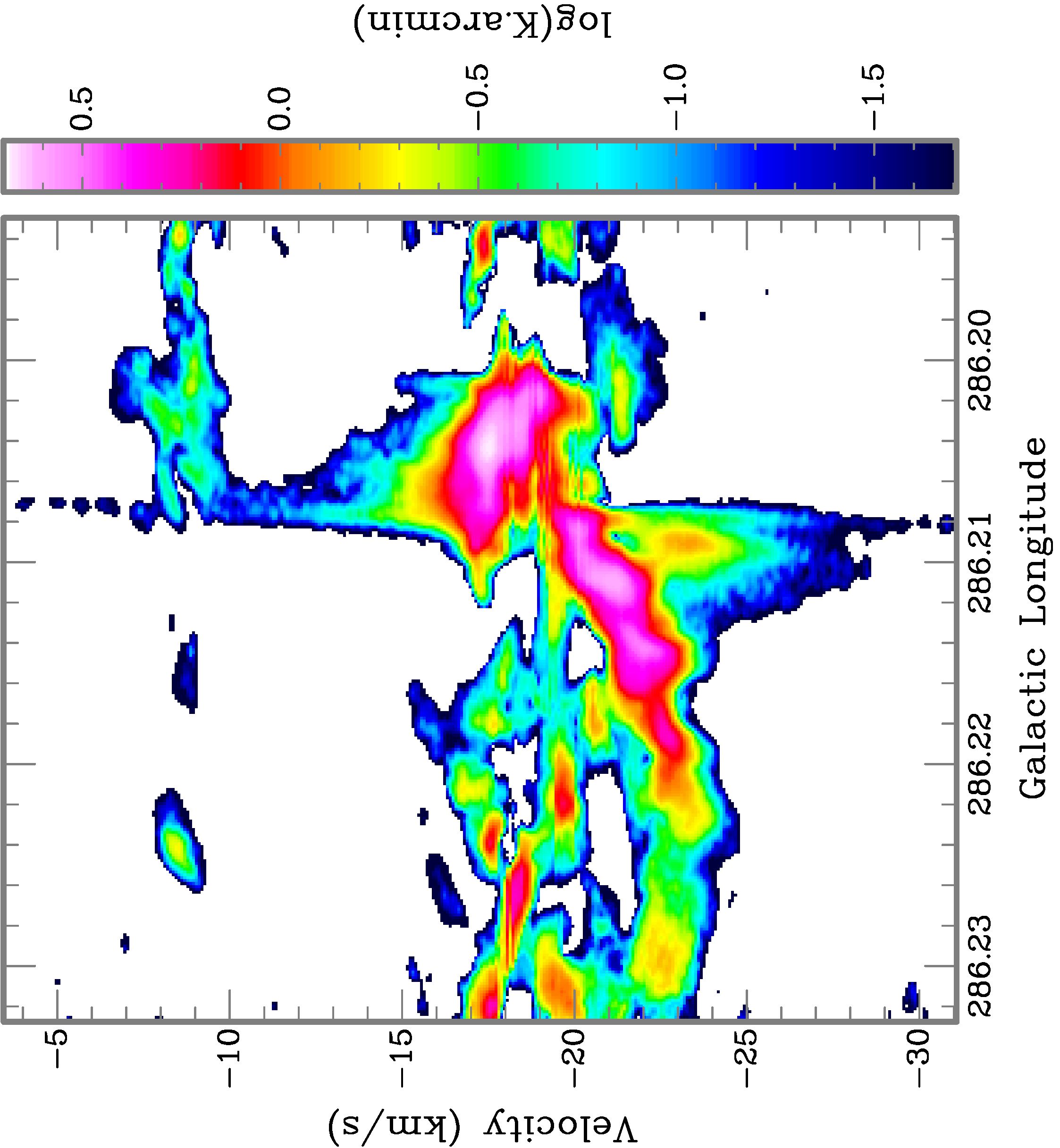}\hspace{5mm}
		\includegraphics[angle=-90,scale=0.101]{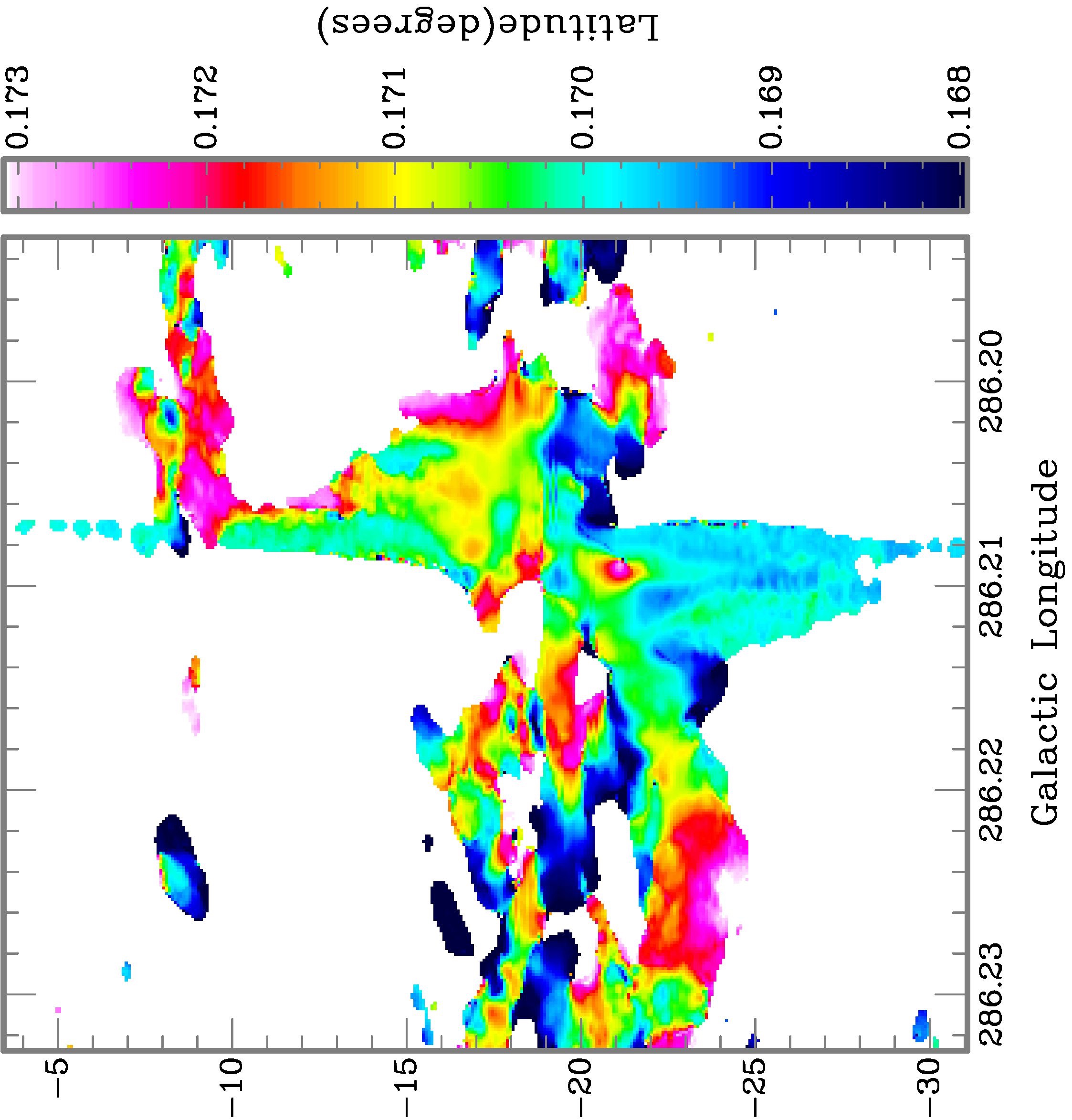}}\vspace{-5mm}
\hspace{90.4mm}\includegraphics[angle=-90,scale=0.42]{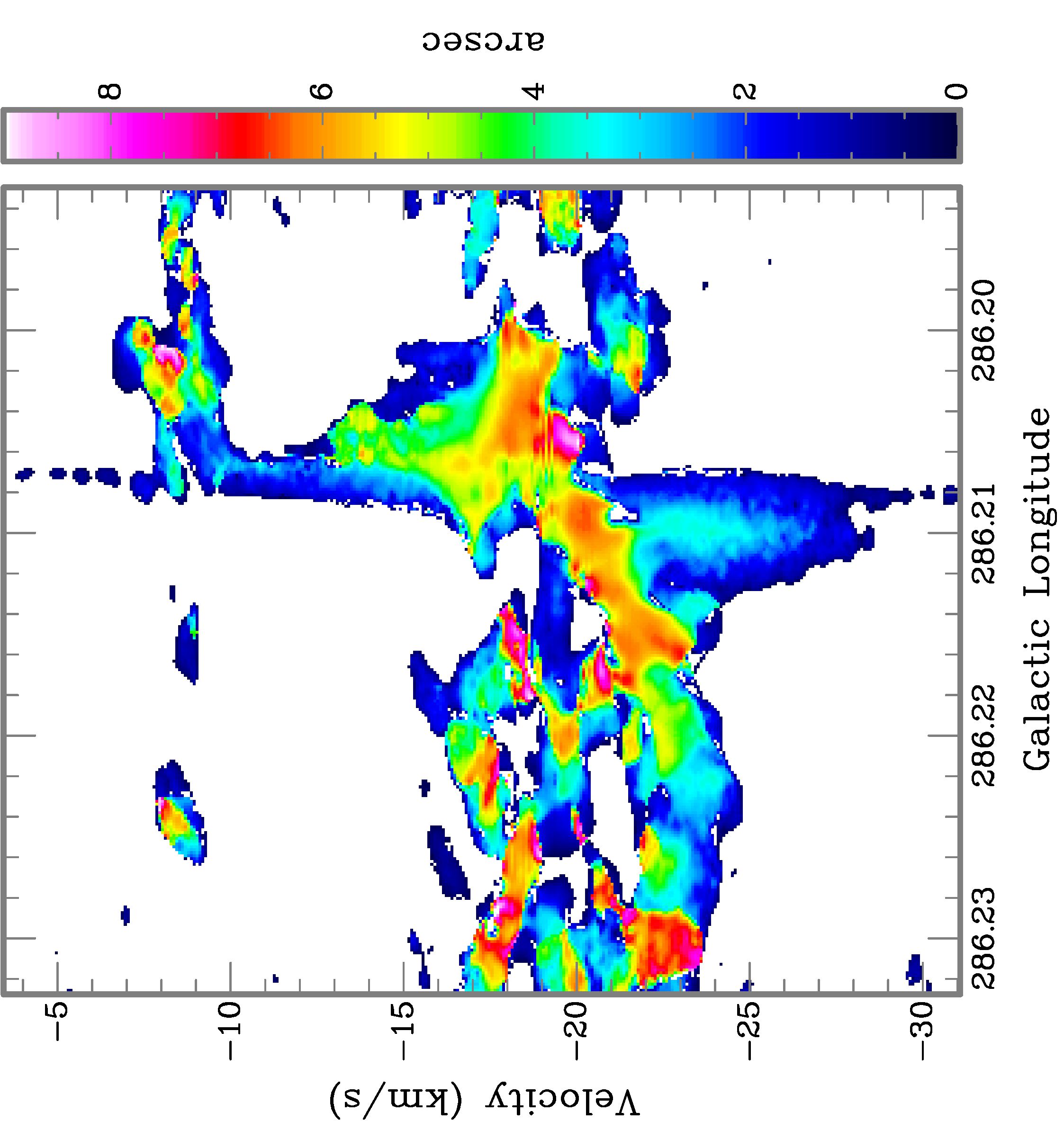}

\vspace{-160mm}\hspace{2.5mm}\includegraphics[angle=0,scale=0.423]{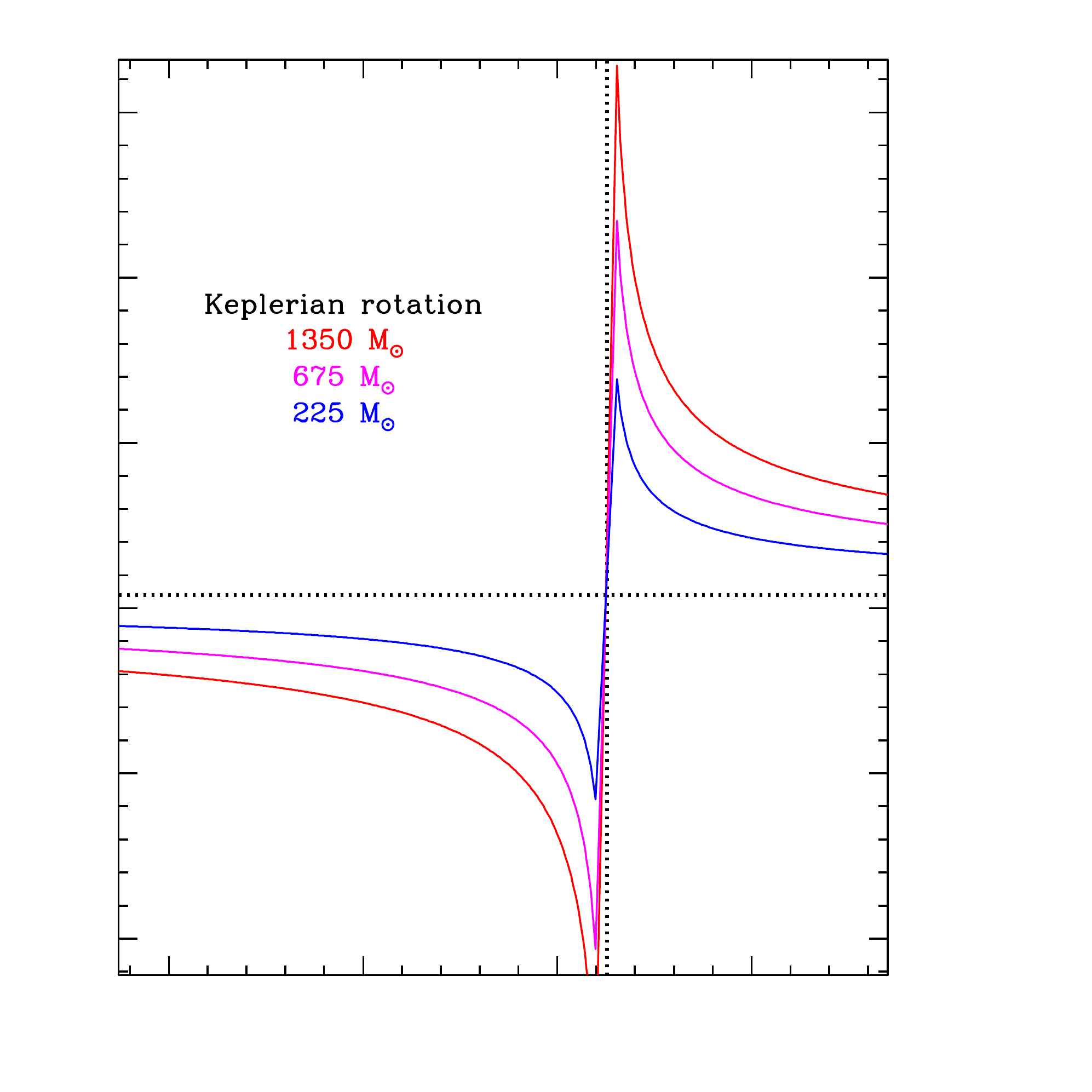}
			\hspace{1.6mm}\includegraphics[angle=0,scale=0.423]{MIR2rotcurve1-eps-converted-to.pdf}

\vspace{-10.6mm}\hspace{90.8mm}\includegraphics[angle=0,scale=0.423]{MIR2rotcurve1-eps-converted-to.pdf}

{\color{cyan}\vspace{-27mm}\hspace{103mm}\rule{25mm}{0.7pt} \\
		\vspace{3mm}\hspace{110mm}Streamer}

{\color{red}\vspace{3mm}\hspace{108.4mm}\tco\ Outflow
		\begin{picture}(1,1)\thicklines \put(3,-5){\line(2,6){20}}\end{picture}}

{\color{blue}\vspace{-16mm}\hspace{140mm}{\Huge \}}Rotation \\
		\vspace{-15mm}\hspace{140mm}{\Huge \}}} \\

\vspace{-54mm}\caption{ }
\vspace{-3mm}\hspace{0mm}\parbox{87mm}{ 
\hspace{17mm}{\bf({\em Top Left})}  Integrated longitude-velocity diagram (zeroth moment) of \ttco\ emission across the Streamer, 
latitudes 0\fdeg1677--0\fdeg1733. 
The log$_{10}$ brightness scale (needed to display the image's dynamic range of $\sim$200) peaks at 4.92$\pm$0.02\,K.arcmin.  The emission at --8.5\kms\ is presumed to arise in a diffuse foreground cloud unrelated to BYF\,73.  Overlaid are coloured curves representing Keplerian rotation for 3 sample masses contained within 1\farcs8 of MIR\,2's position, each half joined by a straight line inside that radius.  Dotted lines indicate MIR\,2's longitude ($l$ = 286\fdeg20745 from T-ReCS astrometry; P18) and BYF\,73's $V_{\rm sys}$ = --19.6\,\kms. \\
{\bf({\em Top Right})}  Longitude-velocity first moment map of the same data as in the top panel.  The colours represent the intensity-weighted mean latitude of the integrated emission, e.g., the highest-velocity emission lies at the same latitude as MIR\,2 (the cyan colour, $b$ = 0\fdeg16975$\pm$0\fdeg00004 in T-ReCS' 0\fdeg00007 = 0\farcs25 PSF/beam). \\
{\bf({\em Bottom Right})}  Longitude-velocity 
second moment (intensity-weighted latitude dispersion, or latitude width of the emission) of the same data as in the above panels, with additional labels to distinguish between the \tco\ and \ttco\ LV patterns.  On this scale, unresolved features smaller than the ALMA beam ($\sim$2\farcs6) are a medium blue or darker, such as the highest-velocity emission, all velocities along the inner edge of the disk, and along a good portion of the 1350\,M\solar\ curve.  Most of the interior of the disk, i.e., the portions between the 1350\,M\solar\ curve and \vlsr\ = --19.6\,\kms, have widths \lapp\ 5$''$, while the super-rotating and solid-body portions of the Streamer (the brighter features in the top panel) have widths up to $\sim$7$''$.
}\label{innerdisk} \vspace{2mm}
\end{figure*}

This compact configuration is completely different to the massive, more extended bipolar outflow clearly visible in \tco\ (\S\ref{outflow}), which is indicated in the bottom panel of {\color{red}Figure \ref{innerdisk}} to help distinguish the LV patterns of the two species.  In particular, the \tco\ outflow emission that extends beyond an area $\pm$10$''$ around MIR\,2 must be relatively low opacity $\tau$ and high excitation \tex, since it is much brighter than the superimposed \ttco\ emission, where the latter is even detectable at the same ($l$,$V$) coordinates.  In contrast, the high-velocity emission coincident with MIR\,2 has \ttco\ almost as bright as \tco, suggesting a much higher $\tau$ and lower \tex.  The respective excitation conditions are consistent with gas being entrained by a powerful mechanical outflow, and gas responding to the local gravitational potential.

Finally, the extended, low-velocity \ttco\ wings are not visible in \tco\ due to the latter's high $\tau$, although much of the \ttco\ line wing emission (Fig.\,\ref{COkep}) is spatially oriented similarly to the inner 
\tco\ outflow, including the 43\degr\ bend in the blue wing and the deviations around the Streamer-west in the red wing (Fig.\,\ref{almaCOjet}).  Line wings similar to either the \tco\ or \ttco\ high-velocity patterns are not detectable in either the \ceto\ or CN cubes.

Thus, while the brightest \ttco\ emission is probably also tracing the outflow, the small extent of the high-$\Delta$$V$ emission is more peculiar.  It becomes progressively tinier as the velocity channel being viewed moves further from the cloud's systemic value, contracting to within a beamwidth of MIR\,2 at the highest velocities.  This is the {\em opposite} of what is typically seen in protostellar jets, where the highest velocities are usually at the most distal parts of the outflow \citep{L00}.

Instead, the LV-moment diagrams in Figure \ref{innerdisk} suggest this pattern might arise from a Keplerian disk.  Sample rotation curves $V_{\rm rot}$ = $\sqrt{GM/R}$ are overlaid for 3 different central masses in Figure \ref{innerdisk} as well.  The only free parameter in fitting such curves is the central mass:\footnote{Of course, the distance (2.50$\pm$0.27\,kpc) also matters.  If this is changed, the linear scale and mass will change proportionately.  However, with an 11\% uncertainty, the implied mass of MIR\,2 remains above 1200\,M\solar\ for rotation, or above 850\,M\solar for infall.} the position of MIR\,2 is well-constrained, as is the velocity extent of the emission.  Only the highest mass curve of the 3 examples fits the high-$\Delta$$V$ \ttco\ emission envelope adequately.  The lower mass curves and P18's mass estimates are all much too small, and are strongly ruled out under this interpretation.

Indeed, a 1350$\pm$50\,M\solar\ curve fits the data remarkably well over a longitude extent of 0\fdeg008 = 0.35\,pc, or a radius of 36,000\,AU, which is about half the longitude extent (286\fdeg203--216) of the IBL as measured in the HAWC+ data (Fig.\,\ref{pplobes}).  Such a disk would also be $\sim$half the size of the Keplerian disk seen in another massive cloud, K3-50 \citep{h97}, and about half or less of K3-50's disk mass \citep{BL15}, so these numbers are not unheard of in high-mass SF.  But it means that the central mass of MIR\,2 (presumably comprising a massive protostar and its envelope) is 5--6$\times$ larger than P18's preferred estimate, and that it dominates the dynamics of the gas over that span.

Alternatively, the kinematics can also be explained by gas in free fall towards a 950$\pm$35\,M\solar\ object, since $V_{\rm ff}$ = $\sqrt{2}V_{\rm rot}$ decreases the central mass required to produce the curves seen in Figure \ref{innerdisk} by $\sqrt{2}$.  Reference to either scenario hereafter is meant to include both as feasible physical settings near MIR\,2.

In our view, such rotation/freefall curves are so distinctive that there is no other reasonable explanation for the motions of the gas, at least within the IBL (i.e., excluding a much smaller amount of apparently counter-rotating gas in the same window). 
Outside the IBL, the deviations from Keplerian rotation are stronger, and may be due to more typical, modestly turbulent cloud motions and/or internal structures. 

\subsection{The Eastern Polarisation Lobe (EPL)}\label{EPLcont}

Even if we have constructed a plausible picture of 
BYF\,73's internal structure and mass flows, the EPL is a distinctive feature in both the SOFIA and ALMA polarisation maps (Fig.\,\ref{IBLpolo}) with an as-yet undetermined role or import.  It lies north of the plane of the Streamer/disk around MIR\,2, yet the inferred $B$ field directions through it still point towards/away from MIR\,2.  It is bright in line emission as well (Fig.\,\ref{almaCOrgb}), particularly \ceto, but also \ttco\ and CN in turn around its arc.  The velocity fields (Fig.\,\ref{almaLVrgb}) reveal little except that it is close to systemic (--20\,\kms) in \ceto\ at its northern apex, and slightly blueshifted (--21.5\,\kms) in both CN and \ceto\ at its base near the Streamer, but blending smoothly with the Streamer's rotational pattern.  Scanning through the channels in the \ttco\ cube, the changing pattern gives the impression of a splash effect or prominence-like tendril, driven by the outflow off ambient Streamer material downstream of the 47\degr\ bend in the blue wing, with a relatively gentle relative blueshift of 1.5\,\kms.

If true this would be quite interesting: does it signify the expulsion of surface material from the Streamer?  Or is it a wisp from the wider cloud, falling in at a slightly higher speed, unconstrained by the Streamer due to its northerly approach?  While at a lower surface density than the disk, 
perhaps 10$^{27}$\,m$^{-2}$ or a tenth of the Streamer (\S\ref{HROstuff}), the continuum data show that its mass is not insignificant, perhaps 15\,M\solar\ in total.  Higher resolution polarisation data would be desirable to determine exactly what the EPL signals.

\subsection{A Second Massive Protostar}\label{mir11}

\begin{figure}[t]
\vspace{2mm}
\includegraphics[angle=-90,scale=0.32]{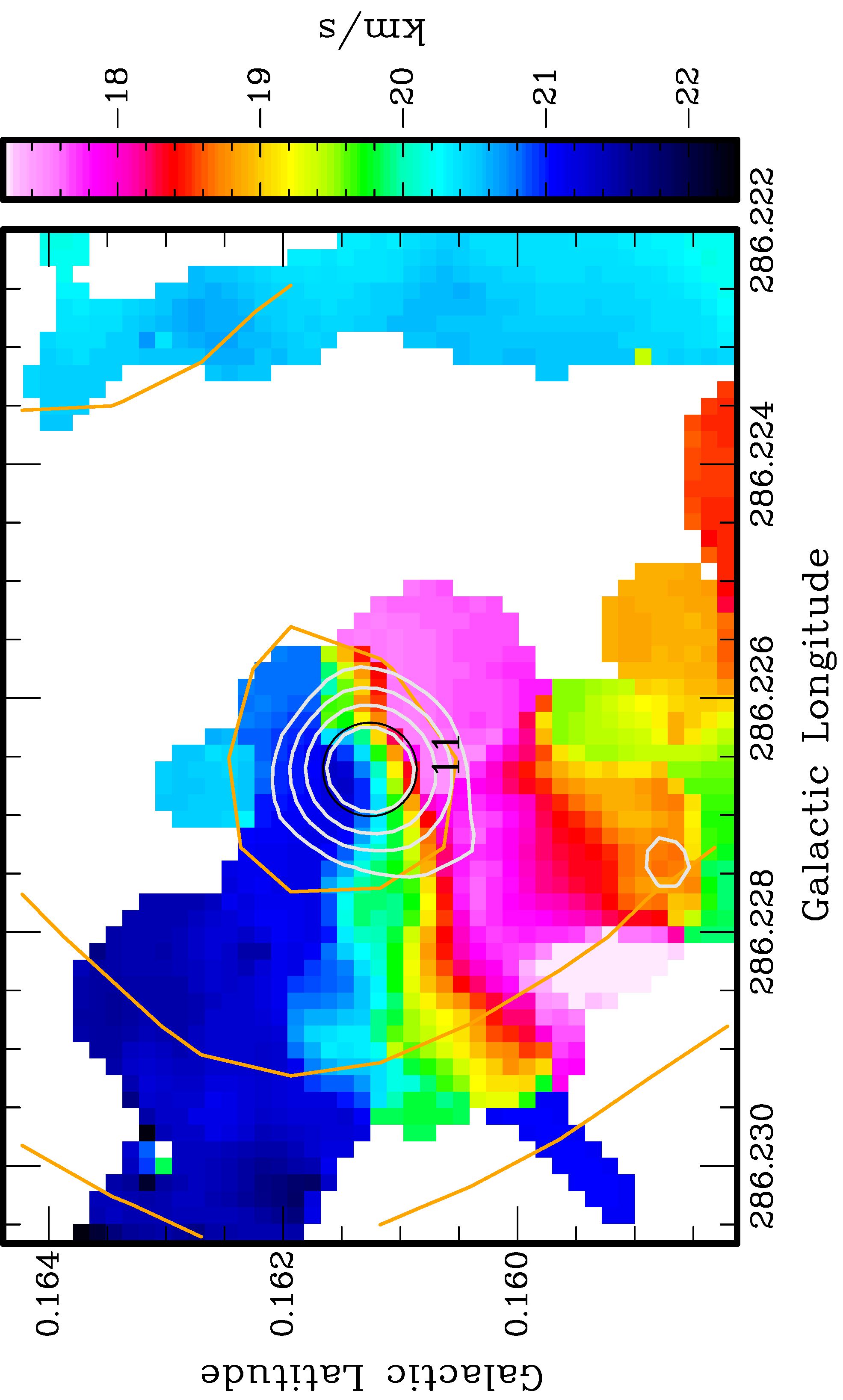}
\vspace{-2mm}\caption{ 
ALMA mosaic velocity field (first moment) of main hyperfine component of CN around MIR\,11 (black), overlaid by white ALMA 3mm $I$ continuum contours at 0.4, 1, 2, 4 mJy/beam and orange HAWC+ $I$ contours at 0.64(0.10)0.84 Jy/pix (compare with Figs.\,\ref{irac}$e,f$).  
}\label{mir11grad}\vspace{0.5mm}
\end{figure}

Around MIR\,11 there appears to be a second strong velocity gradient, similar to that straddling MIR\,2 (\S\ref{outflow}).  
This gradient is somewhat vague in \ttco, clearer in \ceto, but most obvious in CN (see {\color{red}Fig.\,\ref{mir11grad}}), and is oriented with the strongest $\Delta$$V$ along a similar axis ($\sim$N40{\degr}E) as the biconical mid-IR nebula (Figs.\,\ref{irac}$e,f$): blue-shifted emission on the more IR-visible side to the NE, and red-shifted emission to the SW.  The various line profiles show only a narrow velocity range, however, $\pm$2--5\,\kms, rather than a strong outflow signature, and for \ttco\ and \ceto, the red- and blue-shifted emission overlaps somewhat on the sky.  Both lines have strong self-absorption around the line centre of 
at --19.8\,\kms.  The CN is better separated on the sky into blue and red lobes, with no self-absorption.  These biconical characteristics suggest the possibility of MIR\,11 being in an even earlier, pre-outflow stage of evolution than MIR\,2.

\begin{figure*}[t]
\centerline{\includegraphics[angle=-90,scale=0.5]{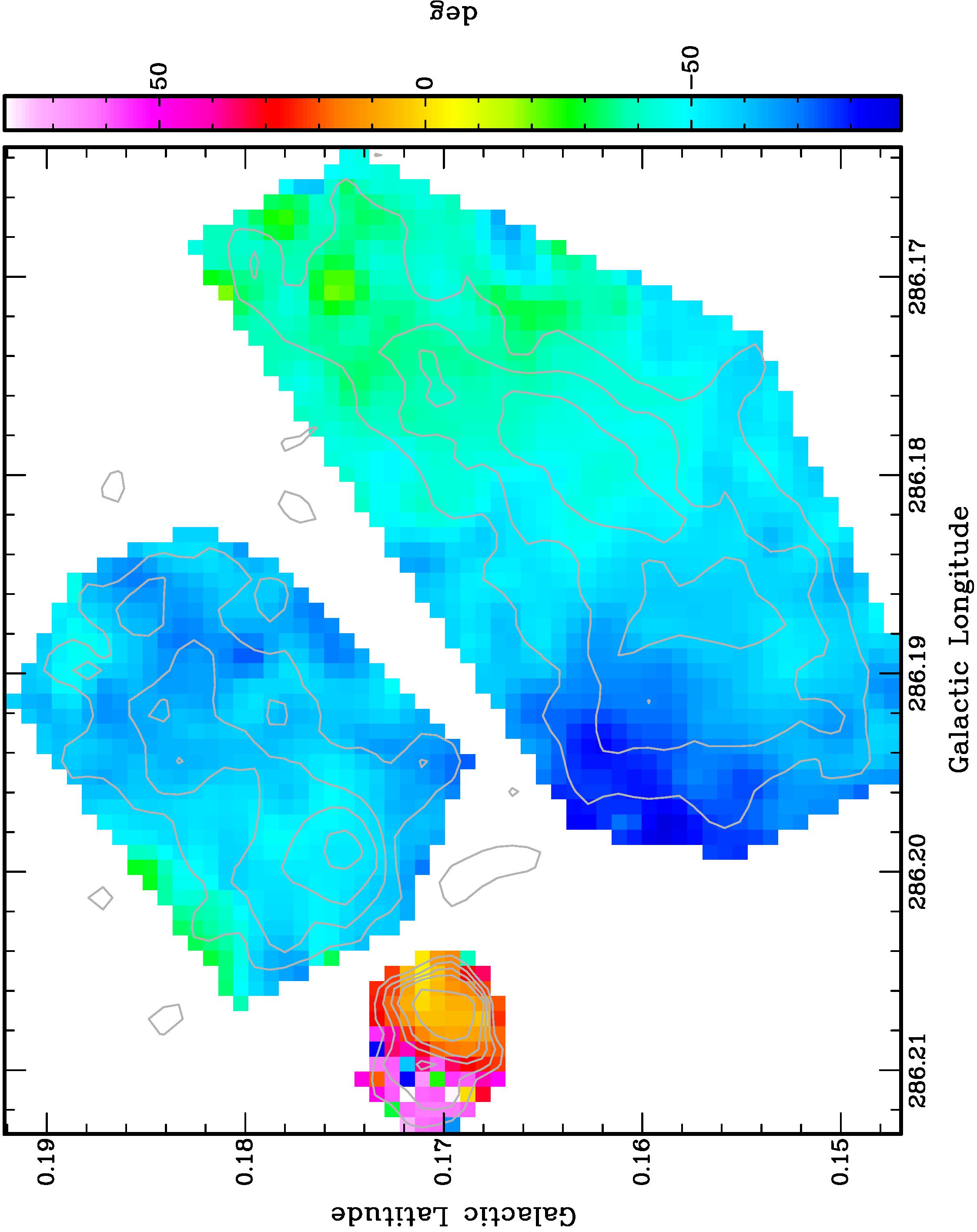}} 

\vspace{-88mm}\hspace{45mm}{\small

\vspace{18mm}\hspace{65mm}{\hii\-north}

\vspace{21mm}\hspace{44mm}{EPL +}

			\hspace{41mm}{MIR 2 core}

\vspace{2mm}\hspace{95mm}{\hii\-south}}

\vspace{32mm}\caption{ 
Cutouts of the $\theta_{B_{\perp}}$ distribution in HAWC+ 
ROIs with S/N($P'$) \gapp 5, overlaid by $P'$ contours 50(16)98,140 mJy/pixel (the same as in Figs.\,\ref{irac},\,\ref{pplobes}).  These ROIs correspond to arcs of polarised emission in the \hii\ region, and to the EPL+MIR 2 core within the IBL.  In the analysis shown in Figs.\,\ref{Hthetas} and \ref{HCFstats}, however, the ROI enclosing MIR 2 excludes the pixels covering the EPL.  
}\label{Hregs}\vspace{0mm}
\end{figure*}

\section{Magnetic Field Analysis}\label{analysis}

\subsection{Davis-Chandrasekhar-Fermi (DCF) Method}\label{DCFstuff}

\subsubsection{Preamble}

We start with the method of \citet{BL15} to make some reasonable estimates of $B$ field strength, based on the dispersion $s$ in inferred field directions from the polarisation data.  The basic DCF approach \citep{d51,cf53} assumes a statistical connection between turbulent motions in the gas and the dispersion in $B$ field direction in the presence of a transverse MHD wave. 
Such analysis is necessarily approximate, since other  
thermal, rotational, gravitational, or even magnetic effects may affect the two processes treated by DCF. 
On the other hand, even for supersonic ({\bf M} $\sim$ 5--9) MHD gases (as in \hii\ regions), \citet{osg01}'s numerical simulations showed that the DCF method can give some useful information, despite not being developed for such a setting.

One approach to evaluating the behaviour of $s$ is that of \citet{mg91}.  In their language, the goal is to identify a ``correlation length'' in the implied $B$ 
field orientation, within which the $B$ field directions are correlated and aligned with each other, and outside of which they are not.

Using the formalism of \citet{mg91}, we fit the distributions of polarisation position angle $\theta$ with a simple gaussian $e^{-\theta_B^2/2s^2}$ to obtain a best-fit value for the dispersion $s$ in $\theta_B$ (measured in radians).  Our method is a simplified version of \citet{mg91}'s analysis, since they showed that this approach gives very reliable results even for their comprehensive data (hundreds of stellar polarisation measurements) on the Taurus molecular clouds.  We have a smaller data set of $\theta_B$, so will not need the full \citet{mg91} treatment.

\subsubsection{HAWC+ Data Analysis}

\begin{figure}[b]
\centerline{\hspace{2mm}\includegraphics[angle=0,scale=0.23]{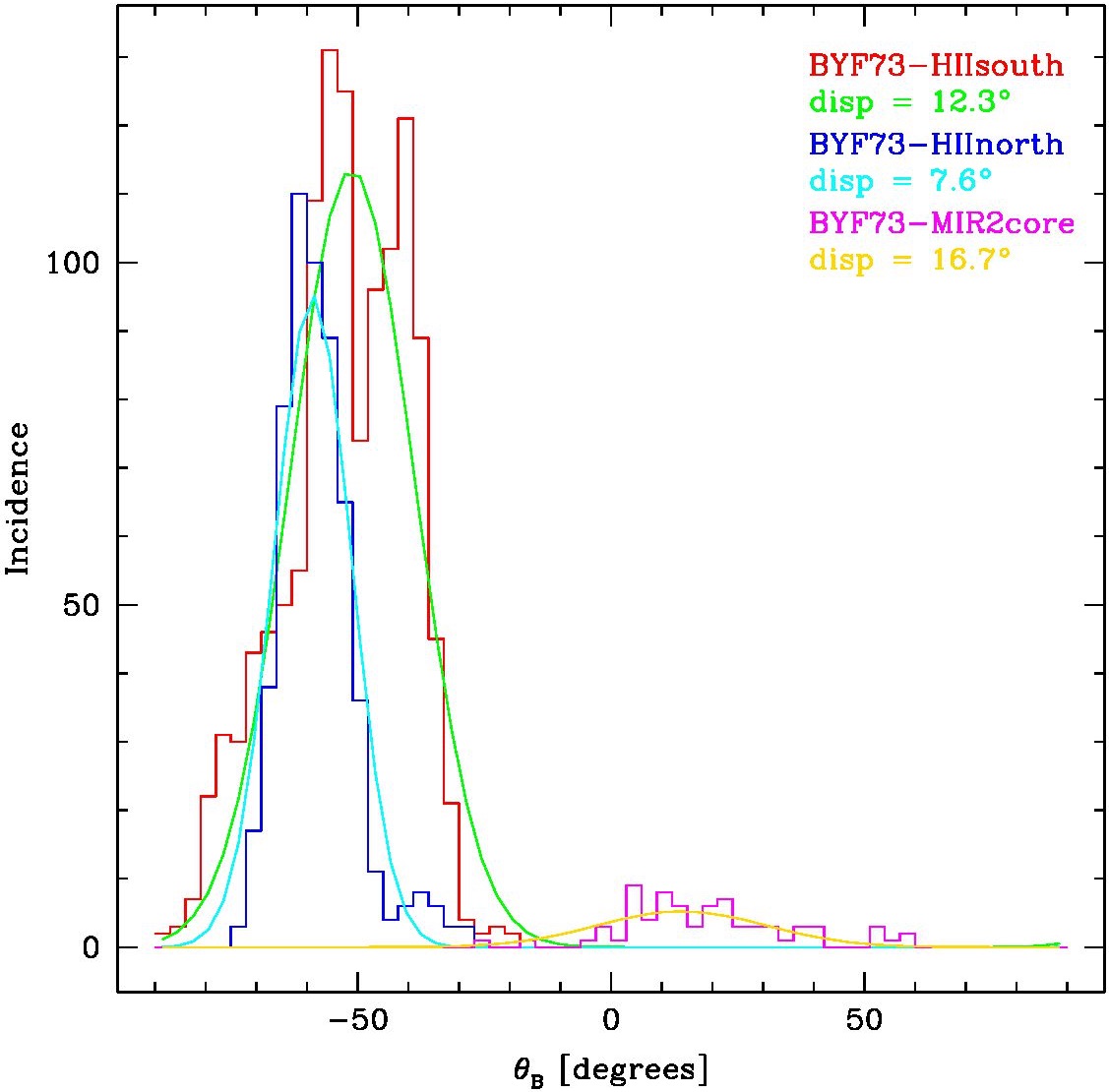}}
\vspace{-1mm}\caption{ 
Histograms of $\theta_{B_{\perp}}$ at all pixels within each of the ROI cutouts from Fig.\,\ref{Hregs}, as labelled.  Also shown are gaussian fits to, and the dispersions in, each $\theta_{B_{\perp}}$ distribution.  
}\label{Hthetas}\vspace{0mm}
\end{figure}

In the HAWC+ data, the measured $\theta_{B_{\perp}}$ in any given region with S/N \gapp 5 will have an uncertainty in orientation dominated by instrumental noise, $\Delta\theta_{B_{\perp}}$ \lapp 3\degree.  This occurs at a level slightly less than the $P'$ = 50\,mJy/pixel contour in Figures \ref{hawcmaps} and \ref{irac}, and we show the corresponding $\theta_{B_{\perp}}$ maps in {\color{red}Figure \ref{Hregs}}.  There are three regions of interest (hereafter ROI) satisfying these criteria: two arcs of polarised emission in the \hii\ region (labelled north \& south), and the area within the IBL containing the MIR 2 core (but excluding the EPL due to the paucity of statistics, $\sim$20 pixels).  To begin the DCF analysis, we show in {\color{red}Figure \ref{Hthetas}} the $\theta_{B_{\perp}}$ distribution in all pixels within each ROI.  None of these is really gaussian, but we overlay such fits in order to compute effective dispersions in $\theta_{B_{\perp}}$ for reasons which will become clear shortly.

We next construct histograms for subsets (comprised of square boxes of area $A$) of each ROI, computing a dispersion $s$($A$) in $\theta_{B_{\perp}}$ for each subset.  The smaller the boxes, the more choice we have of where to fit them inside each ROI.  We then compute a mean dispersion $<$$s_{\theta_{B_{\perp}}}(A)$$>$ ($\pm$ a standard deviation) in the field orientation for all small boxes of a given area $A$, no matter where they are placed within the ROI.  Finally, in {\color{red}Figure \ref{HCFstats}} we plot all such results as a function of box size $A^{1/2}$, ranging from a minimal useful size of $A$ = 3$\times$3 pixels (roughly one Nyquist sample given the HAWC+ band D beam) to the full size of each ROI.  

In each ROI, the mean dispersion $<$$s$$>$ within all boxes of area $A$ rises as the box size increases, meaning that $\theta_{B_{\perp}}$ is more correlated on small scales (e.g., $s$ $<$ 5\degr\ within the \hii\ region over spans $<$ 0.5\,pc), and becomes less correlated over longer distances ($s$$\sim$10\degr across \gapp1\,pc within the \hii\ region).  The scale at which $s$ seems to plateau is where we identify the correlation length as per \cite{mg91}.  Thus, the HAWC+ polarisation data suggest that, as far as the $B$ field is concerned, the \hii\ region consists of one or perhaps two coherent structures, with a correlation length $\sim$0.5\,pc as evidenced by the plateauing of $s$ at 6\degr--7\degr\ in the smaller \hii-North (Fig.\,\ref{HCFstats}), and the slow rise of $s$ across \hii-South as the $B$ field orientation slowly changes across the 1\,pc arc of the \hii\ region.

In contrast, the interior of the IBL (already much smaller than the \hii\ region) shows two closely-spaced and distinct $B$ field configurations, namely the EPL and MIR\,2 core.  While no useful statistics could be compiled for the EPL, even the MIR\,2 core has insufficient resolution to discern more than a strongly rising $s$ at all measured scales, and no correlation length can be defined beyond the $\sim$0.2\,pc extent of the core itself, as in Figure \ref{IBLpolo}.

\subsubsection{ALMA Data Analysis}

\begin{figure}[b]
\centerline{\includegraphics[angle=0,scale=0.23]{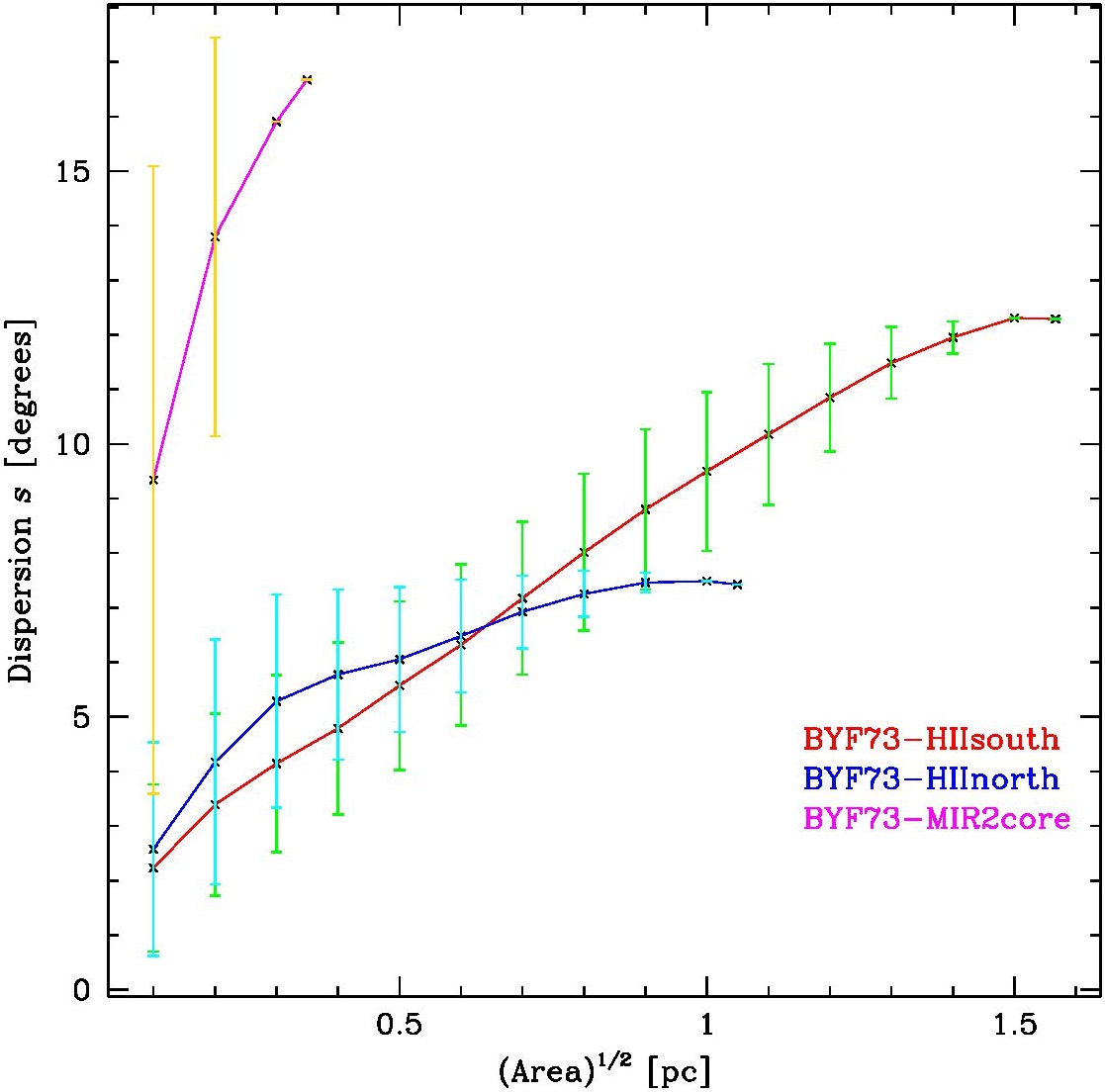}\hspace{2mm}}
\vspace{-1mm}\caption{ 
Mean dispersion of polarization position angles $\theta_{B_{\perp}}$, with the dispersion in the dispersion shown as error bars, as a function of box size within each ROI shown in Figs.\,\ref{Hregs},\,\ref{Hthetas}.
}\label{HCFstats}\vspace{3mm}
\end{figure}

\begin{figure*}[t]
\hspace{-1mm}
\centerline{\includegraphics[angle=-90,scale=0.55]{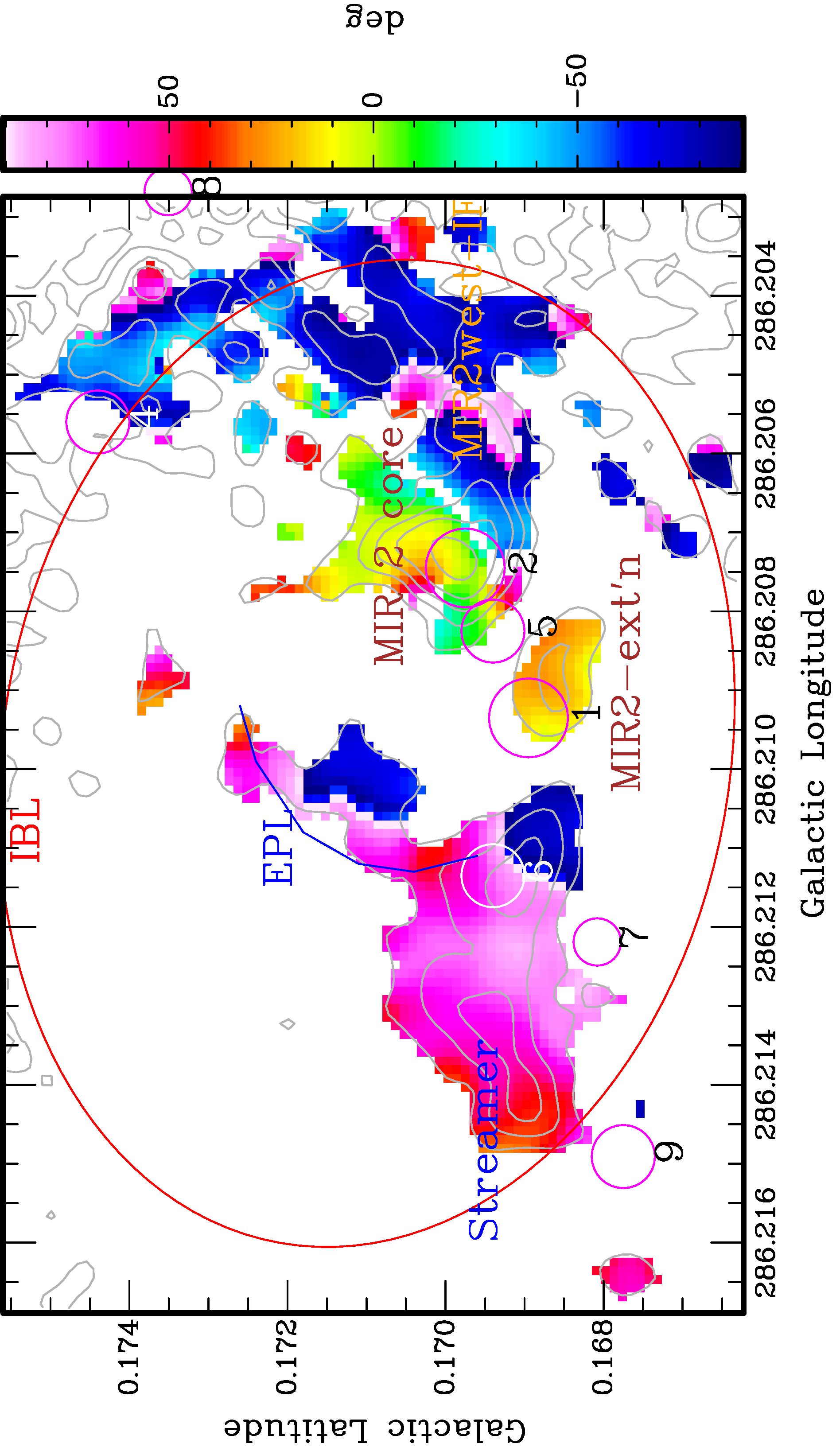}}

\vspace{-74mm}\hspace{35mm}


\vspace{72mm}\caption{ 
Cutouts of the $\theta_{B_{\perp}}$ distribution in ALMA 
ROIs) with S/N($P'$) \gapp 3, overlaid by $P'$ contours 50(16)98,140 mJy/pixel (the same as in Figs.\,\ref{almaCont},\,\ref{IBLpolo}).  These ROIs correspond to polarised emission from the Streamer, EPL, and parts of the MIR\,2 core within the IBL.  
}\label{Aregs}\vspace{1mm}
\end{figure*}

\begin{figure}[b]
\centerline{\hspace{2mm}\includegraphics[angle=0,scale=0.23]{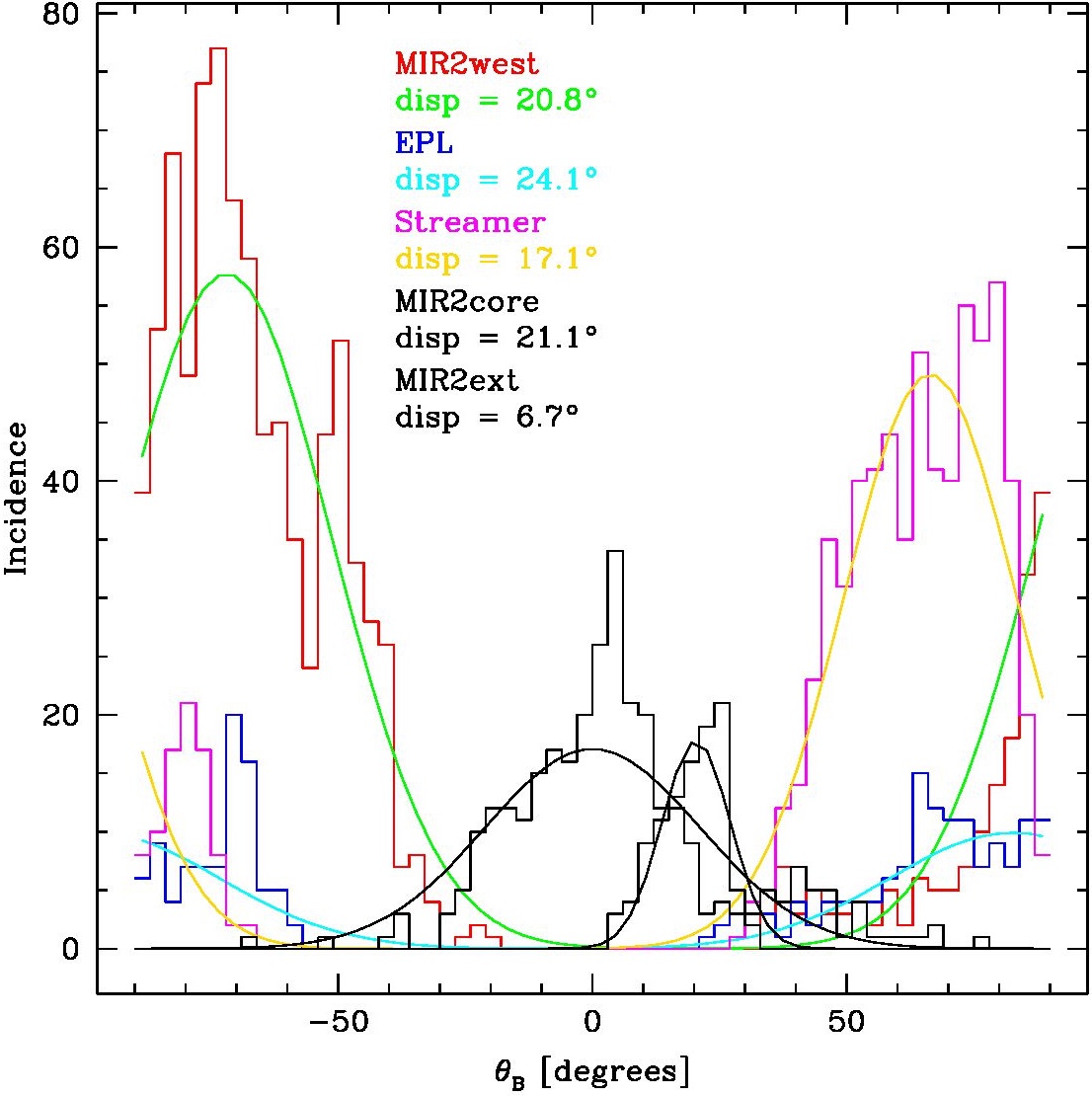}}
\caption{ 
Histograms of $\theta_{B_{\perp}}$ at all pixels within each of the ROI cutouts from Fig.\,\ref{Aregs}, as labelled.  Also shown are gaussian fits to, and the dispersions in, each $\theta_{B_{\perp}}$ distribution.  
}\label{Athetas}\vspace{0mm}
\end{figure}

\begin{figure}[b]
\centerline{\includegraphics[angle=0,scale=0.23]{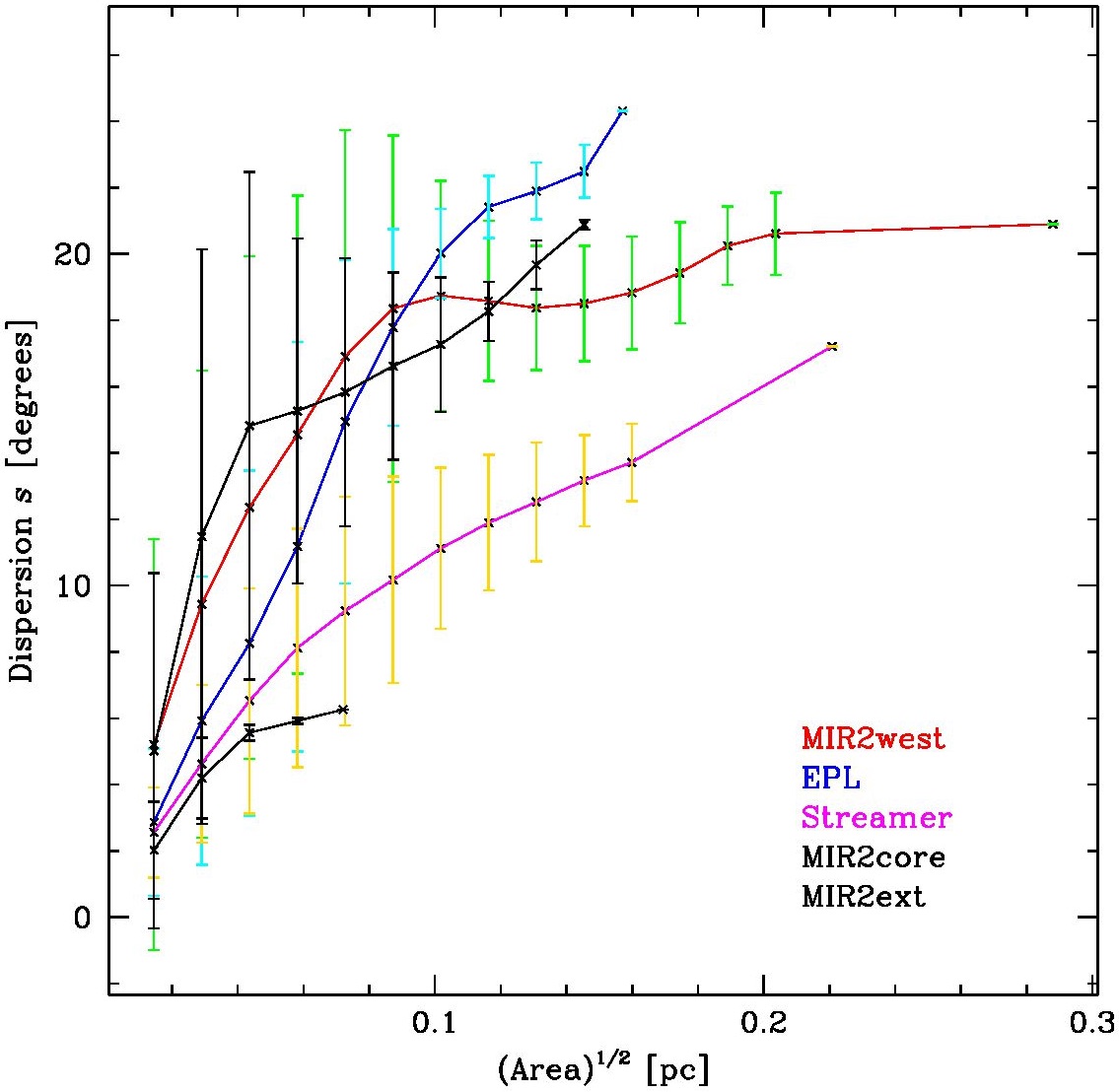}\hspace{2mm}}
\caption{ 
Mean dispersion of polarization position angles $\theta_{B_{\perp}}$, with the dispersion in the dispersion shown as error bars, as a function of box size within each ROI shown in Figs.\,\ref{Aregs},\,\ref{Athetas}.
}\label{ACFstats}\vspace{3mm}
\end{figure}

The same approach can be used with the ALMA polarisation data, which has high enough resolution to resolve the DCF analysis for the MIR\,2 core and its environment, unlike its minimal representation in Figures \ref{Hregs}--\ref{HCFstats}.  We therefore define the ALMA ROIs in {\color{red}Figure \ref{Aregs}} where $\theta_{B_{\perp}}$ has S/N $>$ 2.5 but is typically 5--10, as in Figures \ref{almaCont} and \ref{IBLpolo} and conforming to the description in \S\ref{almaIBL}.  Then in {\color{red}Figure \ref{Athetas}} we show the ALMA $\theta_{B_{\perp}}$ distributions, compiling the ALMA DCF statistics in {\color{red}Figure \ref{ACFstats}}.

We first notice that, in the overlapping range of scales (0.1--0.3\,pc), the dispersion $s$ in the IBL from both ALMA and HAWC+ data are in agreement, rising approximately from 10\degr\ to 20\degr\ when averaging broadly over the 5 structures in Figure \ref{ACFstats}.  Focusing next on the smallest ALMA scales (0.02\,pc), $s$ in the IBL also starts out at a few degrees, then gradually rises.  In each of the 5 ROIs, Figure \ref{ACFstats} hints at an $s$ plateau for each structure, before rising further as other uncorrelated structures are included.  In {\color{red}Table \ref{corrstats}} we compile the ($A_{\rm corr}^{1/2}$, $s_{\rm corr}$) pairs that can be read off the trends in Figure \ref{ACFstats}, and for completeness also include the values for the \hii\ region ROIs from Figure \ref{HCFstats}. 
The EPL seems to have the fastest-rising $s$ and the least well-defined plateau in Figure \ref{ACFstats}, suggesting that it has not mapped out 
a single correlation length in its structure.

Therefore, the measurable correlation lengths in the IBL are perhaps only a tenth those in the \hii\ region, 0.05--0.15\,pc compared to $\sim$1\,pc.  The dispersions in $B$ field direction at those lengths are respectively $\sim$6\degr--20\degr, compared to 7\degr--12\degr for the \hii\ region.  Such lengths and dispersions are the scales above which $B$ field directions are not correlated with each other between neighbouring areas, and are therefore the scales we should explore for other physical thresholds, and particularly for any constraints on $B$ field strength.

\subsubsection{Numerical Results}

As described by \citet{osg01}, \citet{cn04}, \citet{BL15} and many other workers, the standard DCF analysis \citep{d51,cf53} directly links the dispersions in polarisation angle $\delta\theta$ = $s$ (tracing variations in the $B$ field orientation) to three other physical parameters that simply and naturally describe the propagation of a transverse MHD wave in a turbulent plasma: the gas density $n$, the line-of-sight velocity dispersion $\delta$$V$, and the plane-of-sky magnetic field strength $B_{\perp}$.  That is, $s$ will increase as (1) $B$ decreases, since then the magnetic restoring forces are reduced; (2) $n$ increases, since then the medium's inertia to the MHD wave is greater; or (3) $\delta$$V$ increases, since that describes the strength of the MHD wave.  

According to \cite{cn04}, with appropriate SI unit conversions  (1\,nT = 10\,$\mu$G) the projected $B$ field strength 
\begin{equation} 
	B_{\rm \perp,DCF} = 0.543\,{\rm pT}~\sqrt{\mu n}~(\Delta V/s)~,
\end{equation}
where 
$n$ (m$^{-3}$) is the gas density with mean molecular weight $\mu$=2.35, $\Delta V$ = $\sqrt{8{\rm ln}2}~\delta V$ is the velocity FWHM (\kms) in the cloud, and $s$ is measured in degrees.  Included in the constant is a numerical factor $Q$\,=\,0.5 from \cite{cn04} to correct for various smoothing effects \citep[e.g., see][]{osg01}.  

For an illustrative example, consider the region of densest gas inside the IBL.  From modelling of \hcop\ emission by \cite{b10} at the 40$''$ Mopra resolution (roughly 3$\times$ the HAWC+ beam), we take $\delta V\sim$ 1\,\kms\ as a median intrinsic value and the estimated peak $n\sim$ 5$\times10^{11}$\,m$^{-3}$ ostensibly near MIR\,2, to connect $s_{\rm corr}$ in our polarisation maps to the 
field strength $B_{\perp}$.  Eq.\,(1) then becomes
\begin{equation} 
	B_{\rm \perp,DCF}({\rm MIR\,2,\,Mopra}) \sim 92\,{\rm nT}~(s/15^{\circ})^{-1}
\end{equation}  
at these values of $n$ and $\delta V$ around MIR 2 (or 0.9\,mG in cgs units).  Indeed, the value for $n$ may well be higher in the smaller HAWC+ beam, and is certainly much higher in the 0.03\,pc structures revealed by ALMA (see \S\ref{HROstuff}), peaking at 3.6$\times$10$^{13}$\,m$^{-3}$.  
Then, 
\begin{equation} 
	B_{\rm \perp,DCF}({\rm MIR\,2,\,ALMA}) \sim 1.18\,{\rm \mu T}~(s/10^{\circ})^{-1}
\end{equation}
(12\,mG), a value which has not been observed in {\em any} star-forming region outside of maser spots.  Even the smaller value in Eq.\,(2) is among the highest non-maser $B$ field strengths in similar massive star-forming clouds, according to the compilation of \cite[][his Fig.\,7]{cru12}.

Despite the possibly record-setting value for $B_{\rm \perp}$ near MIR\,2, it is commensurate with MIR\,2's high gas density, i.e., together they indicate a mass-to-flux ratio that is very close to critical (see below).  In other words, any field strength much less than this (or density much greater) would probably not provide sufficient support against gravitational collapse (allowing, of course, for the likelihood of $|${\bf B}$|$ $>$ $B_{\rm \perp}$).  However, this density is derived from SED fitting which, as we have already noted, may be significantly underestimating the gravitational potential near MIR\,2, based on the apparently Keplerian or infalling motions seen in the \ttco\ data (\S\ref{kepler}).  In that case, even this high $B$ field cannot avoid criticality.

As a contrasting example, we also consider the \hii\ region ROIs.  In such regions, bulk expansion speeds are typically 2--3$\times$ the velocity dispersion (= sound speed) in the roughly 8000\,K ionised gas, $\sim$12\kms\ \citep{hi79,ftb90}.  Such flows are thought to dominate the energetics in the gas.  For BYF\,73, the 
\hii\ region 
has a total flux density at 843\,MHz of only 60\,mJy \citep{mgps}.  This corresponds to a small emission measure EM = $n^2$$D$ = 9.5$\times$10$^{15}$\,pc\,m$^{-6}$ \citep{mh67}.  
With an apparent diameter 2$R$=$D$$\sim$0.5\,pc, this yields a much lower electron density 
$n_e$ $\approx$ 1.4$\times$10$^8$\,m$^{-3}$ than in the molecular cloud,\footnote{From these parameters, one can also derive an excitation parameter $U$ = $Rn^{2/3}$ = 6.7$\times$10$^{4}$\,pc\,m$^{-2}$ \citep{mst67} for the \hii\ region, needing only a single $\sim$B1 star \citep{p73} to power it, and confirming its modest impact on the molecular cloud.} but we also have a somewhat larger dust-based estimate of $n_d$ $\approx$ 7$\times$10$^9$\,m$^{-3}$ from SED fitting (\S\ref{HROstuff}) 
which may average in material from outside the \hii\ region.  
We bracket this uncertainty by combining these values in Eq.\,(1) with two estimates (e.g., using a lower $\mu$=1.28 in the ionised gas), 
\begin{eqnarray} 
	B_{\rm \perp,DCF}({\rm H\,{\scriptsize II}, ions}) \sim 21\,{\rm nT}~(s/5^{\circ})^{-1}~~~{\rm and} \nonumber \\
	B_{\rm \perp,DCF}({\rm H\,{\scriptsize II}, dust}) \sim 195\,{\rm nT}~(s/5^{\circ})^{-1}~~,
\end{eqnarray} 
for the \hii\ region ROIs, depending on which parts of the line of sight through the \hii\ region are being sampled by the HAWC+ polarisation data.

\begin{deluxetable}{lcccc}
\tabletypesize{} 
\tablecaption{ 
Davis-Chandrasekhar-Fermi correlation statistics for \\ BYF\,73 polarisation structures from SOFIA \& ALMA data\label{corrstats}}
\tablewidth{0pt}
\tablehead{
\colhead{Structure} & \colhead{Correlation} & \colhead{$B$ Field Angular} & \colhead{$\sigma_B$/$\bar B_{\perp}$} \vspace{-0.8mm} \\
\colhead{} & \colhead{Scale $A_{\rm corr}^{1/2}$} & \colhead{Dispersion $s_{\rm corr}$} & \colhead{} \vspace{-3mm} \\ 
}
\startdata
\hii-North & 0.5\,pc & 6\degr\ & 0.1 \\
\hii-South & 1.5\,pc & 12\degr\ & 0.3 \\
Streamer-W & 0.14\,pc & 18\degr\ & 0.3 \\
MIR\,2 core & 0.08\,pc & 16\degr\ & 0.4 \\
EPL & 0.12\,pc & 22\degr\ & 0.6 \\
Streamer & 0.14\,pc & 13\degr\ & 0.3 \\
MIR\,2 extn. & 0.05\,pc & 6\degr\ & 0.1 & \vspace{-3mm} \\
\enddata
\vspace*{-0.1mm}
\end{deluxetable}

Even the lower (pure-\hii) value seems somewhat high compared to a more typical 1\,nT in other \hii\ region studies \citep{cru12,BL15}; whether this value is reasonable is unknown, but $B$ field measurements could also be obtained, for example, via high-resolution HI Zeeman observations.  The higher dust-based estimate would apply if the polarisation contribution is predominantly from outside the \hii\ region, but then the $B$ field configuration still suggests a connection to the \hii\ expansion.  This could be reconciled with an origin in a sheathing, higher-density PDR layer rather than the ionised cavity.

\subsubsection{Energetic Considerations}

For our purposes, though, the point is whether the energy density ${\mathfrak M}$ in these somewhat strong $B$ fields exceeds or is less than the kinetic energy density ${\mathfrak K}$ in the ionised outflow.  Borrowing from Eq.\ (15) in \S\ref{GKstuff}, we can write this ratio as 
$\frac{\mathfrak M}{\mathfrak K}$ = 5.2\%~$\left(\frac{\Delta V/V_{\rm rel}}{s/6\degr}\right)^2$.  From the flow in the \hii\ region, we have that $\Delta V/V_{\rm rel}$ $\sim$ 1, while from Figure \ref{HCFstats} we have $s$ = 6\degr, 12\degr\ in the \hii\ region-north and -south, respectively.  Thus, the $B$ field energy density in the \hii\ region is probably still small compared to the kinetic energy in the ionised flow.

This approach is only valid, however, if the situation of the DCF analysis holds, namely the presence of an MHD wave with turbulent motions. 
If other processes operate, 
then the $B$ field strength may be indeterminate without direct measurements, either smaller or larger than the DCF 
value.  For example, if other motions enhance variations in $\theta_{B_{\perp}}$, $s$ may be larger than the DCF-only value, underestimating $B_{\perp}$.

In expanding \hii\ regions, the kinetic energy density of the expansion ${\mathfrak K}$ often exceeds the thermal energy density ${\mathfrak T}$ by a wide factor (i.e., in addition to exceeding ${\mathfrak M}$): ${\mathfrak K}$/${\mathfrak T}$ = $\frac{2}{3}${\bf M}$^2$ \citep{BL15}, where {\bf M} (= 2--3 in the example above) is the Mach number of the flow.  For star formation in cold molecular gas, the most interesting question is the relationship between $B$ fields and gravity.  If ${\mathfrak M}$ $\ll$ ${\mathfrak G}$, gravity dominates and the gas is considered ``supercritical;'' if ${\mathfrak M}$ $\gg$ ${\mathfrak G}$, it is ``subcritical.''  We discuss this question further in \S\S\ref{HROstuff}, \ref{GKstuff}, and \ref{critical}. 

However, there is an additional factor in criticality: we need to understand not only the {\em value} of the dispersion $s$, but also its behaviour on different length scales.  As explained by \citet{mg91}, where the DCF method applies, these dispersions are related to the ratio of the disordered vs.\ ordered $B$ field strengths via
\begin{equation} 
	s_{\rm corr} = \frac{\sigma_B}{L^{1/2}{\bar B_{\perp}}}~.
\end{equation}
Here $s_{\rm corr}$ is measured in radians, $L$ is the number of $B$ field correlation lengths (assumed $\approx$ $A_{\rm corr}$) in the line of sight, $\bar B_{\perp}$ is the strength of the ordered component of the projected $B$ field, and $\sigma_B$ is the dispersion in the strength of the random component of the projected $B$ field.  For now, we estimate $L$ from the behaviour of $s$ as seen in the DCF structure functions (Figs.\,\ref{HCFstats}, \ref{ACFstats}), ie., where $s$ plateaus in each structure at some size scale $A$, and so constrain somewhat the ratio $\sigma_B$/$\bar B_{\perp}$.

In the \hii\ region, $L$ = 1--2, $s_{\rm corr}$\,$\approx$\,0.1, so we estimate ($\sigma_B$/$\bar B_{\perp}$)$_{\rm HII}$ $\approx$ 0.1--0.3.  This is another way of describing the highly ordered appearance of the $B$ field vectors over large scales (Fig.\,\ref{hawcmaps}).  Likewise, for the 5 structures in the IBL, effectively $L$ = (1--3) $\times$ $A_{\rm corr}^{1/2}$ by construction, and we estimate the random:ordered $B$ field strength ratios for all 7 structures described here in the same way, and list them in Table \ref{corrstats}.

\begin{figure}[t]
\centerline{\includegraphics[angle=0,scale=0.215]{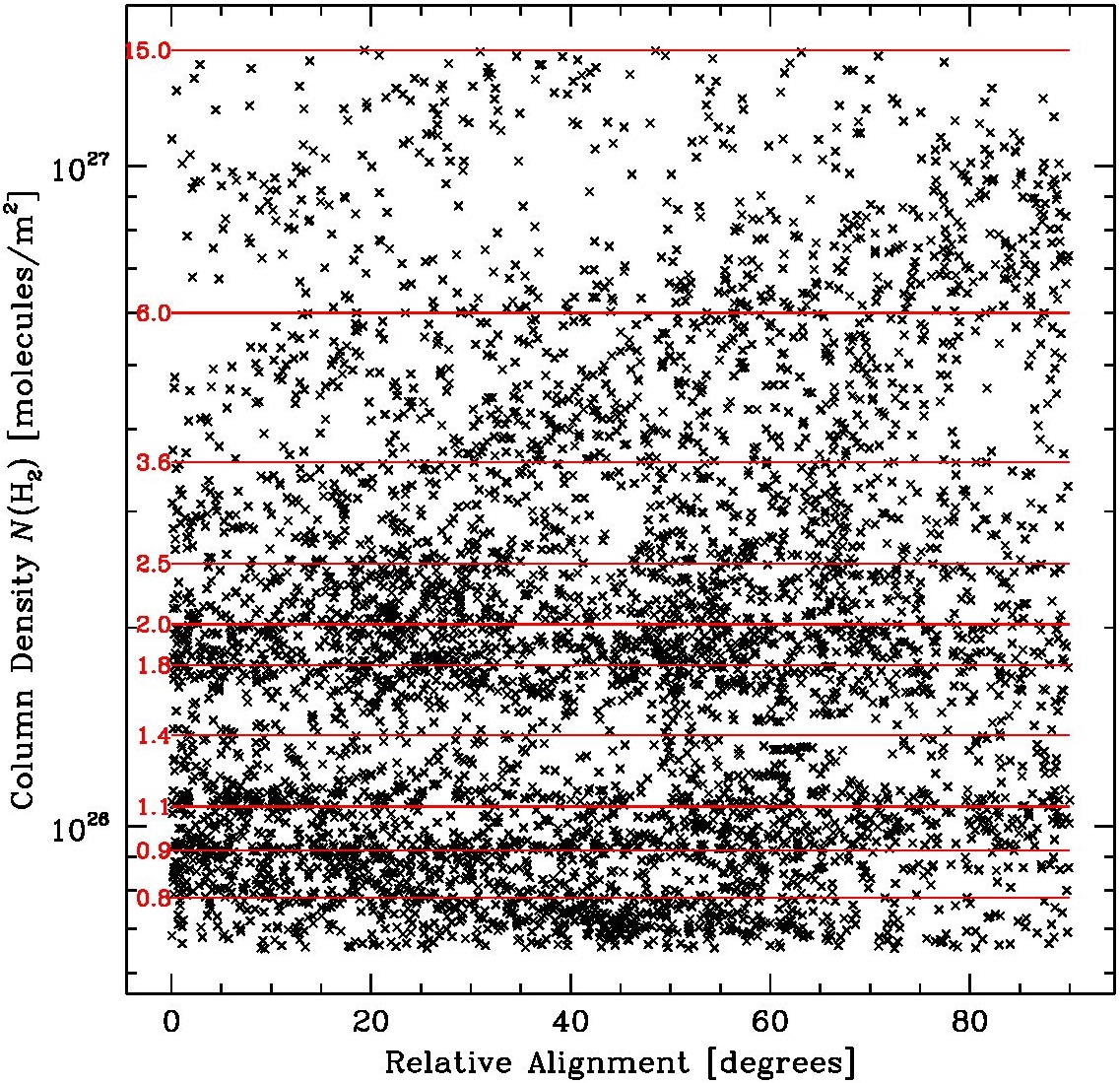}}
\vspace{-1mm}\caption{ 
Relative alignment between polarization position angle $\theta_{B_{\perp}}$ and the tangent to iso-column density $N$ contours, as a function of $N$ across the HAWC+ field.  An angle of 0\degr\ means the $B$ field is oriented along the iso-$N$ contours, while at 90\degr\ the field is perpendicular to the contours and aligned with the gradient $\nabla$$N$.  Also shown as red lines and labelled in $N$, in units of 10$^{26}$m$^{-2}$, are the boundaries of the separate bands in $N$ for which each histogram in Figure \ref{hawcHRObins} was computed.  The boundaries were chosen to ensure histogram equalisation, i.e., to divide all $N$ data into 10 equally-populated bins with comparable statistical noise in each $N$-bin.
}\label{hawcNtheta}\vspace{0mm}
\end{figure}

\begin{figure}[t]
\centerline{\includegraphics[angle=0,scale=0.22]{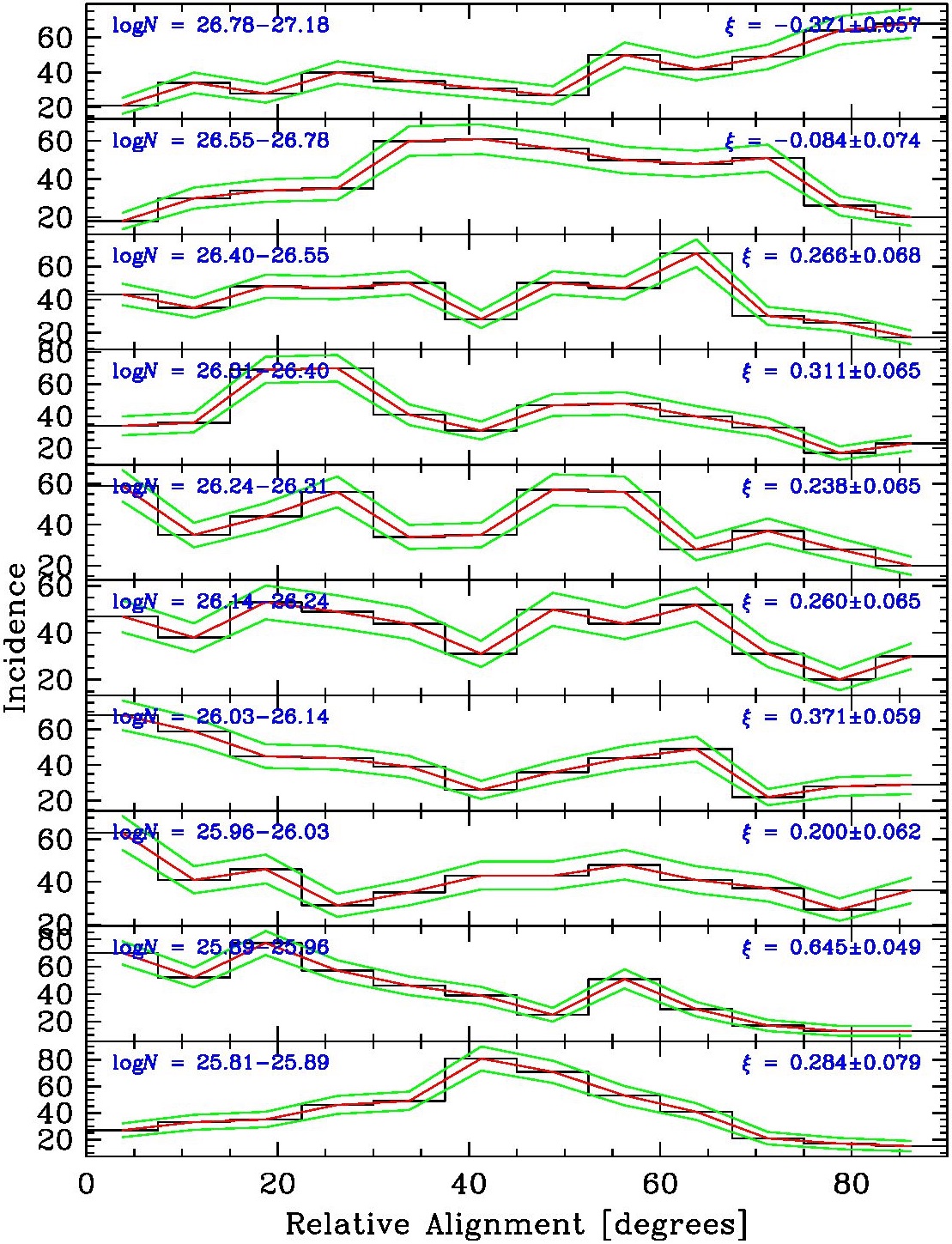}}
\vspace{-1mm}\caption{ 
HRO plots in each of the $N$-bins shown in Figure \ref{hawcNtheta}.  Each window is labelled by the range of column density $N$ in that bin and its fitted HRO shape parameter $\xi$ $\pm$ uncertainty, as defined by \citet{saa17}.
}\label{hawcHRObins}\vspace{-1mm}
\end{figure}

\begin{figure}[h]
\centerline{\includegraphics[angle=0,scale=0.108]{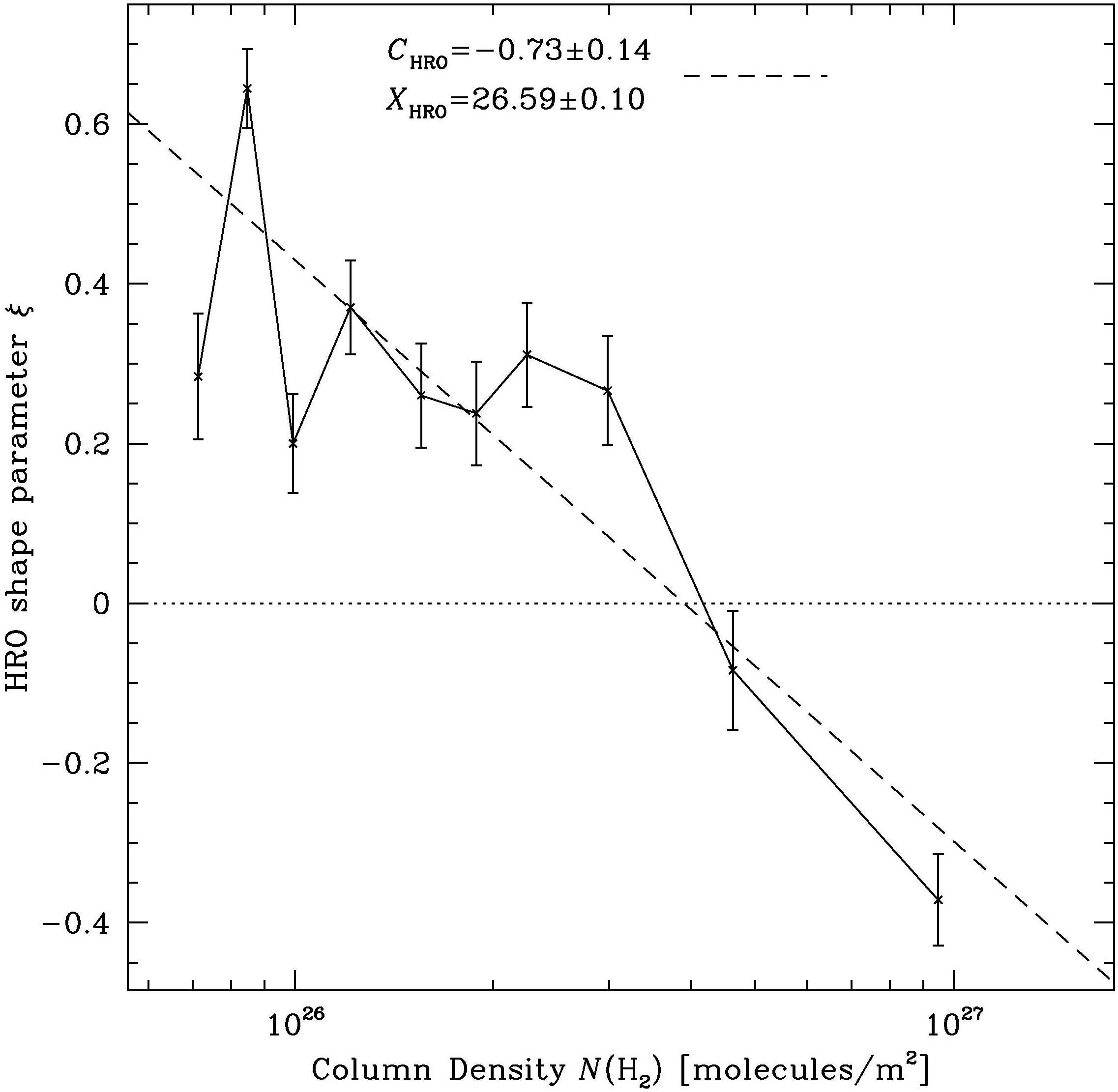}}
\vspace{-1mm}\caption{ 
HRO shape parameter $\xi$ as a function of column density $N$ as fitted in Figure \ref{hawcHRObins}.  The black labels and dashed line are solutions to the parameters $C$ (the slope) and $X$ (log of the $N$-axis intercept) of a linear regression to all the $\xi$ data. 
}\label{hawcHROxi}\vspace{-3mm}
\end{figure}

\begin{figure*}[t]
\centerline{\includegraphics[angle=-90,scale=0.13]{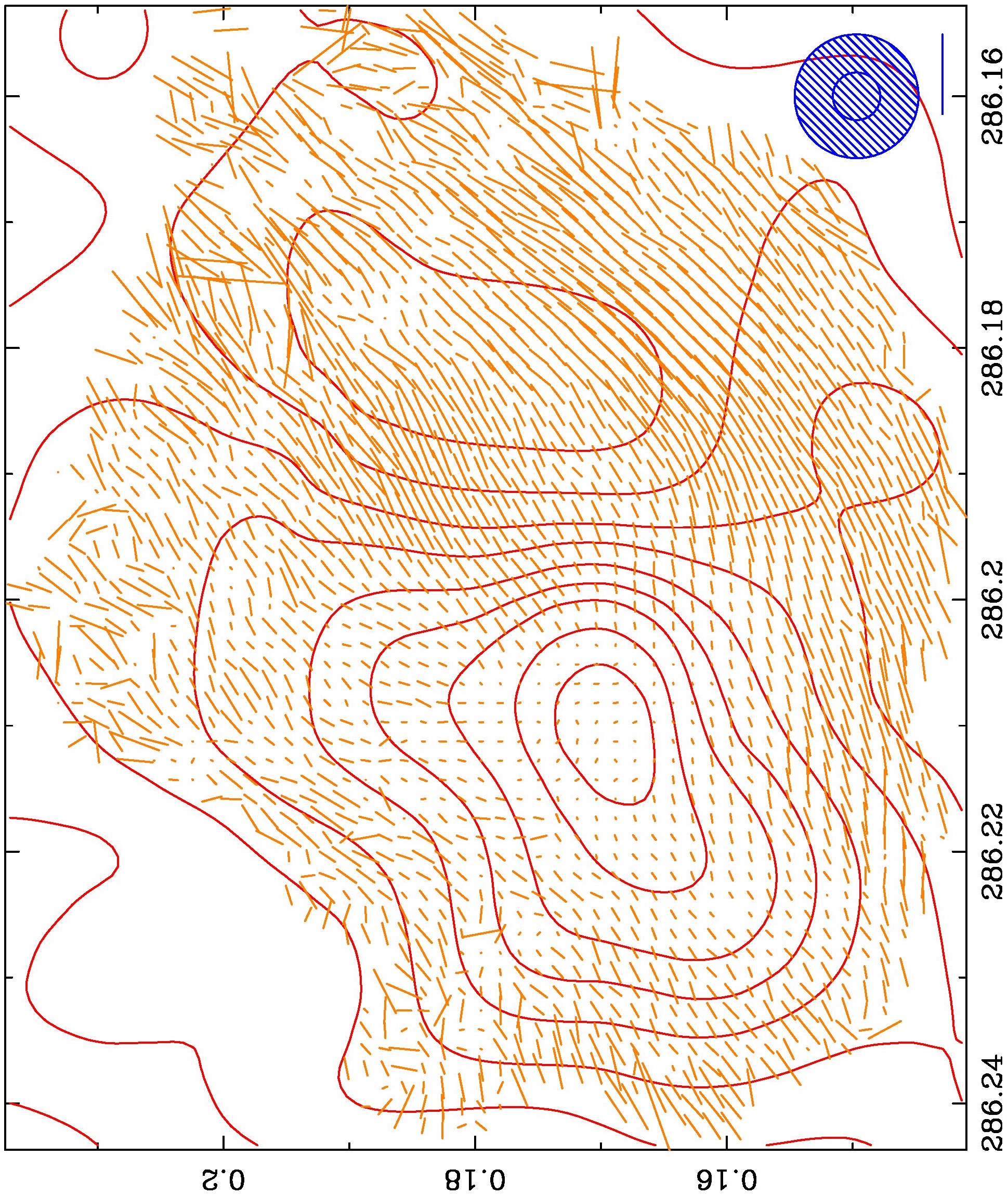}}
\caption{ 
Overlay of the HAWC+ $B$ field vectors (similar to Fig.\,\ref{hawcmaps} but including all vectors with $P'$/$\sigma_{P'}$ $>$ 1.4, 
with a 20\% $p'$ scale in the bottom-right corner) and a column density map (contours 0.8, 1.1, 1.58, 2.27, 3.5, 5.1, 7, 10, 13$\times$10$^{26}$\,m$^{-2}$) derived from the SED-fit $N_{\rm H_2}$ map from {\em Herschel} data \citep{p19} (hence the two beams).
}\label{hawcNH2-B}\vspace{0mm}
\end{figure*}

\subsection{Histogram of Relative Orientations (HRO)}\label{HROstuff}

The HRO method of analysing $B$ field orientations in star-forming gas is by now a fairly standard technique \citep[e.g.,][and references therein]{saa17}.  In the Vela C molecular complex, for example, \citet{saa17} used BLASTPol data with a resolution 3\farcm0 = 0.6\,pc at Vela to examine how the $B$ field orientation changes with column density.  They found that at lower molecular gas column densities $\sim$10$^{26}$\,m$^{-2}$, the $B$ field direction tends to be mostly parallel to or not show any preferred direction relative to gas structures, whereas the $B$ field is mostly perpendicular to higher column density structures $\sim$10$^{27}$\,m$^{-2}$.  This generally confirms a series of results by the lower resolution ($\sim$10$'$) Planck Collaboration \citep[e.g.,][]{pc16} over a wider range of molecular gas column densities.

In the various structural components of Vela C, the transition from mostly parallel to mostly perpendicular can be rather sharp at a certain column density for each structure, but this transition density is different in each structure.  This is widely attributed to a transition from subcritical gas at lower densities, where the flow is at least guided to some extent by the $B$ field, to near-critical or slightly supercritical gas at higher densities, where gravity is capable of overwhelming the magnetic pressure, allowing stars to form.

\subsubsection{HAWC+ Data}

With our substantially higher resolution ALMA and SOFIA/HAWC+ data, it should be instructive to conduct a similar HRO analysis from several pc down to $<$0.1\,pc scales in the massive star formation environment of BYF\,73.  We show first in {\color{red}Figure \ref{hawcNtheta}} the relative alignment of $B$ field vectors with the SED-fit column density $N$ map \citep{p19} as a proxy for ``structure'' in the molecular gas, in all the HAWC+ data as shown in Figure \ref{hawcmaps}.  That is, where the rotated polarisation vectors $\theta_{B_{\perp}}$ are aligned with the tangent to the iso-$N$ contours, the relative angle is close to 0\degr\ and the field is considered to be ``parallel'' to the gas structures.  Where $\theta_{B_{\perp}}$ is perpendicular to the contours and aligned with the column density gradient $\nabla$$N$, the relative angle is close to 90\degr\ and the field is considered to be ``perpendicular'' to the gas structures.  

This approach has the advantage of not imposing any preconceived interpretation of whether the gas structures represent ``clumps,'' ``cores,'' ``filaments,'' or any other potentially subjective term \citep[see, e.g.,][]{pc16,saa17}.  The distribution is quantified by computing histograms on each $N$-bin separately as in {\color{red}Figure \ref{hawcHRObins}}, including the HRO shape parameter --1 $<$ $\xi$ $<$ 1 computed on each $N$ bin's HRO, as described by \cite{saa17}.  This parameter objectively indicates whether there is a preponderance of parallel ($\xi$ $>$ 0) or perpendicular ($\xi$ $<$ 0) alignments in the data, and can be plotted as a function of $N$ ({\color{red}Fig.\,\ref{hawcHROxi}}) to reveal any trends via linear regression,
\begin{equation} 
	\xi = C_{\rm HRO}~({\rm log}N-X_{\rm HRO})~.
\end{equation}

Already in Figure \ref{hawcNtheta} we can see that the distribution of relative alignments has definite patterns in various column density ranges.  These observations are reflected numerically in Figures \ref{hawcHRObins} and \ref{hawcHROxi}.  Thus, in the lower-$N$ ranges, there is an overabundance of parallel alignments (relative PA $<$ 20\degr) between the inferred $B$ field orientation $\theta_{B_{\perp}}$ and the iso-$N$ contours, and $\xi$ $>$ 0 at high significance, 3--13$\sigma$ each across 8 $N$-bins.  In the top two $N$ ranges, however, there is a sudden transition to clearly more perpendicular alignments, PA $\sim$ 40\degr--70\degr\ in the penultimate $N$ bin ($\xi$ $\approx$ 0), and 60\degr--90\degr\ in the top bin ($\xi$ $<$ 0 at 6$\sigma$).  Indeed, compared to Vela C \citep{saa17}, the slope $C_{\rm HRO}$ is substantially closer to --1 in the HAWC+ data for BYF\,73, indicating an even stronger alignment trend with increasing $N$ and a sharper transition ($X_{\rm HRO}$ intercept) than in the Vela cloud, at $N_{\rm crit}$ = (3.9$\pm^{1.0}_{0.8}$)$\times$10$^{26}$\,m$^{-2}$.  Interestingly, the steepness of $C_{\rm HRO}$ as seen in {\em Planck}+BLASTPol large scale maps may be correlated with the inclination angle of the mean $B$ field \citep[e.g.,][]{sfk21}.  One sees a shallower slope in clouds where the polarisation fraction levels indicate that the $B$ field is inclined closer to the line of sight.  Thus, the steeper slope in BYF\,73 may be related to its $B$ field lying close to the plane of the sky (see \S\S\ref{outflow}, \ref{zeeman}, \ref{disc}).

The distribution of points in Figure \ref{hawcNtheta} can be more intuitively understood in {\color{red}Figure \ref{hawcNH2-B}}, which overlays the HAWC+ $p'$ vectors and $N$ contours from the {\em Herschel}-based SED fits \citep{p19}.  This map shows that the large number of points with $N$ \lapp\ 10$^{26}$\,m$^{-2}$ and preferentially parallel alignments arises in the \hii\ region, while the other large concentration of points with $N$ $\approx$ 2$\times$10$^{26}$\,m$^{-2}$ arises mostly from the extended IF to the north and the similar arc bounding the \hii\ region to the southwest of MIR\,2.  The transition between these two column density levels contains relatively few points due to the sharp density gradient across the IF.  For $N$ $\ge$ 3$\times$10$^{26}$\,m$^{-2}$ and progressively closer to MIR\,2, the alignments become preferentially more perpendicular.

\subsubsection{ALMA Data}

\begin{figure}[t]
\centerline{\includegraphics[angle=0,scale=0.215]{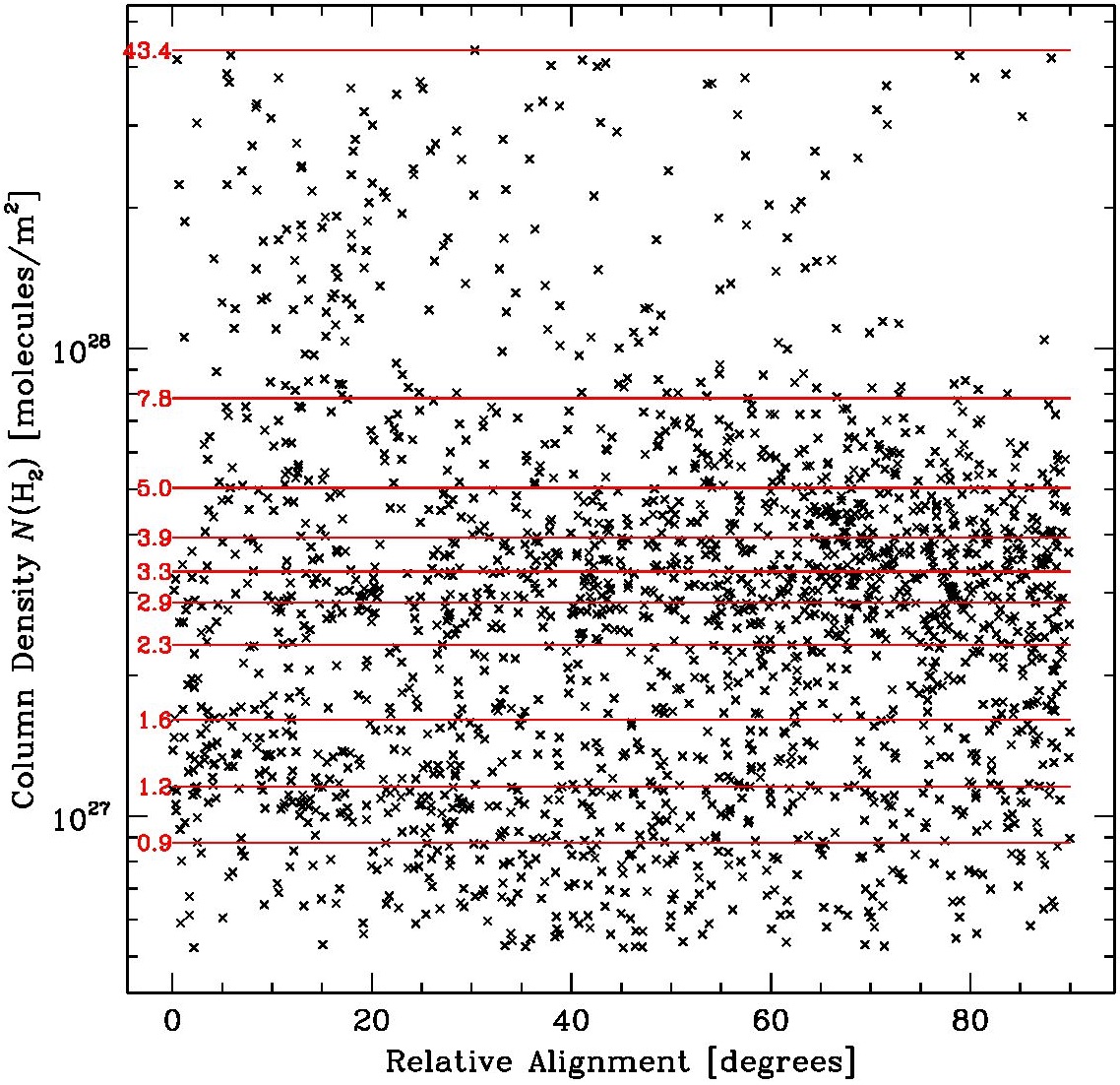}}
\vspace{-1mm}\caption{ 
Similarly to Fig.\,\ref{hawcNtheta}, this shows the relative alignment between polarization position angle $\theta_{B_{\perp}}$ and the tangent to iso-column density $N$ contours, as a function of $N$, except here across the ALMA field.  The red $N$-bin boundaries for each histogram in Figure \ref{almaHRObins} are labelled here in units of 10$^{27}$m$^{-2}$.
}\label{almaNtheta} \vspace{0mm}
\end{figure}

\begin{figure}[t]
\centerline{\includegraphics[angle=0,scale=0.22]{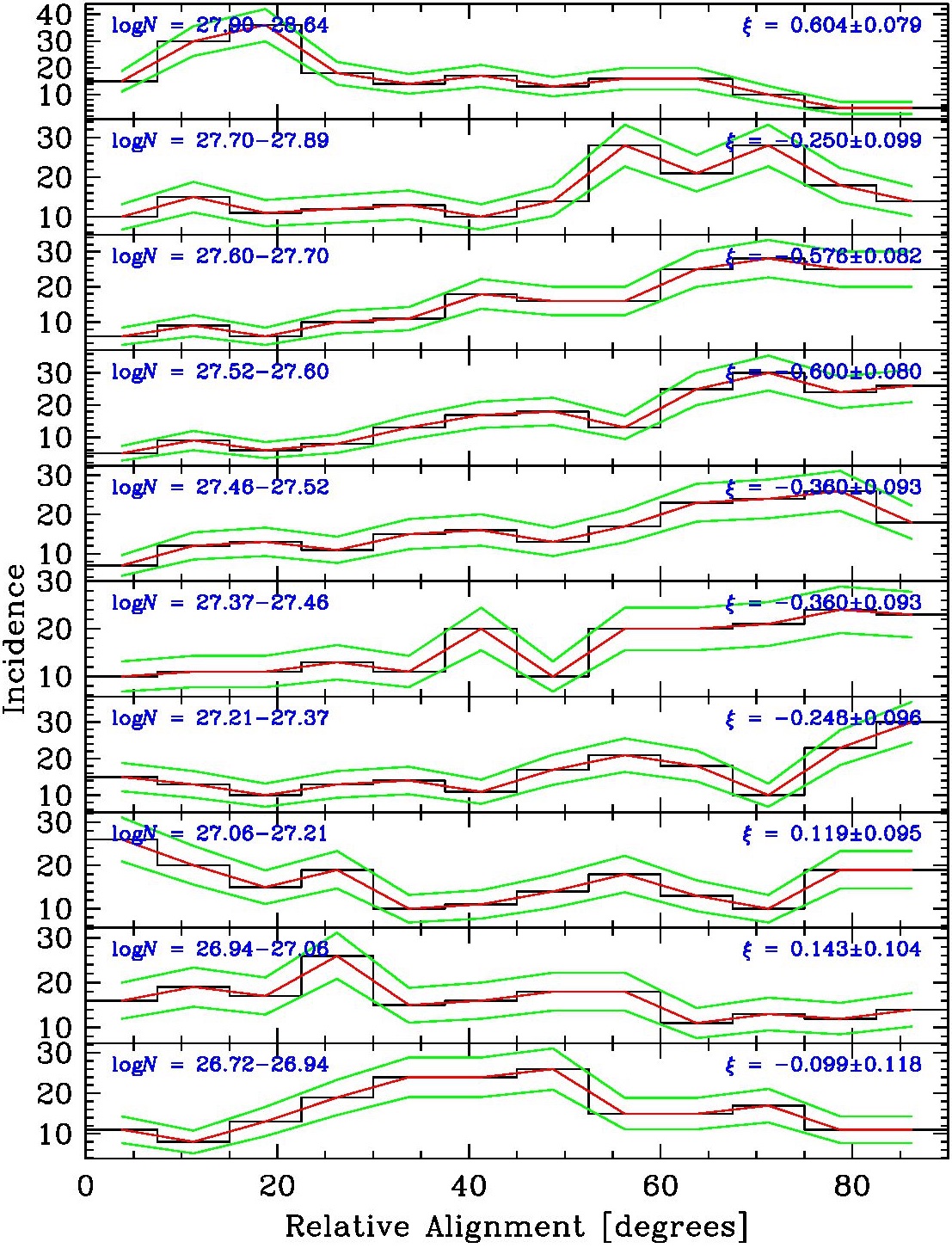}}
\vspace{-1mm}\caption{ 
HRO plots in each of the $N$-bins shown in Figure \ref{almaNtheta}.  Each window is labelled by the range of column density $N$ in that bin and its fitted HRO shape parameter $\xi$ $\pm$ uncertainty, as defined by \citet{saa17}.
}\label{almaHRObins}\vspace{0mm}
\end{figure}

\begin{figure}[h]
\centerline{\includegraphics[angle=0,scale=0.215]{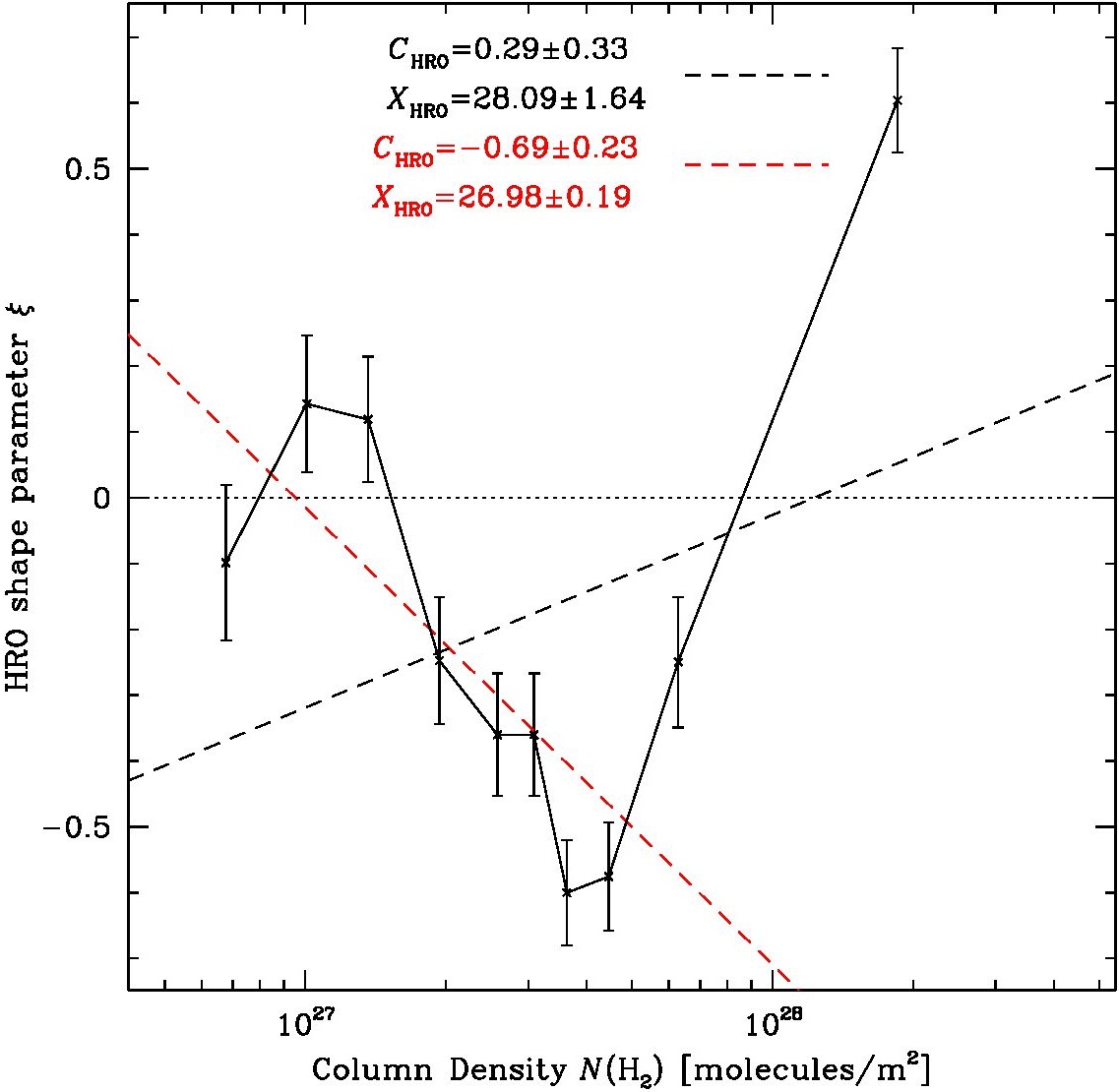}}
\vspace{-1mm}\caption{ 
HRO shape parameter $\xi$ as a function of column density $N$ as fitted in Figure \ref{almaHRObins}.  The black labels and dashed line are solutions to the parameters $C$ and $X$ of a linear regression to all the $\xi$ data, while the red labels and dashed line are for a fit to all data except the highest column density bin with log$N$ $>$ 28.
}\label{almaHROxi}\vspace{0mm}
\end{figure}

We can repeat the HRO analysis on the smaller scale of the ALMA field.  {\color{red}Figures \ref{almaNtheta}--\ref{almaHROxi}} similarly show the overall relative alignment distribution as a function of $N$, the HROs in separate $N$-bins, and the $\xi$ vs.\ $N$ plot for all ALMA data.  The relative alignment distribution shows similarly striking changes with $N$ as in the HAWC+ data. 
In the three lowest-$N$ bins, the $B$ field shows no particular preference for parallel or perpendicular alignments in the ALMA maps ($\xi$ $\approx$ 0 within the uncertainties).  In the middle six $N$ bins, though, the distribution changes to show a substantial preference for perpendicular structures ($\xi$ $<$ 0 at a S/N of 2.5--5$\sigma$ in each bin).  These 9 bins together behave similarly to the BLASTPol results in Vela C, again including the existence of a sharp transition from positive to negative $\xi$, but now at a 4$\times$ higher $N$ $\approx$ 1.6$\times$10$^{27}$\,m$^{-2}$ than in the HAWC+ data.

\begin{figure*}[t]
\centerline{\includegraphics[angle=-90,scale=0.13]{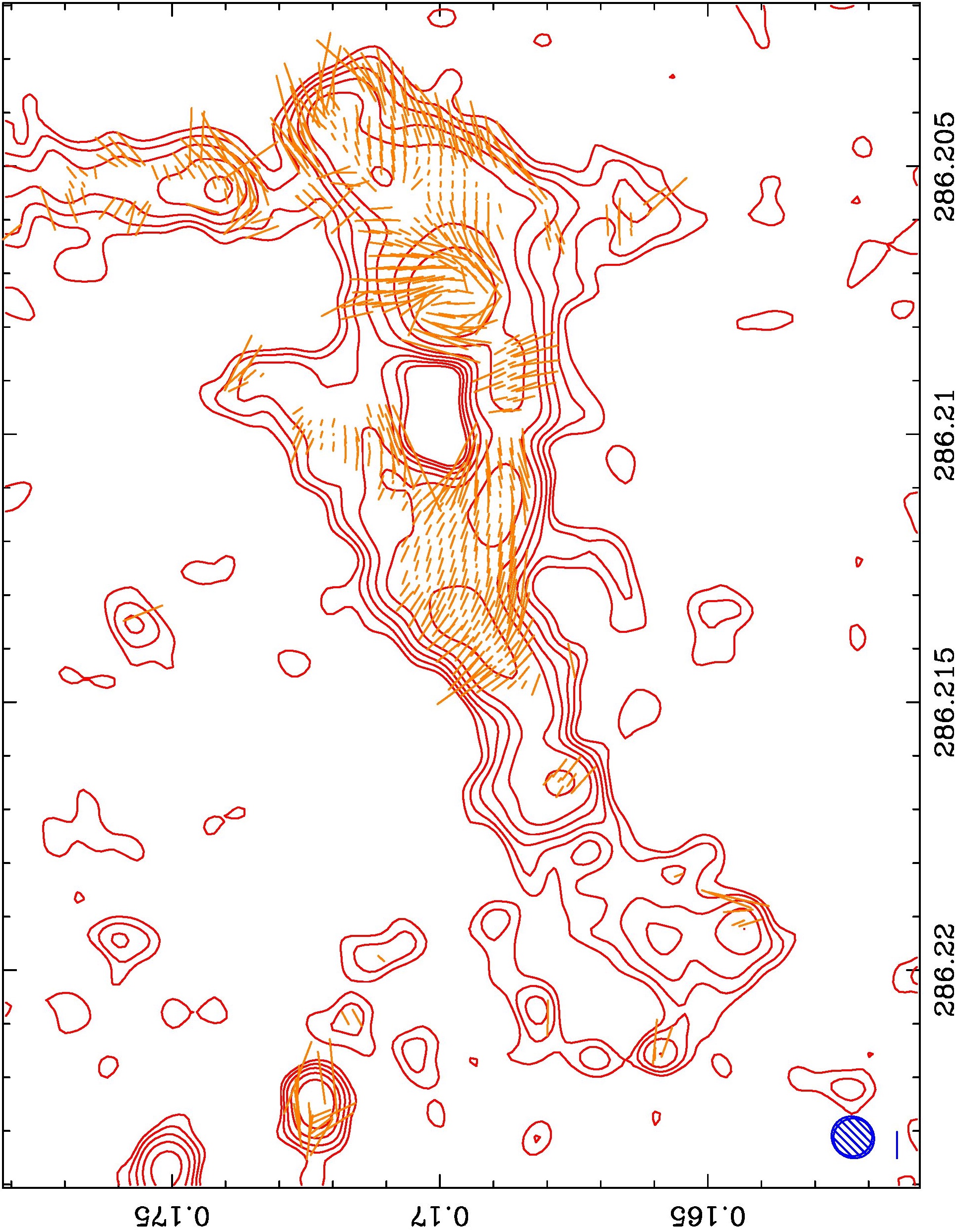}}
\caption{ 
Overlay of ALMA $B$ field vectors (similar to Figs.\,\ref{almaCont}, \ref{IBLpolo}, \ref{Aregs} but including all vectors with $P'$/$\sigma_{P'}$ $>$ 1.4; 20\% $p'$ scale in bottom-left corner) with a column density map (contours 0.3, 0.6, 1, 1.6, 2.5, 5, 10$\times$10$^{27}$\,m$^{-2}$) derived from scaling the ALMA $I$ mosaic (Fig.\,\ref{almaCont}) to an SED-fit $T_{\rm dust}$ map from {\em Herschel} data \citep{p19}.  
}\label{almaNH2-B}\vspace{0mm}
\end{figure*}

However, in the highest-$N$ bin, the distribution changes back to a very strong (7$\sigma$) parallel signature, $\xi$ = 0.60$\pm$0.08.  This produces a very atypical non-negative result in Figure \ref{almaHROxi} for the fitted HRO parameter $C_{\rm HRO}$ (black labels and dashed line).  In Vela C and elsewhere, a negative $C_{\rm HRO}$ means that $\xi$ changes systematically from weakly positive to definitely negative values as $N$ rises, meaning a transition from parallel or random $B$ field alignments to perpendicular ones, often with a sharp transition across $\xi$ = 0 at a particular $N$.  In this context, the last data point in Figure \ref{almaHROxi} may be anomalous, but as it turns out, this may not be that significant.  

To see this, consider the $\theta_{B_{\perp}}$ and $N$ maps together ({\color{red}Fig.\,\ref{almaNH2-B}}), where one can see where each of the ten $N$ bins are located.  The three lowest-$N$ bins with $\xi$ $\approx$ 0 arise in the weaker emission features of the Streamer to the west and farthest east, and southern parts of the IF.  The middle six $N$ bins with $\xi$ $<$ 0 arise in the brighter emission of the EPL, MIR\,2-ext, and the main part of the Streamer.  The highest-$N$ bin arises exclusively from the brightest parts of the MIR\,2 peak, where the structure is actually not well-resolved in the 2\farcs6 ALMA beam.  This would not only preclude accurate $\theta_{B_{\perp}}$ measurements at MIR\,2, but also might include $Q$ and $U$ cancellation within the ALMA beam, underestimating $P'$.  Resolution alone would make any alignment inferences questionable, but in addition we recall that MIR\,2 is near the limit of the reliably-calibrated window of the ALMA polarisation field.  Therefore, we cannot accurately quantify the alignment measurements or their uncertainties right at the MIR\,2 peak, and conclude that the $\xi$ value in this $N$ bin should be discounted.

\begin{figure}[t]
\centerline{\includegraphics[angle=0,scale=0.108]{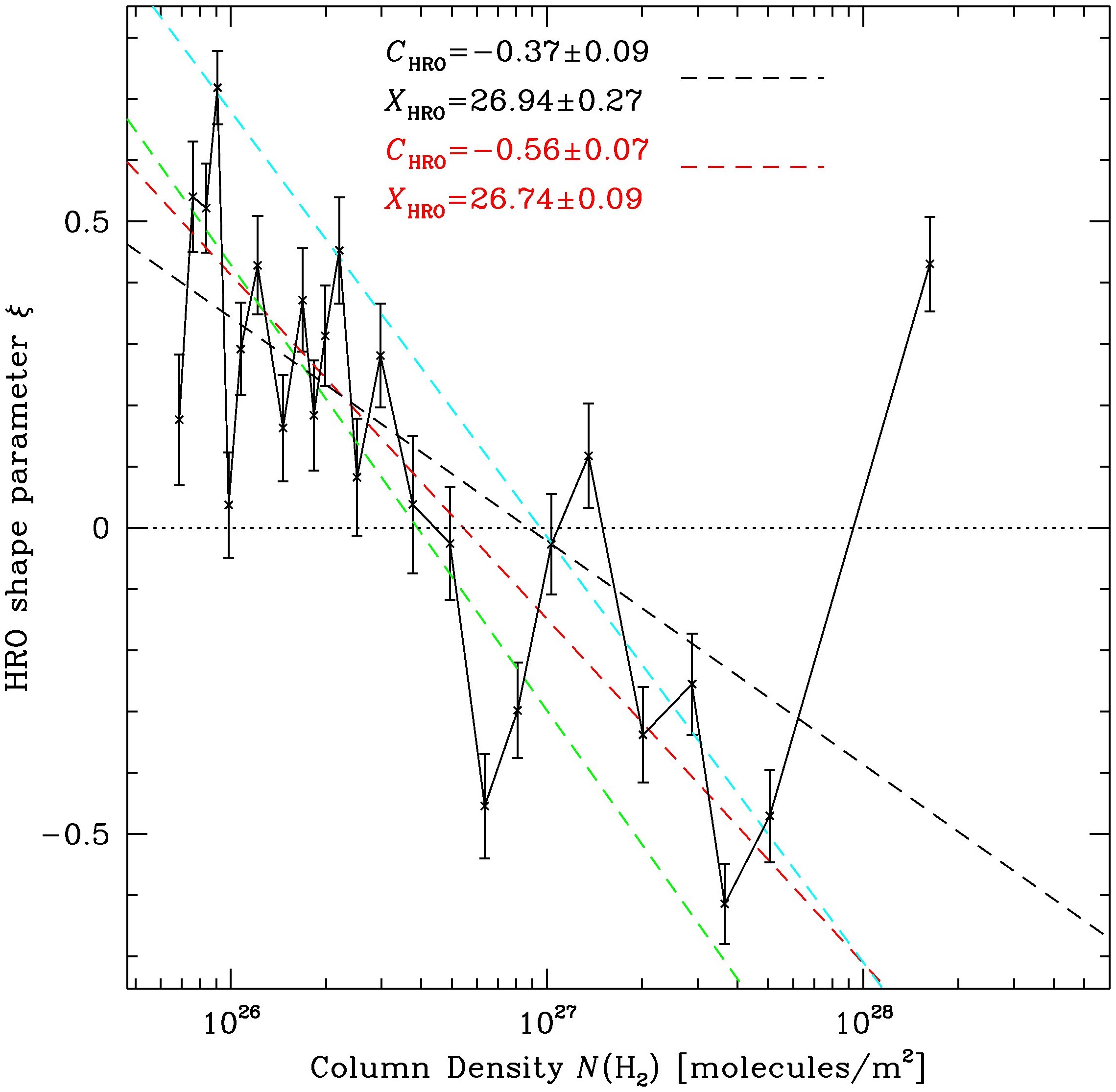}}
\vspace{-0.5mm}\caption{ 
Combined HRO $\xi$ vs.\ $N$ plot from both HAWC+ and ALMA data (Figs.\,\ref{hawcHROxi},\ref{almaHROxi}), where the two data sets overlap in the $N$-bins from 0.5 to 1.5$\times$10$^{27}$\,m$^{-2}$, and we have increased the number of $N$-bins to 25 because of the roughly doubled number of points.  The overall results of the fitting, however, are very similar for any number of $N$-bins between 10 and 30.  As in Fig.\,\ref{almaHROxi}, the black labels and dashed line are solutions to the parameters $C$ and $X$ of a linear regression to all the $\xi$ data, while the red labels and dashed line are for a fit to all data except the highest column density bin with log$N$ $>$ 28.  
Overlaid in green and cyan are the respective fits from Figs.\,\ref{hawcHROxi} and \ref{almaHROxi} for comparison. 
}\label{comboHROxi}\vspace{0.3mm}
\end{figure}

As an exercise, therefore, we also computed the regression parameters $C$ and $X$ for the nine lower-$N$ bins in Figure \ref{almaHROxi}, and show these as red labels and a dashed line.  In this case $C$ is definitely negative (3$\sigma$) and more in line with the Vela C results, while the $X_{\rm HRO}$ intercept gives $N_{\rm crit}$ = (9.5$\pm^{5.3}_{3.4}$)$\times$10$^{26}$\,m$^{-2}$.  Based on this alone, it seems desirable to obtain an even higher-resolution polarisation map of MIR\,2 and its immediate surroundings.  Such a map would allow us to explore the massive protostellar core's $B$ field in much finer detail and track the $\xi$ trend to even higher $N$, not to mention better resolving the core itself (e.g., 250\,AU at 0\farcs1).

\subsubsection{Combined Data}

The $\xi$--$N$ trends in Figures \ref{hawcHROxi} \& \ref{almaHROxi} overlap nicely in column density, and we present a combined plot in {\color{red}Figure \ref{comboHROxi}}.  There, the slope $C$ is slightly shallower compared to either the HAWC+ or ALMA-only results, but the overall trend is firmer (the uncertainties in $C$ \& $X$ are smaller) due to the wider range of $N$ being sampled.  
In combination, the data suggest that the transition to perpendicularity in BYF\,73's Streamer and MIR\,2 core occurs (from the red intercept $X$ in Fig.\,\ref{comboHROxi}) at $N_{\rm crit}$ = (6.6$\pm^{1.2}_{0.9}$)$\times$10$^{26}$\,m$^{-2}$, near the geometric mean of the transitions from the individual instruments.  This can also be converted into an equivalent critical gas density if we assume a line-of-sight depth to the Streamer approximately equal to its projected width, $n$ $\sim$ $N$/$D$, where $D$ $\approx$ 0.087\,pc.  Then, $n_{\rm crit}$ = (2.0$\pm^{0.5}_{0.4}$)$\times$10$^{11}$\,m$^{-3}$. 

This is actually a rather suggestive threshold: in Figure \ref{almaNH2-B}, it is the column density of the second-lowest contour, and includes the MIR\,2 core, $\sim$all of the Streamer-main and -west, and much of the IF.  It suggests that the Streamer's width may be related, locally at least, to the transition between MHD forces governing the gas dynamics and self-gravity, as seen in \S\ref{cubes}.

It is instructive to compare this result with other HRO studies.  For example, \citet{pc16} used 10$'$ resolution {\em Planck} data to study $B$ field orientations in 10 nearby (150--450\,pc) Gould Belt clouds, with a finest physical resolution similar to our HAWC+ data and ranging up: $\sim$0.4--40\,pc.  At this scale the median gas densities are $n$ $\approx$ 5$\times$10$^{8}$\,m$^{-3}$, substantially less than is typical in the Streamer as estimated above.  The threshold column densities in these 10 clouds are also lower than that for BYF\,73, by a factor of 10 on average, $X$ $\approx$ 6$\times$10$^{25}$\,m$^{-2}$.  Similarly in Vela-C, a massive but relatively unevolved cloud at 700--900\,pc, \citet{saa17,zss20} used {\em Herschel} and BLASTpol data at 3$'$ resolution for their HRO analysis \citep[i.e., with a similar physical resolution to][]{pc16}, and found a typical $X$ $\approx$ 3$\times$10$^{26}$\,m$^{-3}$ or about half BYF\,73's value.  Finally, in two portions of L1688 in Ophiuchus at 140\,pc, \citet{lbc21} combined HAWC+ and {\em Planck} data (giving similar physical resolutions to our ALMA data) to confirm a column density threshold similar to \citet{pc16}'s 10 clouds, and a volume density threshold $n$ $\approx$ 10$^{10}$\,m$^{-3}$.

We can relate this column density threshold to the equivalent $B$ field threshold if gravity and the magnetic pressure were critically balanced.  Using the mass-to-flux ratio approach of \citet{cn04} as adapted by \citet{BL15}, we have
\begin{equation} 
	\lambda = \frac{(M/\Phi)_{\rm obs}}{(M/\Phi)_{\rm crit}} = 0.064~\frac{N_{\rm H_2}/10^{24}{\rm m}^{-2}}{B_{\rm TOT}/{\rm nT}}~~,
\end{equation}
or
\begin{equation} 
	B_{\rm crit}~({\rm nT}) = N_{\rm H_2}/(1.57 \times 10^{25}{\rm m}^{-2})~~,
\end{equation}
when $\lambda$ = 1.  With the above threshold, we obtain $B_{\rm crit}$ = 25$\pm$6\,nT for the HAWC+ measurements, $B_{\rm crit}$ = 61$\pm$27\,nT for the ALMA data, and $B_{\rm crit}$ = 42$\pm$7\,nT 
averaged over the mapped HAWC+ and ALMA emission in BYF\,73, based on the combined HRO analysis.

Again, we can compare this result to equivalent $B_{\rm crit}$ for the nearby clouds of $\sim$4\,nT \citep{pc16}, $\sim$20\,nT for Vela-C \citep{saa17}, and $\sim$4\,nT for L1688 \citep{lbc21}, placing BYF\,73 at a higher-$N_{\rm crit}$ and -$B_{\rm crit}$ level than these other clouds.  Our $B_{\rm crit}$ result for BYF\,73 generally is also about half the DCF value at the peak of MIR\,2 with Mopra's resolution (Eq.\,2), which is not unreasonable given the respective density levels.  Thus, both the DCF and HRO analyses give us mutually consistent clues about the $B$ field strengths in BYF\,73, which seem to be significantly stronger than in other clouds.


\subsection{The Goldreich-Kylafis (GK) effect in \tco}\label{GKstuff}

\subsubsection{Widespread Polarisation in the Outflow}

The GK effect can arise when $B$ fields (even weak ones) and velocity \& excitation gradients in molecular gas combine to produce imbalances from thermal equilibrium in populations of magnetic sublevels $M$ of spectral lines with opacity $\sim$1 \citep{gk81,gc04,cru12}.  This can produce linearly polarised spectral line emission that is either aligned with ($\pi$ transitions) or perpendicular to ($\sigma$ transitions) the local $B$ field, depending on the radiative transfer circumstances, namely the unknown angles between the radiation anisotropy, the line of sight, and the $B$ field direction.  In addition, the classical Zeeman effect can give rise to circularly polarised $\sigma$ transitions parallel to $B$, observable for that component of $B$ oriented along the line of sight.

\begin{figure*}[ht]
\includegraphics[angle=-90,scale=0.167]{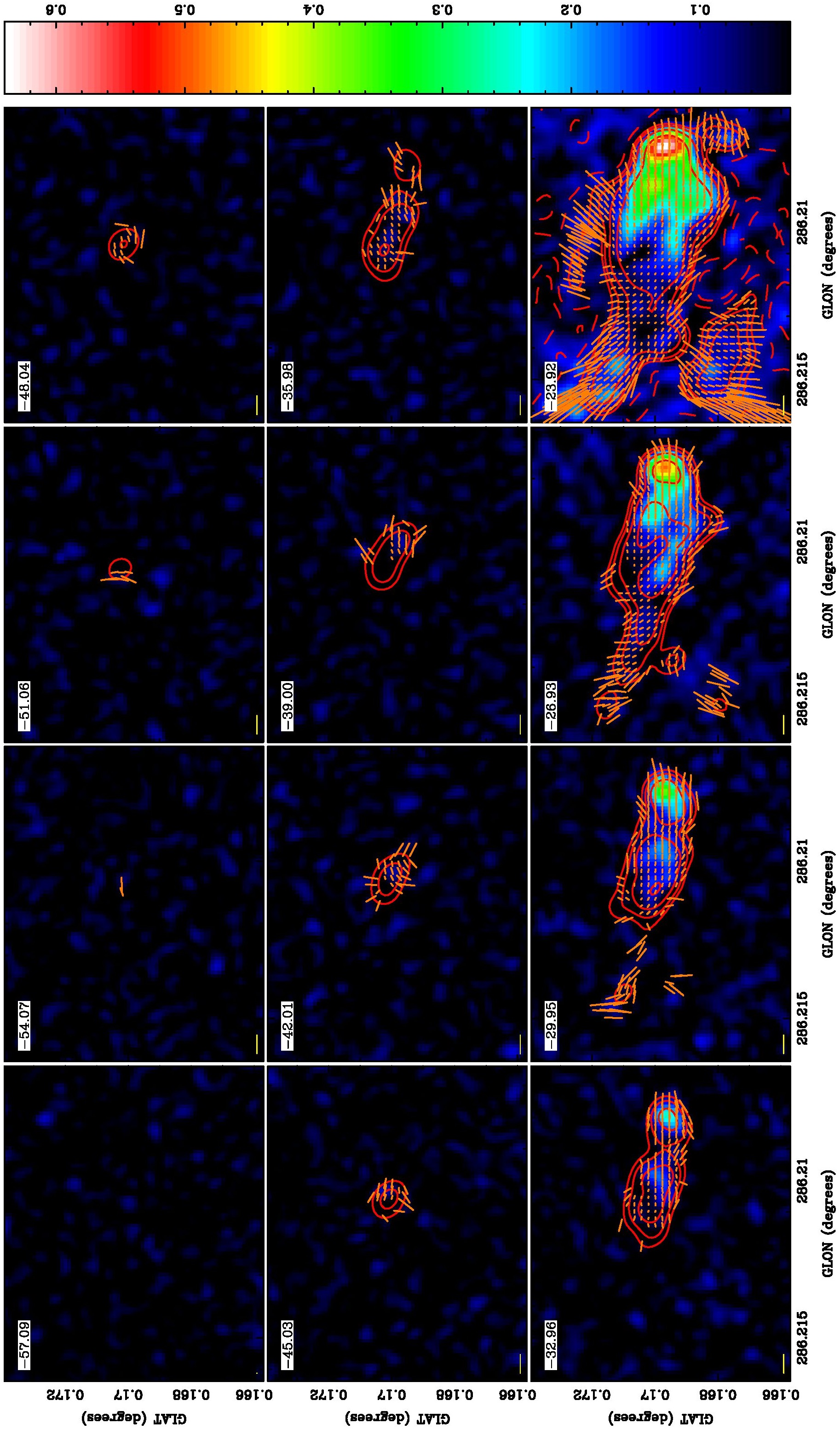}\vspace{-4mm}
\includegraphics[angle=-90,scale=0.167]{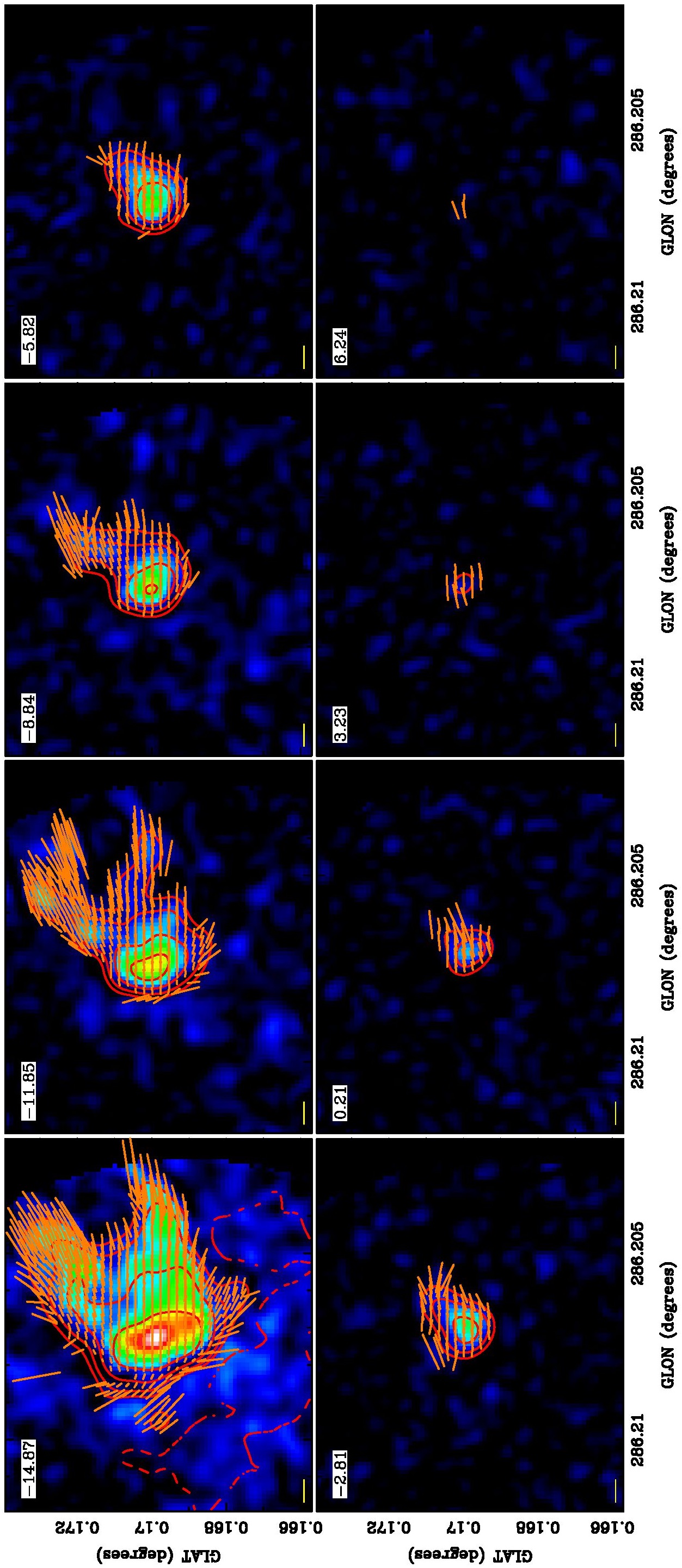}
\vspace{-1mm}\caption{ 
BYF\,73 \tco\ outflow polarisation maps shown in 3\,\kms-wide panels, each labelled by their centre velocities in the top-left corner and a 5\% polarisation vector scale (yellow bar) in the bottom-left corner.  The panels overlay several components, averaged over the same velocity ranges: polarised flux $P'$ images scaled to the colour bar on the right in K; $I$ contours in red at 2,4,8,16\,K, dashed for negative values from missing short-spacing information; and orange percentage $p'$ vectors at every second pixel above 4$\sigma$, with PAs {\em rotated by 90\degr} to indicate $\theta_{B_{\perp}}$. 
For the $P'$, $p'$, and $\theta_{B_{\perp}}$ data in each panel, they were constructed by first binning the native Stokes data by 19 channels, and then forming the products $P'$, $p'$, and $\theta_{B_{\perp}}$ on the new binned cubes.  
In order to better display the high-S/N features, the top 12/bottom 8 panels respectively show the blue-/red-shifted line wings vignetted to the east/west of MIR\,2.
}\label{almaCOpolo}\vspace{0mm}
\end{figure*}

We report here the widespread detection of strong, linearly polarised emission in the \tco\ line wings from BYF\,73 (i.e., its bipolar outflow) that is consistent with the GK effect.  The native Stokes data have high S/N, up to 21 for $P'$ and $p'$, in individual 0.16\,\kms-wide channels and across much of the outflow visible in both line wings.  However, where the polarisation signal weakens, the $p'$ values tend to rise and, with the larger uncertainties, the resulting vector maps become somewhat confusing to look at.  Therefore, we averaged (binned) the native Stokes data into $\sim$3\,\kms-wide channels for display purposes only, and formed the polarisation products on the binned cubes with proper noise weighting: the significant polarisation features are then more easily visualised in the binned data.  {\color{red}Figure \ref{almaCOpolo}} shows overlays of the blue- and red-shifted $I$ and $P'$ emission in these binned data, together with all observed polarisation vectors above 4$\sigma_{\rm rms}$ (and up to 72$\sigma$) across the full velocity range of the line wings.

Interpreting GK polarisation vectors requires some resolution of the 90\degr\ ambiguity described above, depending on whether $\sigma$ transitions from the $M$=$\pm$1 magnetic sublevels (polarised perpendicularly to the $B$ field) will be stronger or weaker than the $\pi$ transitions from $M$=0 sublevels (parallel to $B$).  In general, whether outflows are driven by magnetocentrifugal forces anchored in protostellar disks, or by collimated protostellar jets carrying their own $B$ fields, the nominal expectation is that inferred $B$ fields should be aligned along outflows.  In 
Figure \ref{almaCOpolo} we have {\em rotated} the vectors by 90\degr\ from those observed, and it is this orientation which, remarkably clearly, shows an overwhelming orientation along the outflow direction in each wing, especially for the higher-S/N pixels.  Equivalent plots of the {\em observed} vectors show a near-universal circumferential alignment around MIR\,2, which would seem to be unphysical based on the above understanding.

Indeed, the high-S/N vectors track {\em both} the bend in the blue wing {\em and} the fork in the red wing inferred solely from the $I$ emission pattern (Fig.\,\ref{almaCOjet}).  There are some low-S/N vectors which don't align with this general pattern, however, typically near the $p'$ threshold.  This is most notable in the --24\,\kms\ panel, both north and south of the outflow itself.  In the northern portion of this polarised emission, the alignment is instead approximately 
across the EPL as mapped by HAWC+ (Figs.\,\ref{pplobes}--\ref{almaCOjet}).  South of the outflow, the emission appears to be an artifact of missing short spacings in the field of view, so we discount it.

\begin{figure*}[t]
\includegraphics[angle=0,scale=0.21]{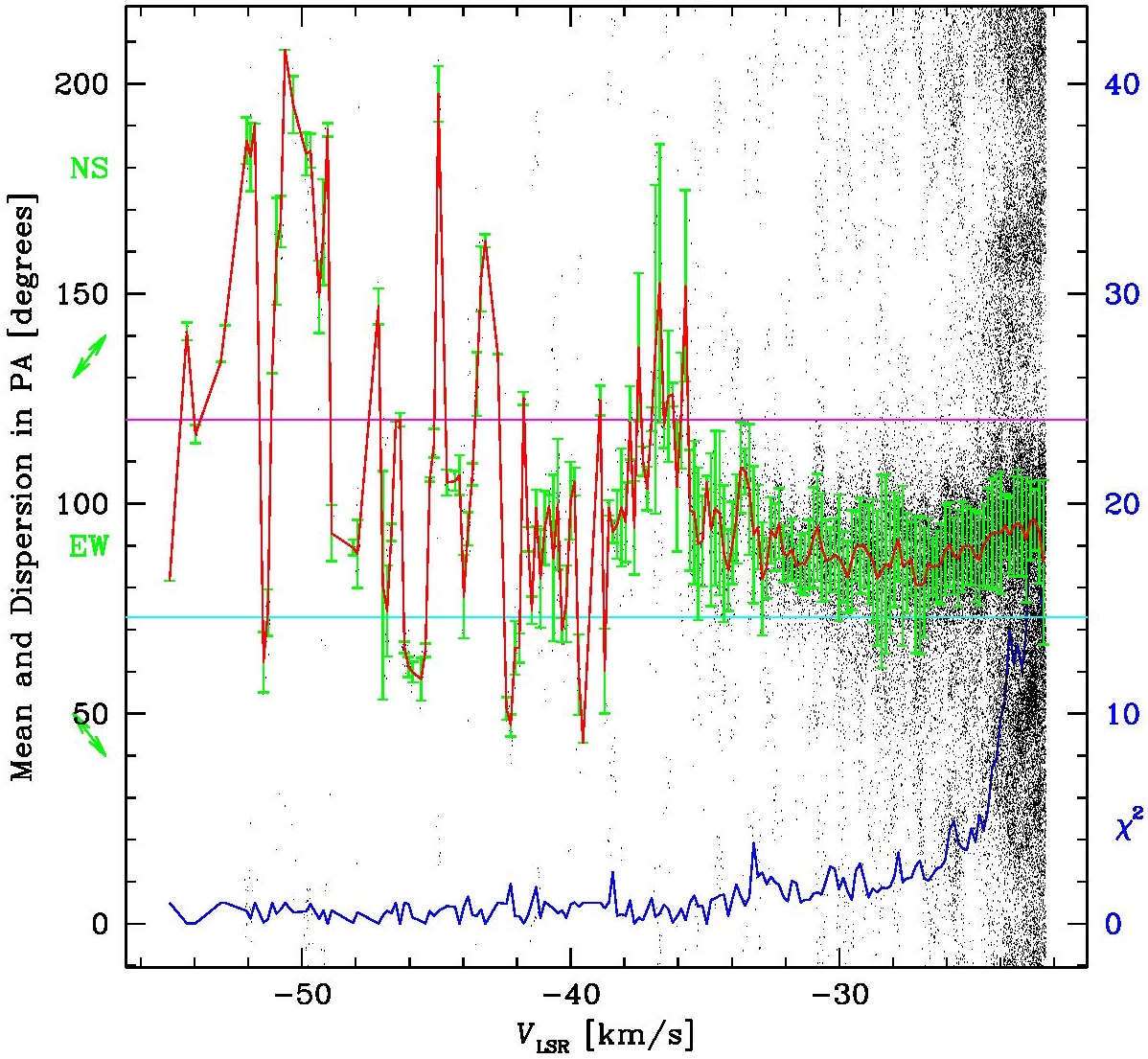}\hspace{2mm}
\includegraphics[angle=0,scale=0.21]{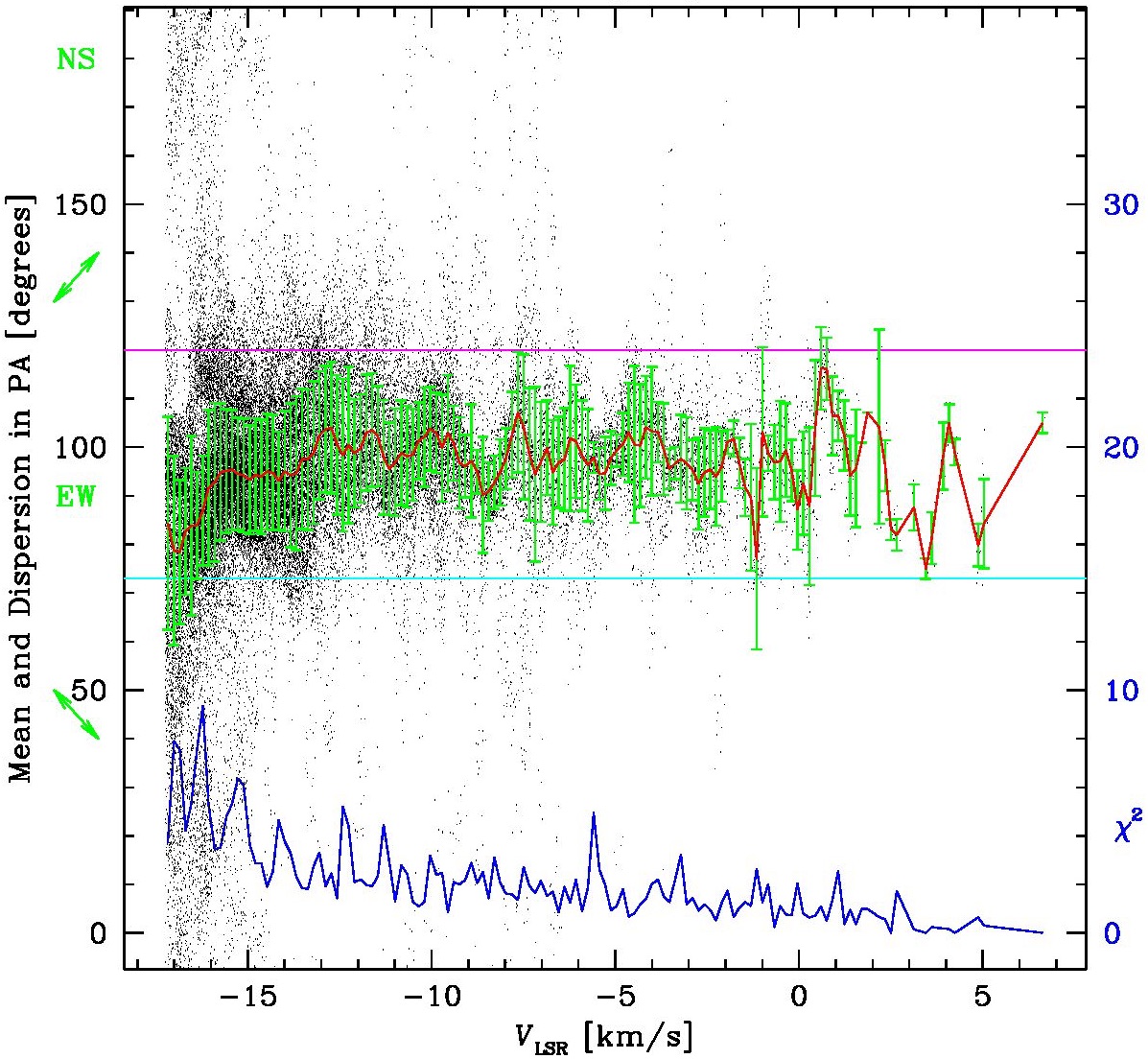}
\vspace{-4mm}\caption{ 
Simplified DCF analysis of polarisation orientations in the ALMA \tco\ data ({\em left}, blue shifted emission; {\em right}, red-shifted emission) as a function of velocity, treated as single boxes encompassing all polarised emission in each channel.  All pixels of $\theta_{B_{\perp}}$ above 4$\sigma$ are shown as black dots; their mean values in each channel are connected by a red line, while the dispersions in each channel's $\theta_{B_{\perp}}$ distributions are drawn as green error bars. The dark blue line shows $\chi^2$ values (on the right ordinate axis of each panel) of gaussian fits to the $\theta_{B_{\perp}}$ distributions in each channel.  For reference, the horizontal magenta and cyan lines respectively show the orientation of the red- and blue-shifted outflow axes, as illustrated in Fig.\,\ref{almaCOjet}, and each left ordinate is additionally labelled with compass directions. 
}\label{CO-DCF} \vspace{0mm}
\end{figure*}

An arrangement with outflows oriented along the $B$ field direction is typical of \cite{cru12}'s summary of outflow studies via GK mapping.  The data for BYF\,73 show that the {\em observed} polarisation is preferentially oriented 90\degr\ from the presumed $B$ field direction down the outflow axes, and so supports an excitation condition in which the $\sigma$ $M$=$\pm$1 transitions are robustly overpopulated in the outflow relative to the $\pi$ $M$=0 transitions.

\subsubsection{Simplified DCF Analysis}

Quantifying this description, however, is challenging due to the sheer volume and effectively 4D nature of the data.  As a first attempt, we perform a simplified DCF analysis {\em per unbinned 0.16\,\kms-wide channel} in the data.  We argue that this is reasonable, even though the original DCF method was not developed for outflows.  Indeed, we believe that DCF analysis of spectral-line linear polarisation will give a better result than for dust polarisation, in the following sense.

A GK-imbued spectral line directly samples the turbulence, density, and polarisation dispersion in the same region.  A problem with application of DCF to dust polarisation is that one needs to estimate the density sampled by the dust polarisation, plus a turbulent linewidth.  Although we have dust-based column density maps of BYF\,73 (Figs.\,\ref{hawcNH2-B}, \ref{almaNH2-B}) from which density estimates can be simply inferred, densities and linewidths are more generally inferred from observations of spectral lines: excitation analysis for density, and directly measured linewidth.  However, different spectral lines sample different density regimes and different lines have different linewidths, so just what is appropriate for the dust polarisation analysis is never clear.  One often ends up with some sort of ill-defined average along the line of sight.  On the other hand, for spectral-line polarisation things are self-consistent.  One can infer the polarisation dispersion, density, and measure the turbulence directly in the same parcel of gas, namely that sampled by the spectral line being observed.  DCF then gives an estimate of the $B$ field strength in that spatial and density parcel, not for some ill-defined and possibly different regions along the line of sight.

If GK polarisation can be detected in multiple spectral lines that sample different density regimes, one can in principle build up a 3D picture of the $B$ field.  So far, however, detections of the GK effect in species other than CO have been rare, 
but presumably that will improve as time goes on.  

For this channel-DCF analysis of BYF\,73, we do not include sub-ROIs of each channel at smaller scales, as in Figs.\,\ref{HCFstats} and \ref{ACFstats}, and assume instead for simplicity that the whole-channel-ROI gives an approximate measure of the $\theta_{B_{\perp}}$ correlation length for that channel, since the polarised emission is dominated by the outflow structure, as seen in Figure \ref{almaCOpolo}.  The results are shown in {\color{red}Figure \ref{CO-DCF}} for both \tco\ line wings.

Several features are immediately evident.  The most significant are the clear trends in $\theta_{B_{\perp}}$, for the blue wing from \vlsr\ = --22 to --36\,\kms, and for the red wing from \vlsr\ = --17 to --2\,\kms, showing a $B$ field orientation that changes gradually, in both cases, from EW to more along the outflow axis and then back to EW, as one looks from the lower to higher outflow speeds.  This internal consistency is not so surprising since the statistics in these channels (a few hundred pixels each) are quite robust.  Observationally, however, there is no reason to expect the polarisation to line up so reliably, channel by channel, unless the polarisation signal in all channels is strongly governed by the intrinsic physics of the outflow.  Thus, over these velocity ranges, the dispersion in $\theta_{B_{\perp}}$ for each channel is quite small, $s$ = 11\degr$\pm$4\degr, 
even where a few pixels appear as outliers in the $\theta_{B_{\perp}}$ distribution of some channels.  This is not much larger than the average noise-derived $\Delta\theta_{\rm rms}$ $\approx$ 7\degr, 
giving an intrinsic mean dispersion $s$ $\approx$ $\pm$8\fdeg6, 
or as little as $\pm$7\degr in some places.  Overall, the polarised emission at these velocities appears very well organised. 

At velocities from \vlsr\ = --36 to --55\,\kms\ and $>$--2\,\kms, however, the mean $\theta_{B_{\perp}}$ direction in each channel becomes more erratic as the outflow speed increases, on average still lying near the outflow directions but with a dispersion {\em among} the channels $s$ $\approx$ $\pm$40\degr\ for the blue wing.  This wider variation probably reflects poorer statistics, with typically only a few, or a few dozen, pixels per channel. 

At velocities closer to the line core, \vlsr\ = --22 to --17\,\kms, aliasing of extended line emission throughout the field of view introduces many polarisation features with probably unreliable $\theta_{B_{\perp}}$, indicated in Figure \ref{CO-DCF} by both the increasing density of dots at all $\theta_{B_{\perp}}$ and higher $\chi^2$ from the non-gaussian $\theta_{B_{\perp}}$ distribution, all becoming more noticeable as the velocity approaches $V_{\rm sys}$ = --19.6\,\kms.

\subsubsection{Magnetic Field Strength Calculations}

As with the continuum data (\S\ref{DCFstuff}), we can use this basic DCF information on the dispersion in $\theta_{B_{\perp}}$ per channel from the GK effect, to make estimates of the $B$ field strength in the outflow.  For Eq.\,(1) from \S\ref{DCFstuff}, we first estimate the \tco\ column density in the line wing emission $I_{\rm ^{12}CO}$ via the velocity-resolved, opacity-corrected conversion law from \citet[][their Fig.\,5b, not their integrated law in Fig.\,9b]{b18}.  Next, we convert that to an H$_2$ column density with \cite{p21}'s dust temperature-dependent abundance law.  Finally, we turn this into a volume density assuming a line-of-sight depth $D$ $\approx$ 0.087\,pc through the outflow (and correlation length, later) equal to the outflow's average projected width:
\begin{equation} 
	n_{\rm ch} = \frac{N_0}{D}~\frac{(I_{\rm ^{12}CO}{\rm d}V/{\rm K\,km\,s^{-1}})^p} 
					{10^{[{-10\,{\rm log}^2(T_d/T_0)+{\rm log}X_0}]}}~. \
\end{equation} 
From \cite{b18}, $N_0$ = 1.27$\times$10$^{20}$\,m$^{-2}$ and $p$ = 1.92; the ALMA channel width d$V$ = 0.159\,\kms\ converts $I$ to the proper units; and from \cite{p21}, $T_0$ = 20\,K is the dust temperature at which the gas phase CO abundance relative to H$_2$ peaks, at a value $X_0$ = 0.74$\times$10$^{-4}$.

Eq.\,(9) thus converts the $I_{\rm ^{12}CO}$ data cube into a cube of H$_2$ density per channel at the observed velocity.  To combine this with Eq.\,(1), we need a turbulent velocity dispersion.  We can choose a velocity FWHM $\Delta$$V$ in the gas corresponding to 1 ALMA channel to be consistent with the above formulation, but in reality it may be several times larger, since the outflow is likely to be turbulent at some level related to the $\pm$25\,\kms\ range of flow speeds.  In that case, the true velocity FWHM in the gas would re-scale the single-channel $B_{\rm \perp,DCF}$ estimated via Eq.\,(1) by $\psi$ 
= ($\Delta$$V$/d$V$)$^{1+p/2}$, since we would need to evaluate $I$ in Eq.\,(9) over the same $\Delta$$V$-wide bins (the column density inferred from $I$, and hence the density, is additive across channels).  With $p$ as above, $\psi$ $\propto$ ($\Delta$$V$/d$V$)$^2$ approximately, or as the $\sim$square of the number of channels in a $\Delta$$V$ bin.

So for molecular gas with $\mu$ as before and $s$ = 7\degr\ in the inner part of the flow, where minimal gas-phase densities inferred from Eq.\,(9) are $n_{\rm ch}$ $\sim$ 10$^{9}$\,m$^{-3}$\,ch$^{-1}$, we obtain 
\begin{eqnarray}  
	B_{\rm \perp,DCF,outflow} &=& 0.61\,{\rm nT} \sqrt{\frac{n_{\rm ch}}{10^{9}{\rm m^{-3}}}}\left(\frac{\psi}{s/7\degr}\right)~~{\rm or} \\
%
	&\approx& 4.5\,{\rm nT}\left(\frac{I_{\rm ^{12}CO}{\rm d}V}{10\,\rm K\,km\,s^{-1}}\right)^{p/2}\hspace{-1.5mm}\left(\frac{\psi}{s/7\degr}\right)~~
\end{eqnarray}
as a minimum for 
$B$ in the molecular outflow, when approximating the dust temperature at 20\,K throughout to compute a minimal \htwo\ density at the peak CO abundance.

As described above, 1 channel probably slices the turbulent structure in the outflow rather finely: for $\Delta$$V$ = 3\,\kms, for example, the scaling would increase the single-channel $B_{\rm \perp,DCF}$ coefficient by $\psi$ = 320$\times$, e.g., to 1.4\,$\mu$T for the same value 
of $I$ in Eq.\,(11).  Indeed, the brightness of the mapped \tco\ outflow emission in Figure \ref{almaCOpolo} ranges up to 90\,K\kms\ in 3\kms-wide bins, suggesting that 
$B$ fields might be stronger still in some locations.  Of course, this scaling may not actually be valid: while simulations of turbulent plasmas suggest that DCF estimates are reasonable up to Mach numbers {\bf M} $\sim$ 5--9 \citep{osg01}, such values may be far exceeded ({\bf M}$>$100) in the outflow.

\subsubsection{Energy Densities}

Despite these somewhat large uncertainties, we at least have rough estimates for $B_{\rm \perp,DCF}$ field strengths in the outflow.  As a final exercise, we compare the energy density ${\mathfrak M}$ that would exist in such $B$ fields with the kinetic energy density ${\mathfrak K}$ of the outflowing gas.  ${\mathfrak M}$ follows directly from Eq.\,(10),
\begin{eqnarray} 
	{\mathfrak M} &=& B^2/2\mu_0 = 1.49\times10^{-13}{\rm Jm}^{-3}~n_9\left(\frac{\psi}{s/7\degr}\right)^2,~~~
\end{eqnarray} 
where $\mu_0$ is the permeability of free space (or the magnetic constant in preferred SI usage) and $n_9$ = $n_{\rm ch}$/10$^9$m$^{-3}$.  Thus, smaller values for ${\mathfrak M}$ will obtain at lower gas densities (and approximately $I$, through Eq.\,(9)), either at the edges of the outflow or at higher velocities, whereas larger ${\mathfrak M}$ will lie closer to the outflow origin where the density or $I$ is higher.

For ${\mathfrak K}$ we can assume a cylindrical outflow geometry (diameter $D$, length $L$) to approximately compute 
\begin{eqnarray} 
	{\mathfrak K} &=& \frac{\frac{1}{2}MV_{\rm rel}^2}{\frac{1}{4}\pi D^2L} = \frac{1}{2}\rho V_{\rm rel}^2 \nonumber \\
	&=& 3.9\times10^{-12}\,{\rm Jm}^{-3}~n_9~(V_{\rm rel}/{\rm km\,s^{-1}})^2
\end{eqnarray}
where $L$ $\approx$ 0.6\,pc is the physical length of the 50$''$-long blue lobe of the outflow.  The second expression is much simpler to use, since the mass density $\rho$ = 2$\mu\,m_Hn_{\rm ch}$, with the number density $n_{\rm ch}$ from Eq.\,(9).  This can actually be done separately for each channel if we use its relative outflow speed $V_{\rm rel}$ as measured from $V_{\rm sys}$ = --19.6\kms.  
Thus, the value of ${\mathfrak K}$ will be larger at higher $\rho$ $\propto$ $I$ (approximately) but especially at higher $V_{\rm rel}$, or smaller at lower $I$ or especially lower $V_{\rm rel}$.

\begin{figure*}[ht]
\centerline{\includegraphics[angle=-90,scale=0.166]{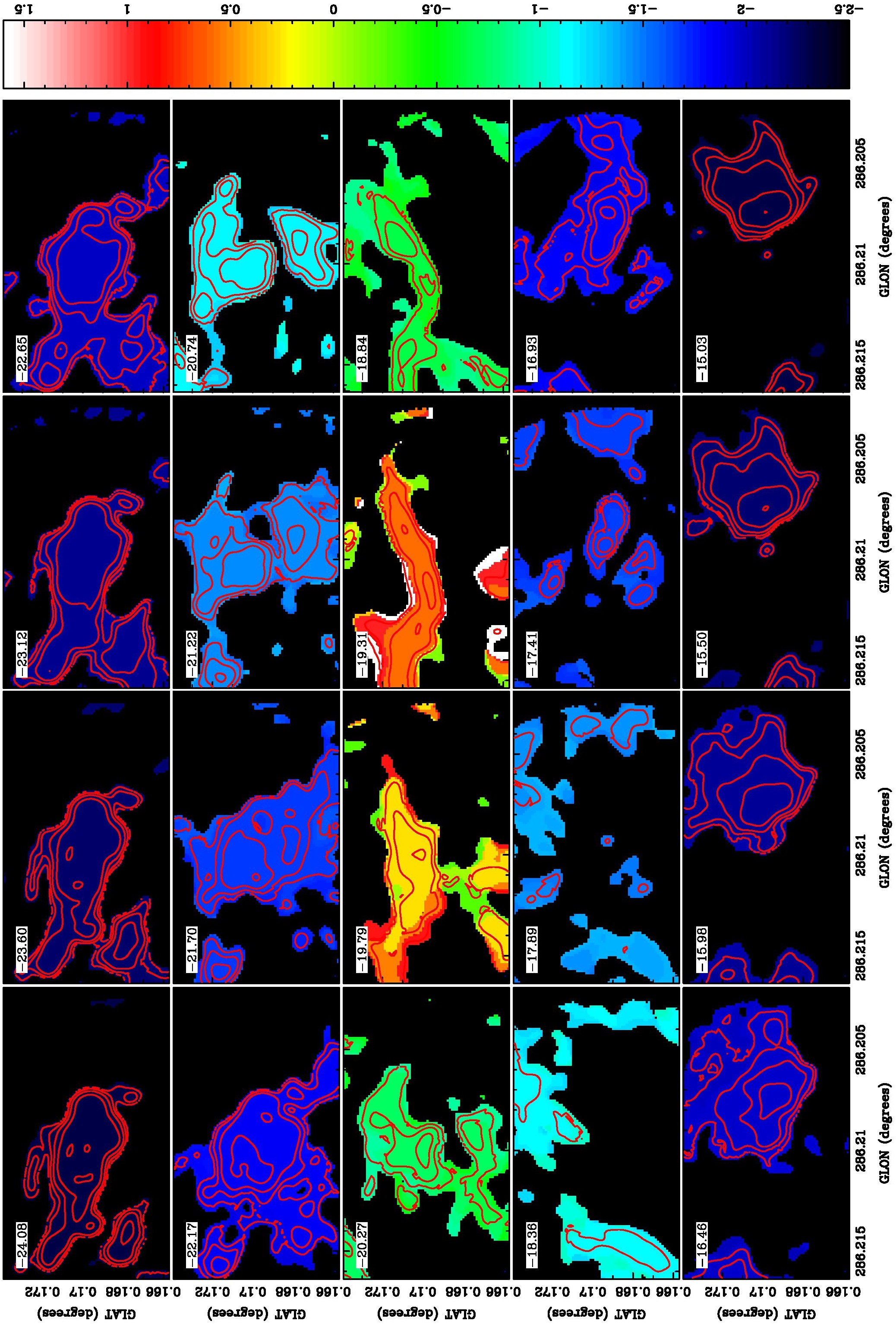}}
\vspace{-1mm}\caption{ 
Logarithm of the ratio of magnetic ${\mathfrak M}$ and kinetic ${\mathfrak K}$ energy densities (colour bar on the right) in the \tco\ outflow wings of BYF\,73.  Each panel is a binned average of 3 channels (0.47\kms\ wide) labelled by their mean \vlsr\ in the top-left corner, and covering both the red \& blue vignettes of Fig.\,\ref{almaCOpolo}.  Red contours in each panel are of the respective binned Stokes $I$ of \tco\ at 2,4,8,16,32\,K.
}\label{MKratio}\vspace{1mm}
\end{figure*}

We can now compute a datacube of the ${\mathfrak M}$/${\mathfrak K}$ ratio,
\begin{equation} 
	\frac{\mathfrak M}{\mathfrak K} = 3.8\%~\frac{n_9\left(\frac{\psi}{s/7\degr}\right)^2}{n_9 (V_{\rm rel}/{\rm km\,s^{-1}})^2}~,
\end{equation}
which turns out to be independent of the density as long as we measure both over the same channels or velocity bins.  Since $\psi$ is part of the density scaling, this simplifies to
\begin{equation} 
	\frac{\mathfrak M}{\mathfrak K} = 3.8\%~\left(\frac{\Delta V/V_{\rm rel}}{s/7\degr}\right)^2~,
\end{equation}
which we can evaluate per native channel of width d$V$ (or any other binning), even while using a larger $\Delta$$V$ to represent the turbulence in the flow.  In other words, the ${\mathfrak M}$/${\mathfrak K}$ ratio can reasonably be estimated with some knowledge of only the gas turbulence $\Delta$$V$ and polarisation dispersion $s$ in the $B$ field directions in each channel at $V_{\rm rel}$, which is ultimately just a restatement of 
the DCF method, per unit volume.

For purposes of illustration, we take $\Delta$$V$ = 3\kms\ and a more conservative $s$ = 13\degr, and present the ${\mathfrak M}$/${\mathfrak K}$ ratio results in {\color{red}Figure \ref{MKratio}} at some representative channels $V_{\rm rel}$.  Different $\Delta$$V$ or $s$ values would obviously scale the ratios as ($\Delta$$V$/$s$)$^2$.  Despite the larger $s$ and smaller ${\mathfrak M}$ in Figure \ref{MKratio} than discussed above, ${\mathfrak M}$/${\mathfrak K}$ peaks at 37, i.e., $\gg$1.  We discuss this further in \S\ref{disc}.

\subsection{The Zeeman effect in CN}\label{zeeman}

As the only observational technique capable of directly measuring $B$ field strengths, the Zeeman effect has been widely utilised over five decades \citep{cru12}.  However, successful detections are notoriously difficult: for extended thermal emission from molecular clouds, only HI, OH, and CN have yielded $B$ field detections, and among these, only CN can provide information on field strengths in dense (\gapp10$^{11}$m$^{-3}$) gas.  Despite considerable effort, there still exist only 14 individual CN Zeeman measurements from a heterogeneous sample of clouds \citep{ftc08}.  But with the advent of full polarisation capability in Cycle 7, anticipation has been high that ALMA might fundamentally change the state of play in this field. 

\begin{figure}[t]
			\includegraphics[angle=-90,scale=0.375]{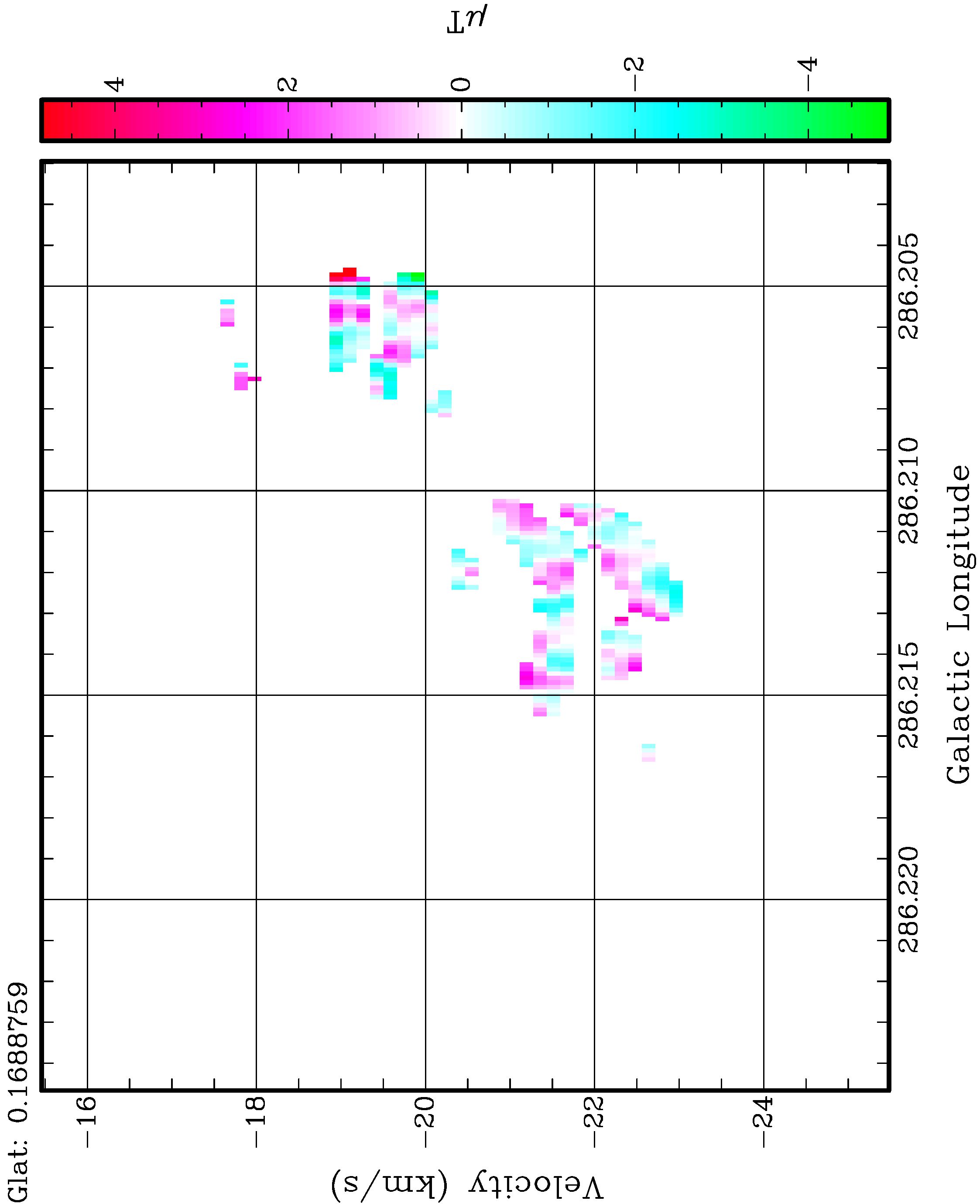} \\
\vspace{-2mm}\includegraphics[angle=-90,scale=0.375]{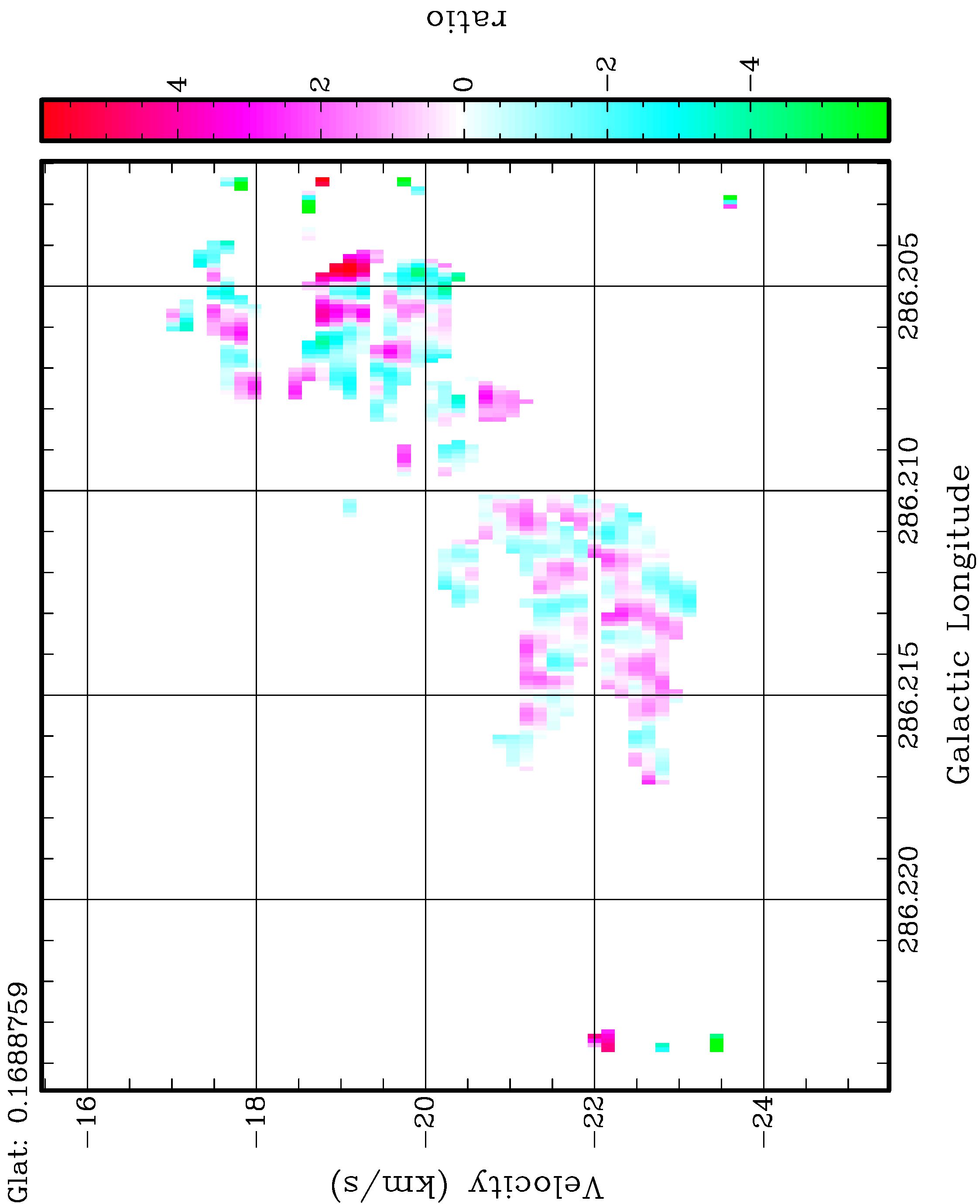}
\caption{ 
({\em Top}) Sample Zeeman calculation of $B_{||}$ at the latitude labelled in the top-left corner, using the {\sc Miriad} task {\sc zeemap} for the brightest CN hyperfine component at 113.490982\,GHz, presented as an LV diagram. ({\em Bottom}) S/N ratio of the data in the top panel.  Note that for Cycle 7, the reliably-calibrated Stokes $V$ data are limited to the inner 10\% of the ALMA primary beam, which encompasses only the area within 286\fdeg214 \gapp\ $l$ \gapp\ 286\fdeg212 ($\sim$6$''$).  Also, to the west (right) of this area, {\sc zeemap} underestimates the noise due to the ALMA primary beam correction, and so the few pixels with apparently larger S/N are actually not.  Effectively, there are no pixels with $B_{||}$ measurements $>$ 3$\sigma$.
}\label{almaZB}\vspace{0mm}
\end{figure}

Unfortunately, despite the very high S/N ($\sim$200) in the ALMA Stokes $I$ data for BYF\,73, covering 8 of the 9 hyperfine transitions of the CN \joz\ line, the $V$ cube shows nothing discernible above the noise.  Computing the ratio of Stokes $V$ to d$I$/d$V$ and scaling this to the Zeeman splitting coefficient of any of the brightest hyperfine transitions \citep[as in Table 1 of][]{ftc08} yields only 3$\sigma$ limits $\sim$1\,$\mu$T (10\,mG), as seen in {\color{red}Figure \ref{almaZB}}.  This is near the upper end of the range of field strengths seen before in dense gas \citep{cru12}, but the noise level would need to be at least halved to obtain reliable measurements even at those levels.  A further issue was the Cycle 7 limit for accurately-calibrated $V$ data lying within the inner 10\% of the primary beam.

This also means we can't use Zeeman data to distinguish between the scenarios (a pure $B_{\perp}$-twist or an additional $B_{||}$ component) put forward to explain the HAWC+ $P'$ null on the western edge of the IBL (\S\ref{hawcIBL}).

While somewhat discouraging, the non-detection may partly be due to the cloud's orientation.  That is, the Zeeman effect can only measure the line-of-sight component $B_{||}$.  The fact that the outflow is viewed close to side-on (\S\ref{outflow}) suggests that most organised structures in the molecular cloud, such as an accretion disk around MIR\,2, would also probably be presented edge-on to us, as might any structures being accelerated away from it, thus possibly maximising $B_{\perp}$ and minimising $B_{||}$.

\section{Discussion}\label{disc}

\subsection{Dynamics: ALMA Reveals the Outflow \\ and Isolates the Inflow}


Based on 40$''$ resolution Mopra \hcop\ maps, \cite{b10} first described a massive infall of dense, cold material within the wider BYF\,73 cloud, without any evidence of an outflow characteristic of lower-mass YSOs.  This suggested an extremely early evolutionary state for a very massive protostar, which seemed to be confirmed by the mid- and far-IR data of P18.  With the higher-resolution SOFIA and ALMA data presented here, particularly the strong bipolar \tco\ outflow, we see that the original appearance of outflow-free, extremely young massive star formation may have been something of a masquerade.\footnote{Inspired by the ALMA results, we re-examined the Mopra \tco\ data \citep{b18} to see if we could tease out hints of the outflow, but still found no clear evidence of the strong red- and blue-shifted emission so easily visible in the ALMA maps.  However, convolving the ALMA data to the Mopra resolution, and adding in the missing short-spacing \tco\ information plus the higher Mopra noise per 40$''$ beam, we found that the outflow became invisible to Mopra at that sensitivity.  So the two instruments' results are consistent, and provide an object lesson against similar masquerades in other sources.}  Nevertheless, through the ALMA \ttco\ data, we are able to discern more specific clues to the configuration of the inflow originally seen in \hcop.

However, the 3D relationship between the Streamer, outflow, and disk or infall as described in \S\ref{cubes} remains puzzling.  The disk can be traced from an outer radius of 0.18\,pc = 36,000\,AU to an inner radius no larger than the limit of the ALMA resolution, 1\farcs8 = 4500\,AU.  Apparently, this disk is close to edge-on based on the sharp velocity gradient across MIR\,2, so the filamentary impression of the Streamer may be an illusion.  For example, we see that both the outflow and rotational/infall patterns are oriented EW, but this arrangement would seem physically counterintuitive.  Undoubtedly, there is some depth to these features in the line of sight, and it is possible that any inflow to the disk might be approaching MIR\,2 from behind its eastern side, even while the blue jet is receding from MIR\,2 on the same side.  Likewise, inflow overlying the western Streamer may be from the front, while the red jet recedes from MIR\,2 as it encounters the \hii\ region.  Separating these features in the line of sight requires only a few $\times$10$^4$\,AU $\sim$ 0.1\,pc, so we consider this scenario reasonable.  Moreover, the Streamer/disk is clearly not flat; there is a measurable width and curvature to its structure.

What are the dimensions of the disk/infall zone centred on MIR\,2?  The modal value of the latitude across all the disk emission, from the LV-1st moment map in Figure \ref{innerdisk} (middle panel), is $b$ = 0\fdeg16997$\pm$0\fdeg00005, only 0\farcs8 = 0.3 ALMA beamwidths north of MIR\,2 itself.  Three-quarters of this emission lies within 0\fdeg169 $<$ $b$ $<$ 0\fdeg172, a span of only 11$''$ or $\sim$4 ALMA beams, strongly suggesting a somewhat narrow structure for the high-$\Delta$$V$ material.  We also computed an LV-2nd moment map (Fig.\,\ref{innerdisk}, bottom panel) to examine the latitude width, confirming that it is indeed thin, from only $\sim$3 ALMA beams = 7$''$ thick to $<$1 beam.  From an inspection of all three LV moment maps, we see that the eastern side of the disk at the higher velocities (\vlsr\ $<$ --25\,\kms) seems to lie mostly at one latitude close to that of MIR\,2, $b$ = 0\fdeg1698$\pm$0\fdeg0005, and so is indeed quite flat in the EW plane to within 1 ALMA beamwidth.  This is across an extent of 0\fdeg01 = 36$''$, for an aspect ratio of 15--20:1 oriented EW.

Given this, it is hard to imagine a disk oriented in the {\em same direction} as the outflow it is supposed to be driving.  This favours the \ttco\ data tracing free-falling material onto a 950\,M\solar\ MIR\,2 within a 36$''$$\times$2$''$ structure, rather than a Keplerian disk, since such a disk ought to be oriented close to NS, parallel to the sharpest velocity gradient across MIR\,2.  If this is indicative, the gradient suggests a disk thickness perhaps \lapp2$''$ or 5000\,AU, but possibly even narrower.  On the other hand, even if we separate the infall from the outflow along our line of sight, it is equally hard to see how a predominantly EW infall (i.e., along a polar direction) produces a disk oriented NS.  So the puzzle persists.

Additionally, if both the MIR absorption and FIR emission mass estimates at MIR\,2's peak position are 5--12$\times$ too small (P18), this is possible evidence for significant grain growth in MIR\,2's protostellar envelope; any free-free emission from MIR\,2 (\S\ref{comps}) would make this discrepancy worse.  P18's gravitational energy release luminosity also scales with the mass, raising it to perhaps 20--33\% of MIR\,2's total luminosity.  Indeed, if future higher-resolution observations revealed an impact radius for the inflow only 5$\times$ smaller at 1000\,AU, this could not only account completely for MIR\,2's brightness via gravitational energy release, but also possibly reveal it to be the first example of a massive ``first hydrostatic core.''

At velocities closer to systemic, the brightest LV emission in the eastern disk, corresponding to the Streamer at 286\fdeg22 $>$ $l$ $>$ 286\fdeg21, also lies very close to this EW plane, although slightly north of it.  However, it is not distributed along any of the Kerplerian curves: instead, its velocity drops linearly from --23\,\kms\ to systemic over its 36$''$ length, with kinematics mimicking that of solid-body rotation.  This portion of the eastern disk is thicker, 5$''$--7$''$, than the high-velocity emission there, $<$3$''$, giving it an aspect ratio around 6:1.  Continuing the non-Keplerian behaviour, east of $l$ = 286\fdeg22 or outside the Streamer's distance from MIR\,2, there appears to be some material in ``super-rotation'' in the bottom-left quadrant of the curves, i.e., with \vlsr\ exceeding the rotational curve for 1350\,M\solar.  This lies at $b$ = 0\fdeg172 (red in the middle panel of Fig.\,\ref{innerdisk}), or 7$''$ north of MIR\,2, but is again about as thin (1 ALMA beam) as the high-velocity disk material.  However, the envelope of this material's super-rotation is moving at close to $\sqrt{2}\times$ this curve, suggesting either free-fall of ambient material towards the Streamer/disk from the rear of the cloud, or that the enclosed mass at this radius has $\sim$doubled.

In contrast, the western side of the disk seems to curve somewhat north of the EW plane of the eastern disk, to a latitude as far as 12$''$ north of MIR\,2 at 0\fdeg173, and at a moderately high velocity (\vlsr\ = --10\,\kms) from systemic.  The rest of the western disk ranges in latitude from MIR\,2's value up to this limit, and the line of maximum velocity in the red-shifted wing map (Fig.\,\ref{COkep}, right panel) is clearly curved to the north-west from MIR\,2.  The western disk's thickness (Fig.\,\ref{innerdisk}) is also broader than for the eastern disk overall, up to 7$''$, but is also thin ($<$3$''$) in many places, even where it curves to the NW.  It is worth noticing that the solid-body portion of the Streamer/eastern disk seems to continue part-way (0\fdeg006 = 22$''$) into the western disk in both the 1st- and 2nd-LV-moment maps, but then seems to reverse bluntly back to MIR\,2's longitude at \vlsr\ = --16\,\kms, while thickening to a width of almost 10$''$ just west of MIR\,2, almost as if the Streamer's infall (if that's what it is) were being deflected from the EW plane by some obstacle west of MIR\,2.


What of the counter-rotating parts of the LV diagrams (i.e., the ``empty'' top-left and bottom-right quadrants of the rotation curves)?  Much of this emission, especially the brighter portions thereof, lie north ($b$ $>$ 0\fdeg172, magenta) or south ($b$ $<$ 0\fdeg168, black) of the disk, and outside the longitude range of the IBL (286\fdeg216 $>$ $l$ $>$ 286\fdeg203): they appear to be associated with other internal structures of the cloud, supporting the rotational interpretation for the inner parts of the Streamer.

While MIR\,2's mass seems dynamically dominant within the IBL, the mass in the Streamer/disk must nevertheless also be significant.  In \S\ref{HROstuff} 
we find a mean column density 3$\times$10$^{27}$\,m$^{-2}$ $\approx$ 6000\,M\solar/pc$^2$ along the Streamer, or roughly 15\,M\solar\ per 4$''$ box, assuming there is not the same mass deficit/degree of grain growth as in the MIR\,2 core.  A rough total along the full 72$''$ length and 8$''$ width of the streamer is then perhaps 500\,M\solar.  This would explain why MIR\,2's gravitational influence seems to drop beyond the outer Keplerian radius determined above, since there the gas mass of the broader cloud starts rivalling MIR\,2's effects.

In such a disk, the 0.034\,M\solar/yr mass accretion rate determined by \cite{b10}, still a record as far as we know, can be supported by a merely 0.01\%/yr ``leakage'' of mass through the disk onto MIR\,2's core, or alternatively, that the Streamer can supply this accretion rate for another 10$^4$\,yr.  On the other hand, the time required to build up the more massive MIR\,2 core at this accretion rate is closer to 40,000\,yr instead of the 7,000\,yr estimated by P18, assuming that \cite{b10}'s accretion rate is correct.  Without detailed modelling, we cannot refine the accretion rate value here beyond an approximate calculation below, but the true rate seems unlikely to be too much less than this, with such a massive reservoir available for accumulation.

As one example, consider the velocity difference between the brightest disk material within the bottom-left quadrant of the rotation curves, and the 1350\,M\solar\ curve itself, as seen in Figure \ref{innerdisk} between 286\fdeg217 \gapp $l$ \gapp 286\fdeg208 and at latitudes $\sim$4$''$ north of MIR\,2.  This difference runs from 0\,\kms\ at the eastern end of this window to $\sim$+10\,\kms\ at MIR\,2, in the sense of being ``sub-rotational'' in our line of sight.  If we suppose that the rotational motion here is being translated into proper motions inward to MIR\,2, the effective accretion speed can be taken (very approximately!) as $V_{\rm accr}$ $\sim$ 5\,\kms\ = 5\,pc/Myr.  The emission along this feature (which covers perhaps half of the main part of the 8$''$-wide Streamer) averages $\sim$10\,K/ch in brightness, or conservatively, $\sim$20\,K\kms\ integrated.

Using a simple conversion factor $X$(\ttco) = 10$^{26}$\,m$^{-2}$ (K\kms)$^{-1}$ \citep[probably an underestimate;][]{b18}, the column density in this feature alone runs around $N_{\rm accr}$ $\sim$ 2$\times$10$^{27}$\,m$^{-2}$ = 4000\,M\solar/pc$^{2}$, or perhaps $\frac{2}{3}$ of the Streamer's total column density as seen in Figure \ref{almaNH2-B}.  This translates to a linear density $\Lambda_{\rm accr}$ $\sim$ 400\,M\solar/pc within the 0.1\,pc width of the presumed accretion stream.  Therefore we have a mass flux in this accretion stream of $\dot{M}_{\rm accr}$ =  $\Lambda_{\rm accr}$$V_{\rm accr}$ $\sim$ 2$\times$10$^{-3}$\,M\solar/yr.

This very rough estimate is still on the large side compared to other massive protostars \citep{r13}, but smaller than \cite{b10}'s rate of 0.034\,M\solar/yr.  The true value is likely a multiple of the above example, however, due to several factors: our conservative starting column density, other accretion flows such as the western side of the Streamer, the super-rotating material, higher density streams traced better by \ceto\ or CN, 2$\times$ faster flow closer to MIR\,2, and a probably 5$\times$ larger effective $X$(\ttco).  Thus, the \cite{b10} rate may still be a reasonable global estimate.


In summary, the complex yet potentially understandable structure of the Streamer near MIR\,2, and Keplerian disk/freefalling infall zone around it, may have much to tell us about heavy mass accretion onto a massive protostar.  Clearly, MIR\,2 is an exceptional and exciting object that demands further study.

\subsection{Magnetic Fields: Driving the Outflow?}\label{Bdriver}

The \tco\ outflow from MIR\,2 is fairly massive: with a typical line brightness \ico\,$\sim$\,10--30\,K per 0.16\kms\ ALMA channel or average integrated intensity $\sim$200\,K \kms, we can use Eq.\,(9) or its ilk \citep{b18} to estimate gas column densities of  
about 8$\times$10$^{25}$\,m$^{-2}$ or 160\,M\solar\,pc$^{-2}$ in the outflow.  Inside the flow dimensions of roughly 1$\times$0.1\,pc, this gives a total outflow mass of perhaps 16\,M\solar.  This mass is being driven to speeds of 10s of \kms, so the kinetic energy of the flow is similarly large, 
about 1.6$\times$10$^{39}$\,J.  If this emerges over timescales of 10s of kyr, then the mass outflow rate is $\dot{M}_{\rm out}$ = $\Lambda_{\rm out}$$V_{\rm out}$ $\sim$ 5$\times$10$^{-4}$\,M\solar/yr \citep[or $\sim$10\% of the infall rate as estimated above, similar to other outflows;][]{pr19} and the mechanical luminosity of the outflow $L_{\rm out}$ $\sim$ 4\,L\solar.  Could this mechanical power be imparted by $B$ fields?

From Eq.\,(15) and Figure \ref{MKratio}, we see that the energy density in the $B$ field is typically well below the kinetic energy density in the higher-velocity and lower-density gas: 
the $B$ field is therefore likely a passenger in the flow at these points.  On the other hand, ${\mathfrak M}$ may rival or even exceed ${\mathfrak K}$ where $V_{\rm rel}$ is small.  This result, however, should be taken as merely suggestive, since ${\mathfrak M}$/${\mathfrak K}$ $>$ 1 only in the 20 lowest-$V_{\rm rel}$ channels, where the \tco\ opacity is still high and we may not be mapping much of the outflow via \tco.  But $B$ fields do seem to be detected throughout the outflow, at a few $\times$ 10\,nT.  It seems reasonable to suppose that similar $B$ fields (at least!) should exist close to BYF\,73's $V_{\rm sys}$, and specifically close to the base of the flow at MIR\,2.  If this were true, the $B$ field would at least have the potential of being energetically important. 

As such, this is circumstantial yet valuable evidence that the $B$ field may be intimately involved in driving, or at least shaping, the outflow.  \citet{pr19} showcase some other recent observational results making this $B$ field connection to the outflow, typically on $\sim$100\,AU (i.e., disk) scales.  As far as we are aware, this is the first instance where the structure of the whole molecular outflow might at least partially be attributable to the $B$ field at its origin.  The connection is not always clear, however: a rare case where the $B$ field seems to play an important role in massive star formation is the compact \hii\ region K3-50, where a strong {\em ionised} outflow emanates from a high-mass protostellar object, surrounded by a Keplerian disk extending over radii 0.1--0.7\,pc \citep{h97}.  There, the $B$ field inferred at the inner edge of the disk seems to be strong enough to provide support against gravity; however, even in K3-50, the $B$ field does not seem strong enough to influence the outflow \citep{BL15}.

Our results for BYF\,73 seem to provide additional observational support for the picture of magneto-centrifugally powered protostellar outflows \citep{sn94,op97,t11}, as opposed to the main competing model of turbulent entrainment of gas from a bipolar jet \citep[e.g.,][]{r93}.  Recent numerical work on solar coronal mass ejections \citep{j21} may even supply a specific mechanism for the high speeds in such outflows, namely sudden magnetic reconnection in bipolar loops, presumably anchored in the inner parts of a protostellar accretion disk. 
It is tempting to suppose such reconnecting loops drive the vigorous outflows widely seen in other star-forming clouds, as may be happening here with MIR\,2/BYF\,73.  Future work in this area, supported by ALMA+SOFIA observations, should be very interesting.

\subsection{Magnetic Criticality}\label{critical} 
We briefly also note that the critical $N$, $n$, and $B_{\perp}$ values derived from HRO analysis of the SOFIA+ALMA data (\S\ref{HROstuff}) place BYF\,73 just below the maximum $B_{||}$ trend line in \citet{cru12}'s $n$ vs $B$ summary plot (his Fig.\,6), nicely among other dense clouds' CN-Zeeman measurements.  In contrast, BYF\,73 is slightly above the line of criticality in \citet{cru12}'s $N$ vs $B_{||}$ plot (his Fig.\,7), on the side of being subcritical and a little above its supercritical counterparts in this regard.  Given the uncertainties in our results, however, this may not be terribly significant.  Further, BYF\,73 is not the most extreme strong-$B$ outlier compared to the line of criticality, but it may be the highest column density subcritical cloud.  As \citet{cru12} points out, this would be unusual in the sense of at least supporting the possibility of ambipolar diffusion playing an important role in cloud stability \citep{mc99}.  However, our HRO result is not the same as a direct Zeeman strength measurement, and it should probably not yet be overinterpreted without some confirming evidence.

Nevertheless, it is tempting to wonder what higher-resolution and -sensitivity observations might reveal about the material closer in to MIR\,2, where the infall might be more clearly imaged, the outflow may be driven, and its originating disk might be discerned.

\section{Conclusions}\label{concl}

We have presented a range of new observational data exploring details of the massive molecular clump BYF\,73, previously thought to harbour a massive (240\,M\solar), very young (7,000\,yr), Class 0 protostar (MIR\,2) with the largest mass inflow rate (0.034\,M\solar/yr) observed to date.  The new data include far-IR (SOFIA/HAWC+) and 3mm (ALMA) continuum emission, mm-wave spectroscopy of several molecular species, and polarisation maps in both the continuum and spectral lines from both facilities.  The polarisation data in particular have been analysed in order to learn about the structure, strength, role, and significance of the $B$ field in this cloud (as summarised in {\color{red}Table \ref{summtable}}), and the continuum and spectral line data were analysed and interpreted in this context.  Our results include the following.

\textbullet\ The 14$''$ resolution HAWC+ data show a centrally concentrated cloud with generally low polarised emission (a few percent) from the central 0.5\,pc of the molecular clump, but at a relatively high polarisation (10--20\%) extended across the adjacent, 2\,pc wide, low-power \hii\ region.  The polarisation structure east of MIR\,2 shows a second, distinct feature in the form of an arc, the eastern polarisation loop (EPL); there is also a clear, very-low to zero-polarisation boundary layer (IBL) around MIR\,2 and the EPL.

\textbullet\ The 2\farcs5 resolution ALMA continuum data show four main features: a narrow, massive EW Streamer of cold, dense gas; a fainter, NS line of emission coincident with the ionisation front (IF) facing the \hii\ region; another faint spur of emission aligned with the EPL; and a small number (5) of 3mm point sources, of which MIR\,2 is by far the brightest.  These 3mm point sources are far fewer than the number seen at near- or mid-IR wavelengths, suggesting that many of the latter may be relatively low-extinction and/or more evolved objects.  The polarised 3mm emission comes from parts of the Streamer, IF, and EPL; it is somewhat patchy but also mainly oriented EW, switching to a NS orientation across MIR\,2, with very high (20--40\%) fractional polarisation in most locations.

\textbullet\ The ALMA \tco\ Stokes $I$ cube reveals a prominent, powerful, bipolar outflow from MIR\,2, extending over a velocity range almost $\pm$40\kms\ from the cloud's $V_{\rm sys}$.  Both the 0.4\,pc red and 0.6\,pc blue wings of this outflow appear to be deflected from their starting vectors by inertially significant parts of the Streamer.  The EPL may be a result of this deflection in the blue wing of the outflow.

\textbullet\ The wider ALMA \ttco, \ceto, and CN mosaics reveal much more extended emission across the cloud than in the 3mm continuum, with only modest structural or kinematic correspondence to the Streamer, IF, or point sources.  These suggest that the wider cloud is somewhat porous to the UV radiation from the adjacent \hii\ region.  The outflow can, however, be traced closer in to MIR\,2 within the \ttco\ cube.

\textbullet\ In the same area, the \ttco\ also shows clear evidence for material in either Keplerian rotation about, or free-fall onto, MIR\,2; the apparent axial geometry of this material, however, is puzzling.  If Keplerian, it implies a gravitating mass 1350$\pm$50\,M\solar\ within 1\farcs8 = 4500\,AU of MIR\,2 and any envelope; if freely infalling, the implied mass within that radius is 950$\pm$35\,M\solar.  These masses are \gapp5$\times$ larger than from SED fitting, suggesting possibly significant grain growth has occurred in MIR\,2.  The larger mass in a small radius also suggests up to 33\% of MIR\,2's luminosity could be powered by gravitational energy release.  In light of these higher-resolution and -sensitivity data, the prior mass infall rate is found to be reasonable; however with a 5$\times$ larger mass, MIR\,2's age may be more like 40,000\,yr.

\textbullet\ Davis-Chandrasekhar-Fermi (DCF) analysis of the continuum polarisation data suggest relatively strong $B$ fields are present 
in the gas near MIR\,2: 92\,nT at the Mopra scale $\approx$ 2$\times$the HAWC+ scale, and 1.18\,$\mu$T at the ALMA scale, the latter a possible record in cold, non-masering molecular gas.  Despite these high values, they are nominally consistent with critical balance between $B$ fields and gravity.  With the higher central mass for MIR\,2 indicated by the Keplerian pattern in the \ttco\ data, the gas is supercritical in these areas.  In the \hii\ region, the DCF estimate is 21\,nT, also somewhat stronger than typical in such gas, but where the ionised flow still dominates the energetics.

\begin{deluxetable}{lccccc}
\tabletypesize{} 
\tablecaption{ 
Summary of Magnetic Field Results in BYF\,73\label{summtable}}\vspace{-5mm}\tablewidth{0pt}
\tablehead{
\colhead{Structure} & \colhead{Facility} & \colhead{Method} & \colhead{$B$} &\colhead{$n$} \\
\colhead{ } & \colhead{ } & \colhead{ } & \colhead{(nT)} & \colhead{(m$^{-3}$)} \vspace{-3mm} \\
}
\startdata
\hii-N,ions\tablenotemark{a} & HAWC+ & DCF\tablenotemark{c} & 17.5 & 1.4e8 \\ 
\hii-N,dust\tablenotemark{b} & HAWC+ & DCF				& 162 & 7e9 \\ 
\hii-S,ions\tablenotemark{a}& HAWC+ & DCF				& 8.8 & 1.4e8 \\ 
\hii-S,dust\tablenotemark{b}& HAWC+ & DCF				& 81 & 7e9 \\ 
MIR\,2 core			& HAWC+ & DCF				& 77 & 4e11 \\ 
Streamer-W			& ALMA & DCF					& 92 & 8e11 \\ 
MIR\,2 core			& ALMA & DCF					& 740 & 3.6e13 \\ 
MIR\,2 extn.			& ALMA & DCF					& 330 & 1e12 \\ 
EPL					& ALMA & DCF					& 49 & 3e11 \\ 
Streamer				& ALMA & DCF					& 106 & 5e11 \\ 
Streamer				& HAWC+ & HRO  & 25$\pm$6 & 1e11 \\ 
Streamer				& ALMA & HRO					& 61$\pm$27 & 3e11 \\
Streamer			& HAWC+ALMA & HRO				& 42$\pm$7 & 2e11 \vspace{-2.9mm} \\
\enddata
\tablenotetext{a}{Using a mean molecular weight $\mu$ = 1.28 for ionised gas.}
\tablenotetext{b}{Using a mean molecular weight $\mu$ = 2.35 for molecular gas.}
\tablenotetext{c}{DCF $B$ field values (Eq.\,(1)) are scaled to the local dispersion $s$ (e.g., Table \ref{corrstats}) with uncertainties $\sim$ $\pm$30\%.}
\vspace*{1mm}
\end{deluxetable}

\textbullet\ Histogram of relative orientations (HRO) analysis gives a sharper estimate of where the $B$ field might reach criticality in the gas.  In the Streamer, we obtain thresholds for criticality of $B_{\rm crit}$ = 42$\pm$7\,nT at log($N_{\rm crit}$/m$^2$) = 26.74$\pm$0.09 or log($n_{\rm crit}$/m$^{3}$) = 11.31$\pm$ 0.09, where $B$ likely dominates gravity and helps organise the gas structures below these thresholds, and gravity likely dominates above them.

\textbullet\ The \tco\ polarisation cubes reveal the presence of the Goldreich-Kylafis effect almost everywhere in the outflow.  The orientation of the $B$ field is seen to lie closely along the outflow direction, consistent with prior studies but in a far more widespread manner than seen before.  A simplified DCF analysis of the \tco\ emission in each channel shows that, for most of the outflow, the $B$ field does not dominate the kinetic energy of the flow.  However, the two energy densities may be comparable at the lowest outflow velocities, where the $B$ field may even drive the flow close to MIR\,2.

\textbullet\ Despite a peak S/N of 200 in Stokes $I$, the ALMA CN polarisation data detects no Zeeman effect above the noise in the Stokes $V$ cube, with a 3$\sigma$ limit of 1\,$\mu$T.  This may partly be attributable to the outflow's predominant orientation across our line of sight, perhaps organising other cloud structures in a similar direction, and minimising $B_{||}$.

These results suggest that even higher resolution and/or sensitivity data on BYF\,73 and MIR\,2 would produce exciting constraints on early stages of massive star formation. \vspace{-2mm}


\acknowledgments
We thank the SOFIA crew and scientific staff, and the ALMA-North America staff, for outstanding support of their respective facilities.  We also thank the anonymous referee for a careful reading of the manuscript and many helpful suggestions which improved the presentation of the paper.  PJB gratefully acknowledges financial support for this work provided by NASA through awards SOF 07-0089 and 09-0048 issued by USRA.  Based in part on observations made with the NASA/DLR Stratospheric Observatory for Infrared Astronomy (SOFIA).  SOFIA is jointly operated by the Universities Space Research Association, Inc. (USRA), under NASA contract NNA17BF53C, and the Deutsches SOFIA Institut (DSI) under DLR contract 50 OK 2002 to the University of Stuttgart.  This paper makes use of the following ALMA data: ADS/JAO.ALMA.2019.1.01031.S.  ALMA is a partnership of ESO (representing its member states), NSF (USA) and NINS (Japan), together with NRC (Canada), MOST and ASIAA (Taiwan), and KASI (Republic of Korea), in cooperation with the Republic of Chile.  The Joint ALMA Observatory is operated by ESO, AUI/NRAO and NAOJ.  The National Radio Astronomy Observatory is a facility of the National Science Foundation operated under cooperative agreement by Associated Universities, Inc.

Facilities: \facility{SOFIA(HAWC+), ALMA}.




\end{document}